\documentclass[preprint,12pt]{elsarticle}
\usepackage{amssymb}
\usepackage{physics}
\usepackage{float}
\usepackage{amsmath}
\usepackage{enumerate}
\usepackage{caption}
\usepackage{subcaption}
\usepackage[colorlinks,linkcolor=red,anchorcolor=blue,citecolor=green]{hyperref}
\usepackage{geometry}
\usepackage{mathrsfs}

\newcommand*\Laplace{\mathop{}\!\mathbin\bigtriangleup}

\journal{Elsevier}

\begin{document}

\begin{frontmatter}

\title{Unified Gas-Kinetic Wave-Particle Method for Multiscale Flow Simulation of Partially Ionized Plasma}

\author[a]{Zhigang PU}
\ead{zpuac@connect.ust.hk}
\author[a,b]{Kun XU\corref{cor1}}
\ead{makxu@ust.hk}
\cortext[cor1]{Cooresponding author}

\address[a]{Department of Mathematics, Hong Kong University of Science and Technology, Clear Water Bay, Kowloon, Hong Kong, China}
\address[b]{Shenzhen Research Institute, Hong Kong University of Science and Technology, Shenzhen, China}

\begin{abstract}
    The Unified Gas-Kinetic Wave-Particle (UGKWP) method is constructed for partially ionized plasma (PIP). This method possesses both multiscale and unified  preserving (UP) properties. The multiscale property allows the method to capture a wide range of plasma physics, from the particle transport in the kinetic regime to the two-fluid and magnetohydrodynamics (MHD) in the near continuum regimes, with the variation of local cell Knudsen number and normalized Larmor radius.
    The unified preserving property ensures that the numerical time step is not limited by the particle collision time in the continuum regime for the capturing of  dissipative macroscopic solutions of the  resistivity, Hall-effect, and all the way to the ideal MHD equations.
    The UGKWP is clearly distinguishable from the classical single scale Particle-in-Cell/Monte Carlo Collision (PIC/MCC) methods.
    The UGKWP method combines the evolution of microscopic velocity distribution with the evolution of macroscopic mean field quantities, granting it UP properties. Moreover, the time step in UGKWP is not constrained by the plasma cyclotron period through the Crank-Nicolson scheme for fluid and electromagnetic field interactions. The momentum and energy exchange between different species is approximated by the Andries-Aoki-Perthame (AAP) model. Overall, the UGKWP method enables a smooth transition from the PIC method in the rarefied regime to the MHD solvers in the continuum regime. This method has been extensively tested on a variety of phenomena ranging from kinetic Landau damping to the macroscopic flow problems, such as the Brio-Wu shock tube, Orszag-Tang vortex, and Geospace Environmental Modeling (GEM) magnetic reconnection. These tests demonstrate that the proposed method can capture the fundamental features of PIP across different scales seamlessly.
\end{abstract}


\end{frontmatter}


\section{Introduction}
\label{introduction}

Partially ionized plasmas (PIP) represent a fundamental and pervasive form of matter that plays a crucial role in shaping our understanding of the universe and driving technological advancements. The importance of PIP in science and engineering cannot be overstated, as it bridges the gap between neutral gases and fully ionized plasmas, exhibiting unique properties that are essential in various fields. In the realm of astrophysics, PIP is a key component in understanding cosmic phenomena in the solar chromosphere, Earth's ionosphere, molecular clouds, and various interstellar mediums \cite{ballester2018partially,braileanu2019two,khomenko2018three,soler2022theory,kuzma2020numerical,ballai2019linear,ballester2020nonlinear}. The unique physical behaviors exhibited by PIP in these astrophysical contexts, such as the decay of Alfvén waves \cite{balsara1996wave}, cut-off modes \cite{ballester2018partially,wojcik2018acoustic,alharbi2022waves}, and Biermann battery effects \cite{martinez2021simulations}, are essential for developing accurate models of stellar and galactic processes. In the domain of low-temperature plasmas, PIP is at the forefront of numerous technological applications, such as microelectronic fabrication \cite{song2014control,zhang2014pic,yu2021numerical}, material processing, medical treatment \cite{park2019electron,laroussi2020cold}, water purification, and so on. The non-equilibrium nature of low-temperature PIP, where electrons are significantly hotter than ions and neutral particles, gives rise to complex kinetic effects. This characteristic is crucial for many applications and requires sophisticated modeling and diagnostic techniques to fully understand and exploit. In aerospace engineering, PIP plays a pivotal role in advancing space exploration and aeronautics like plasma-based flow control, ion thrusters, interplanetary reentry, and so on \cite{shang2016computational,lefevre2022magnetohydrodynamic,xie2023experimental,otsu2004reentry,katsurayama2011particle}.

The plasma physics community commonly employs two distinct methodologies for solving partially ionized plasma systems: Riemann-solver-based MHD solvers and PIC-based kinetic solvers. The fundamental distinction between these approaches lies in their ability to resolve physics at macroscopic and microscopic scales, respectively.
MHD solvers treat both plasma and neutral species as continuous fluids, utilizing sophisticated Riemann solvers such as Roe \cite{roe1981approximate} or HLL (Harten-Lax-van Leer) \cite{harten1983upstream} to solve the system equations. In contrast, the PIC method represents the distribution function through discrete computational particles, tracking their individual trajectories through phase space. PIC solvers typically incorporate the Monte Carlo Collision method to model inter-particle collisional interactions accurately.
While PIC approaches can provide high-fidelity predictions from the mean-free-path scale upwards, they incur substantial computational costs due to the necessity of tracking individual particles within a Lagrangian framework. This computational intensity becomes particularly pronounced in near-continuum flow regimes. Conversely, MHD solvers operate under the assumption of near-equilibrium distributions for all species, enabling the application of fluid models. This assumption significantly enhances computational efficiency compared to PIC methods. However, the reduction of the distribution function's degrees of freedom to a limited set of variables in fluid models results in an inability to capture critical kinetic phenomena.
The PIC method's strength lies in its capacity to resolve microscopic kinetic effects with high accuracy, while MHD solvers excel in efficiently modeling macroscopic plasma behavior. Given the limitations of each approach, there is a pressing need for the development of a multiscale method. Such a method would ideally leverage the respective advantages of both fluid and kinetic solvers, enabling efficient and accurate simulations across a broad range of spatio-temporal scales in partially ionized plasma systems.

Recent advancements have seen a systematic development of the UGKWP method for modeling multiscale flows, spanning from the rarefied to the continuum regime \cite{xu2001gas,xu2010unified,liu2021unified}.
The UGKWP method is founded on the integral solution of relaxation models, such as the Bhatnagar–Gross–Krook (BGK) model, and employs a physically grounded decomposition of the numerical flux across cell interfaces into two distinct components: the equilibrium wave flux and the non-equilibrium particle flux.
This formulation allows for the deterministic computation of the equilibrium component, circumventing the need for discrete simulation particles in regions where continuum assumptions hold. This feature significantly enhances computational efficiency while maintaining accuracy across diverse flow regimes.
The multiscale formulation in UGKWP has demonstrated its versatility across a range of transport problems, including gas-mixture dynamics, radiative transfer, phonon transport, plasma physics, and granular flow systems \cite{yang2021unified, li2020unified, liu2021unified,li2024unified}. Its application to fully ionized plasma has successfully resolved the plasma’s complex multiscale flow regime \cite{liu2021unified}. In the continuum regime, UGKWP captures ideal MHD, MHD with dissipative terms, and Two-Fluid Models.As the flow transitions to the rarefied regime, UGKWP seamlessly evolves into a particle-based method, effectively characterizing the kinetic behavior of plasma. This smooth transition is facilitated by the method's ability to dynamically adjust the balance between wave and particle components based on the local Knudsen number, a dimensionless parameter quantifying the degree of rarefaction.

The present research aims to extend the previous work of utilizing the Gas-Kinetic Scheme (GKS) to simulate PIP in the continuum regime \cite{pu2024gas} to the whole flow regime by the UGKWP method. These systems exhibit inherently greater complexity compared to their fully ionized counterparts and are more frequently encountered in both scientific investigations and engineering applications.
Our approach begins with the construction of a comprehensive kinetic model for electrons, ions, and neutrals, based on the BGK-Maxwell system and AAP models. Through asymptotic analysis, we demonstrate that this model reduces to multifluid equations when the characteristic timescale exceeds the collision time for each species—i.e. when the frequency is significantly lower than the collision frequency of every species. As the timescale further increases, depending on the collision frequency between charged particles and neutrals, the system may further simplify to either a single-fluid MHD system or an Euler-MHD system. Various non-ideal effects, including the Hall effect and resistivity, can be reproduced by the system. Furthermore, the ambipolar effect, a characteristic phenomenon in weakly ionized plasmas, is introduced.
The UGKWP method is then applied to solve the fluid flow within this complex system. Based on the integral solution of the BGK equation, the numerical flux can be decomposed into wave and particle components. The evolution of the electromagnetic field is computed using a wave-propagating-based finite-volume scheme, while the interaction between fluid and field is resolved through the Crank-Nicolson method. Cross-species collision source terms are addressed using an operator-splitting approach.
The current method is capable of capturing plasma physics across a wide range of Knudsen numbers and Larmor radii, representing varying degrees of rarefaction and magnetization. This multiscale capability enables the simulation of diverse plasma regimes within a unified framework. In the continuum flow regime, the UGKWP method has the unified preserving (UP) properties for capturing the solution of dissipative MHD equations without enforcing the time step being less than the particle collision time \cite{guo2023unified}.
To validate and demonstrate the efficacy of our numerical method, we employ a suite of rigorous numerical tests, including Landau damping, Brio-Wu shock tube, Orszag-Tang vortex, and Geospace Environmental Modeling (GEM) magnetic reconnection problem. These tests demonstrate the method's performance across a spectrum of plasma phenomena, from fundamental wave-particle interactions to complex magnetohydrodynamic processes.

This paper is structured as follows: Section \ref{model} introduces the kinetic model for PIP and analyzes its asymptotic behaviors. Section \ref{numerical} presents the detailed numerical methods employed in this study. Section \ref{results} shows the numerical tests and their outcomes. Finally, Section \ref{conclusions} offers comprehensive conclusions.

\section{Kinetic model and asymptotic behavior}
\label{model}

\subsection{BGK-Maxwell kinetic model}

For partially ionized plasma, the kinetic equation can be written as \cite{liu2017unified}:

\begin{equation}
    \begin{aligned}
    & \frac{\partial f_\alpha}{\partial t}+\boldsymbol{u}_\alpha \cdot \nabla_x f_\alpha+\boldsymbol{a}_\alpha \cdot \nabla_u f_\alpha= Q_{\alpha}, \\
    & \frac{\partial \boldsymbol{B}}{\partial t}+\nabla_x \times \boldsymbol{E}=0, \\
    & \frac{\partial \boldsymbol{E}}{\partial t}-\boldsymbol{c}^2 \nabla_x \times \boldsymbol{B}=-\frac{1}{\epsilon_0} \boldsymbol{J}, \\
    &\nabla_{\boldsymbol{x}} \cdot \boldsymbol{E}=\frac{q}{\epsilon_0},\\
    &\nabla_{\boldsymbol{x}} \cdot \boldsymbol{B}=0,
    \end{aligned}
\end{equation}
where $f_\alpha = f_\alpha(t, \boldsymbol{x}, \boldsymbol{u})$ is the distribution function for species $\alpha$ ( $\alpha=i$ for ion and $\alpha=e$ for electron, $\alpha=n$ for neutral) at space and time $(\boldsymbol{x},t)$ and microscopic translational velocity $\boldsymbol{u}$. $\boldsymbol{a}_\alpha$ is the Lorenz acceleration taking the form
$$
\boldsymbol{a}_\alpha=\frac{q_\alpha(\boldsymbol{E}+\boldsymbol{u}_\alpha \times \boldsymbol{B})}{m_\alpha}.
$$

For neutral species, $q_n=0$, thus $\boldsymbol{a}_n=0$. $Q_{\alpha}=\sum_{k=1}^{m}Q_{\alpha k}(f_\alpha, f_k)$ is the collision operator of species $\alpha$ between species $k$, where $m$ is total number of species in the system. In this work, $m=3$. In the Maxwell equations,  $\boldsymbol{E}$ and $\boldsymbol{B}$ are the electric field strength and magnetic induction, $\boldsymbol{c}$ is the speed of light, and $\epsilon_0$ is the vacuum permittivity. $n_\alpha$ is number density of species $\alpha$. In this work, the charged species in the system are just protons and electrons, then the electric current is $\boldsymbol{J}=e\left(n_i \boldsymbol{U}_i-n_e \boldsymbol{U}_e\right)$ and the charge density is $q=e(n_i-n_e)$, where $\boldsymbol{U}$ is macroscopic velocity and $e$ is the charge of a proton.

The collision term between multiple species is modeled by the relaxation model by Andries, Aoki, and Perthanme \cite{andries2002consistent}, which is,
\begin{equation*}
    Q_\alpha=\frac{g_\alpha^M-f_\alpha}{\tau_\alpha},
\end{equation*}
where $g_\alpha^{M}$ is a Maxwellian distribution,
\begin{equation*}
    g_\alpha^M=\rho_\alpha\left(\frac{m_\alpha}{2 \pi k T_\alpha^*}\right)^{3 / 2} \exp \left(-\frac{m_\alpha}{2 k_B T_\alpha^*}\left(\boldsymbol{u}_\alpha-\boldsymbol{U}_\alpha^*\right)^2\right),
\end{equation*}
and post-collision temperature and velocity are chosen as:
\begin{equation}
    \begin{aligned}
    \boldsymbol{U}_\alpha^* & =\boldsymbol{U}_\alpha+\frac{\tau_\alpha}{m_\alpha} \sum_{k=1}^N 2 \mu_{\alpha k} \chi_{\alpha k} n_k\left(\boldsymbol{U}_k-\boldsymbol{U}_\alpha\right), \\
    T_\alpha^* & =T_\alpha-\frac{m_\alpha}{3 k_B}\left(\boldsymbol{U}_\alpha^*-\boldsymbol{U}_\alpha\right)^2+\tau_\alpha \sum_{k=1}^N \frac{4 \mu_{\alpha k} \chi_{\alpha k} n_k}{m_\alpha+m_k}\left(T_k-T_\alpha+\frac{m_k}{3 k_B}\left(\boldsymbol{U}_k-\boldsymbol{U}_\alpha\right)^2\right),
    \end{aligned}
    \label{eq:aap U and E}
\end{equation}
where $\mu_{\alpha k} = m_{\alpha}m_k/(m_{\alpha}+m_k)$ is reduced mass, mean relaxation time $\tau_\alpha$ is determined by $1 / \tau_\alpha=\sum_{k=1}^m \chi_{\alpha k} n_{k} $, and interaction coefficient $\chi_{\alpha k}$ for hard sphere model is \cite{morse1963energy}:
\begin{equation*}
    \chi_{\alpha k}= \frac{4 \sqrt{\pi}}{3}\left(\frac{2 k_B T_{\alpha}}{m_\alpha}+\frac{2 k_B T_{k}}{m_k}\right)^{1 / 2}\left(\frac{d_\alpha+d_k}{2}\right)^2 .
\end{equation*}
In this above formula, $ d_\alpha, d_k $are the diameters of the particles and can be approximated by
$$(d_\alpha+d_k)^2 = \frac{1}{\sqrt{2}\pi (n_\alpha+n_k) \text{Kn} \text{L}},$$
where $\text{Kn}$ is Knudsen number, $\text{L}$ is reference length. For Coulomb interaction \cite{morse1963energy}:
\begin{equation*}
    \chi_{\alpha k} = \frac{e^4 \ln \Lambda (m_\alpha + m_k)^2}{6\sqrt{m_\alpha m_k}(2\pi k_B m_\alpha T_\alpha + 2\pi k_B m_k T_k)^{3/2}},
\end{equation*}
where the coulomb logarithm is
$$
\Lambda \equiv \frac{12\pi(\epsilon_0k_BT_e/e^2)^{3/2}}{n_e^{1/2}}.
$$

To satisfy the divergence constraint, the Perfect Hyperbolic Maxwell equations (PHM) are used to reformulate the Maxwell equations as
\begin{align}
    & \frac{\partial \boldsymbol{E}}{\partial t}-c^2 \nabla_{\boldsymbol{x}} \times \boldsymbol{B}+\chi c^2 \nabla_{\boldsymbol{x}} \phi=-\frac{1}{\epsilon_0} \boldsymbol{J},\label{eq:PHM Amphere law}\\
    & \frac{\partial \boldsymbol{B}}{\partial t}+\nabla_{\boldsymbol{x}} \times \boldsymbol{E}+\gamma \nabla_{\boldsymbol{x}} \psi=0, \label{eq:PHM Faraday law}\\
    & \frac{1}{\chi} \frac{\partial \phi}{\partial t}+\nabla_{\boldsymbol{x}} \cdot \boldsymbol{E}=\frac{q}{{\epsilon_0}}, \label{eq:PHM E divergence}\\
    & \frac{\epsilon_0 \mu_0}{\gamma} \frac{\partial \psi}{\partial t}+\nabla_{\boldsymbol{x}} \cdot \boldsymbol{B}=0,\label{eq:PHM B divergence}
\end{align}
where $\phi,\psi$ are artificial correction potentials to accommodate divergence errors traveling at speed $\gamma c$ and $\chi c$ \cite{munz2000divergence,munz2000three}.

\subsection{Asymptotic analysis}

The outline of this section is as follows. First, the dimensionless form of the kinetic system is introduced.
Second, a three-fluid system coupled Maxwell equation can be derived using the Chapman-Enskog method. Within this three-fluid model, the electron and ion fluids constitute a two-fluid subsystem, which interacts with the neutral fluid via collision source terms. By varying small parameters such as the electron-ion mass ratio, normalized Larmor radius, and electron-ion collision frequency, the electron-ion two-fluid subsystem can transform into various MHD systems, like resistive MHD, Hall-MHD, and ideal MHD. In the non-viscous limit, the neutral fluid equation becomes the Euler equation.

The reference quantities are defined as follows:
$$
u_0 = \sqrt{\frac{2k_BT_0}{m_0}}, \rho_0 = m_0 n_0, E_0 = B_0 u_0, a_0 = eB_0u_0/m_0, f_0=m_0n_0/u_0^3, c_{s0} = u_0\sqrt{\gamma/2},
$$
where charateristic velocity $u_0$ is thermal velocity. $\rho_0$ is reference density, $E_0$ and $B_0$ are reference electric and magnetic field strength, $a_0$ is reference acceleration, $f_0$ is reference distribution function, $c_{s0}$ is reference sound speed and $\gamma$ is specific heat. Then variables are non-dimensionalized as:
\begin{equation*}
    \begin{aligned}
    &\hat{x}=\frac{x}{l_0}, \hat{\boldsymbol{u}}=\frac{\boldsymbol{u}}{u_0}, \hat{t}=\frac{u_0}{l_0} t, \hat{m}=\frac{m}{m_0}, \hat{n}=\frac{n}{n_0},\hat{\mathscr{E}}=\frac{\mathscr{E}}{m_i n_0 u_0^2}, \hat{f}=\frac{u_0^3}{m_0 n_0} f, \hat{\boldsymbol{B}}=\frac{\boldsymbol{B}}{B_0}, \\
    &\hat{\boldsymbol{E}}=\frac{\boldsymbol{E}}{B_0 u_0}, \hat{\boldsymbol{J}}=\frac{\boldsymbol{J}}{en_0 u_0},\hat{\lambda}_D=\frac{\lambda_D}{r_{L}} = \sqrt{\frac{\epsilon_0m_0u_0^2}{ne^2}}\frac{eB_0}{m_0u_0}, \hat{r_{L}} = \frac{r_{L}}{l_0} = \frac{m_0u_0}{eB_0l_0},\hat{c}=\frac{c}{u_0},
    \end{aligned}
\end{equation*}
where $\hat{\lambda_D}$ is normalized Debye length, $\hat{r}_{L}$ is normalized larmor radius and $\hat{c}$ is normalized speed of light. The normalized plasma skip depth is given as $\hat{d}_S = \hat{\lambda}_D\hat{r}_L\hat{c}l_0$. It is worth mentioning that when choosing $m_0=m_e$, the normalized plasma skip depth is the real plasma skin depth (i.e. $d_S = d_e$) and ion inertia length is given as
\begin{equation}
   d_i = \frac{c}{\omega_{pi}} = d_e \sqrt{\frac{n_em_i}{n_im_e}}=  \hat{c}\hat{\lambda_D}\hat{r_L}l_0\sqrt{\frac{n_e m_i}{n_i m_e}}.
   \label{eq:skin length}
\end{equation}

Inserting the above-normalized variables into the BGK-Maxwell system, the following dimensionless system is obtained:

\begin{equation}
    \begin{aligned}
    &\frac{\partial \hat{f}_{\alpha}}{\partial \hat{t}}+\hat{\boldsymbol{u}_\alpha} \cdot \nabla_{\hat{x}} \hat{f}_{\alpha}+\frac{q_{\alpha}(\hat{\boldsymbol{E}}+\hat{\boldsymbol{u}_\alpha }\times\hat{\boldsymbol{B}})}{m_\alpha\hat{r}}  \cdot \nabla_{\hat{u}} \hat{f}_{\alpha}=\frac{\hat{g}_{\alpha}^M-\hat{f}_{\alpha}}{\hat{\tau}_{\alpha}}, \\
    &\frac{\partial \hat{\boldsymbol{B}}}{\partial \hat{t}}+\nabla_{\hat{x}} \times \hat{\boldsymbol{E}}=0, \\
    &\frac{\partial \hat{\boldsymbol{E}}}{\partial \hat{t}}-\hat{c}^{2} \nabla_{\hat{x}} \times \hat{\boldsymbol{B}}=-\frac{1}{\hat{\lambda}_{D}^{2} \hat{r_L}} \boldsymbol{J},\\
    &\nabla_{\hat{\boldsymbol{x}}} \cdot \hat{\boldsymbol{E}}=\frac{\hat{n_i}-\hat{n_e}}{\hat{\lambda_D}^2 \hat{r_{L}}}, \quad \nabla_{\hat{\boldsymbol{x}}} \cdot \hat{\boldsymbol{B}}=0.
    \end{aligned}
    \label{BGKnondim}
\end{equation}
For simplicity, in the following of this paper, all the hats for nondimensionalized variables are omitted.

Zeroth-order asymptotic solution of $f$ based on Chapman-Enskog expansion is
$
    f=g+O\left(\tau^{1}\right)
$ \cite{xu2001gas}. Substitute the solution into BGK equation and take moments, a three-fluid system composed of neutral species, ions, and electrons is obtained,
\begin{equation}
\begin{aligned}
& \partial_t \rho_n+\nabla_{\boldsymbol{x}} \cdot\left(\rho_n \boldsymbol{U}_n\right)=0, \\
& \partial_t\left(\rho_n \boldsymbol{U}_n\right)+\nabla_{\boldsymbol{x}} \cdot\left(\rho_n \boldsymbol{U}_n \boldsymbol{U}_n+p_n \mathbb{I}\right)=S_n, \\
& \partial_t \mathscr{E}_n+\nabla_{\boldsymbol{x}} \cdot\left(\left(\mathscr{E}_n+p_n\right) \boldsymbol{U}_n\right)=Q_n,\\
& \partial_t \rho_i+\nabla_{\boldsymbol{x}} \cdot\left(\rho_i \boldsymbol{U}_i\right)=0, \\
& \partial_t\left(\rho_i \boldsymbol{U}_i\right)+\nabla_{\boldsymbol{x}} \cdot\left(\rho_i \boldsymbol{U}_i \boldsymbol{U}_i+p_i \mathbb{I}\right)=\frac{q_in_i}{r_{L}}\left(\boldsymbol{E}+\boldsymbol{U}_i \times \boldsymbol{B}\right)+S_i, \\
& \partial_t \mathscr{E}_i+\nabla_{\boldsymbol{x}} \cdot\left(\left(\mathscr{E}_i+p_i\right) \boldsymbol{U}_i\right)=\frac{q_in_i}{r_{L}} \boldsymbol{U}_i \cdot \boldsymbol{E}+Q_i,\\
& \partial_t \rho_e+\nabla_{\boldsymbol{x}} \cdot\left(\rho_e \boldsymbol{U}_e\right)=0, \\
& \partial_t\left(\rho_e \boldsymbol{U}_e\right)+\nabla_{\boldsymbol{x}} \cdot\left(\rho_e \boldsymbol{U}_e \boldsymbol{U}_e+p_e \mathbb{I}\right)=\frac{q_en_e}{r_{L}}\left(\boldsymbol{E}+\boldsymbol{U}_e \times \boldsymbol{B}\right)+S_e, \\
& \partial_t \mathscr{E}_e+\nabla_{\boldsymbol{x}} \cdot\left(\left(\mathscr{E}_e+p_e\right) \boldsymbol{U}_e\right)=\frac{q_en_e}{r_{L}} \boldsymbol{U}_e \cdot\boldsymbol{E}+Q_e, \\
&\frac{\partial {\boldsymbol{B}}}{\partial {t}}+\nabla_{{x}} \times {\boldsymbol{E}}=0, \quad \frac{\partial {\boldsymbol{E}}}{\partial {t}}-{c}^{2} \nabla_{{x}} \times {\boldsymbol{B}}=-\frac{1}{{\lambda}_{D}^{2} {r_L}} \boldsymbol{J},\\
&\nabla_{{\boldsymbol{x}}} \cdot {\boldsymbol{E}}=\frac{{n_i}-{n_e}}{{\lambda_D}^2 {r_{L}}}, \quad \nabla_{{\boldsymbol{x}}} \cdot {\boldsymbol{B}}=0,
\label{threefluid}
\end{aligned}
\end{equation}
where $S_\alpha$ and $Q_\alpha$ are momentum and energy exchange between species $\alpha$ and other species in the system.
\begin{equation}
    \begin{aligned}
    &S_\alpha = \sum_{k=1}^N 2\mu_{nk}\chi_{\alpha k}n_\alpha n_k(\boldsymbol{U}_k - \boldsymbol{U}_\alpha),\\
    &Q_\alpha = \sum_{k=1}^N 4\mu_{nk}\chi_{\alpha k}n_\alpha n_k(\frac{3}{2}k_BT_k - \frac{3}{2}k_BT_\alpha + \frac{1}{2}m_k(\boldsymbol{U}_k-\boldsymbol{U}_\alpha)^2).
    \end{aligned}
\end{equation}

Except for the first three equations for the neutral gas, the rest system in Eq.\eqref{threefluid} forms an ion-electron two-fluid subsystem, based on which Hall-effect MHD and ideal MHD can be derived \cite{liu2017unified,shen2018magnetohydrodynamic}.

\begin{figure}[H]
    \centering
    \includegraphics[width=1.05\textwidth]{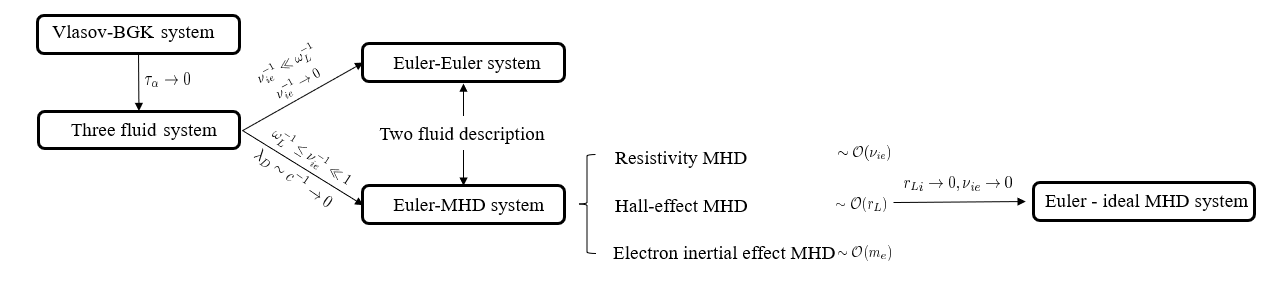}
    \caption{Asymptotic Behavior of the Kinetic System.}
    \label{fig:asymptotic behavior of BGK-Maxwell}
\end{figure}

With the definition of center-of-mass velocity as
$$
\boldsymbol{U}=\frac{m_i \boldsymbol{U}_i+m_e \boldsymbol{U}_e}{m_i+m_e},
$$
with the mass ratio $\epsilon = m_e/m_i$,
$(1+\epsilon)\boldsymbol{U} = \boldsymbol{U}_i + \epsilon \boldsymbol{U}_e $, 
the equation on the $\mathcal{O}(\epsilon^0)$ balance gives,
$$
\boldsymbol{U}_i = \boldsymbol{U} \quad \text{and}\quad \boldsymbol{U}_e = \boldsymbol{U} - \boldsymbol{U}_i + \boldsymbol{U}_e  = \boldsymbol{U} - \frac{\boldsymbol{J}}{ne}.
$$
Substituting the above approximation into an electron momentum equation in the two-fluid subsystem, it gets to 
$$
\boldsymbol{E}+\boldsymbol{U} \times \boldsymbol{B}=\frac{2 m_i m_e v_{i e}}{\left(m_i+m_e\right) n_e e^2} \boldsymbol{J}+\frac{r_{L} }{n_e e} \boldsymbol{J} \times \boldsymbol{B}+\frac{r_{L}}{n_e e} \partial_t\left(\rho_e \boldsymbol{U}_e\right)+\frac{r_{L}}{n_e e} \nabla_{\boldsymbol{x}} \cdot\left(\rho_e \boldsymbol{U}_e \boldsymbol{U}_e+p_e \mathbb{I}\right).
$$
The right-hand side of the above equation contains four terms: the first term represents electric resistivity, the second term corresponds to the Hall effect, and the last two terms describe the effects of electron inertia and pressure. In the limit $\epsilon \rightarrow 0$, the electron's inertia can be neglected and the electron momentum equation gives the generalized Ohm's law
\begin{equation}
\boldsymbol{E}+\boldsymbol{U} \times \boldsymbol{B}=\frac{1}{\sigma} \boldsymbol{J}+\frac{r_{L}}{n_e e} \boldsymbol{J} \times \boldsymbol{B}+\frac{r_{L}}{n_e e} \nabla_{\mathrm{x}} p_e.
\label{eq:general ohm}
\end{equation}
The electron's inertia term cannot be ignored only when the electron speed $U_e$ is much larger than the ion speed $U_i$ \cite{liu2024ohm}. For non-relativistic flows, the displacement current is negligible in view of,
$$
\frac{1}{c^2}|\frac{\partial \boldsymbol{E}}{\partial t}| \sim \frac{U^2}{c^2} \frac{B}{L} \ll |\nabla\times \boldsymbol{B}| \sim \frac{B}{L},
$$
where $U$ and $L$ is the characteristic macroscale velocity and spatial length of plasma, and $U^2/c^2\ll 1$.
When $\lambda_D \sim c^{-1} \rightarrow 0$ (i.e., $\lambda c^{-1}=1$), the Ampère's law then becomes
$$
\boldsymbol{J}=r_{L}  \nabla_{\boldsymbol{x}} \times \boldsymbol{B}+\mathcal{O}\left(U^2 / c^2\right).
$$
The above low-frequency Ampère's law indicates that $\nabla \cdot \boldsymbol{J}=0$, and therefore in this regime, the plasma is quasi-neutral, namely $n_i \approx n_e$.
In such a regime, the two fluid equations reduce to one fluid Hall-MHD equation. The Hall term and the electron pressure term are on the order of Larmor radius. Then Hall-MHD can be written as,
$$
\begin{aligned}
& \partial_t \rho+\nabla_{\boldsymbol{x}} \cdot(\rho \boldsymbol{U})=0, \\
& \partial_t(\rho \boldsymbol{U})+\nabla_{\boldsymbol{x}} \cdot\left(\rho \boldsymbol{U U}+p_i \mathbb{I}\right)=\frac{\rho_i}{m_i r_{L}}(\boldsymbol{E}+\boldsymbol{U} \times \boldsymbol{B}), \\
& \boldsymbol{E}+\boldsymbol{U} \times \boldsymbol{B}=\frac{1}{\sigma} \boldsymbol{J}+\frac{r_{L}}{n_e e} \boldsymbol{J} \times \boldsymbol{B}+\frac{r_{L}}{n_e e} \nabla_{\boldsymbol{x}} p_e, \\
& \partial_t \mathscr{E}_\alpha+\nabla_{\boldsymbol{x}} \cdot\left(\left(\mathscr{E}_\alpha+p_\alpha\right) \boldsymbol{U}_\alpha\right)=\frac{1}{r_{L}} \boldsymbol{J}\cdot \boldsymbol{E}, \\
& \partial_t \boldsymbol{B}+\nabla_{\boldsymbol{x}} \times \boldsymbol{E}=0, \\
& \boldsymbol{J}=r_{L} \nabla_{\boldsymbol{x}} \times \boldsymbol{B}.
\end{aligned}
$$
Hall-MHD is on the ion inertia scale where the ions are demagnetized and electrons are still frozen to the magnetic field lines. Compared to resistive MHD, Hall-MHD can realize fast reconnection as introduced in the numerical test section.
In the limit as the Larmor radius $r_L$ approaches zero, the skin depth $d_s$ also approaches zero. Under this condition, the velocity separation between the ions and electrons disappears, and the strong magnetic field forces the ions and electrons to move together as a single fluid. Therefore, in this limit, the Hall current term in the generalized Ohm's law (Eq.\eqref{eq:general ohm}) becomes negligible. Additionally, since the magnetic force is much larger than the thermal pressure in this regime, i.e. $\beta \ll 1$, the pressure term can also be ignored. If collisions between electrons and ions can also be neglected, i.e. $\nu_{ie}=0$, then the electric resistivity can be dropped. Under these assumptions, the generalized Ohm's law reduces to the ideal Ohm's law,
$$
\boldsymbol{E}+\boldsymbol{U} \times \boldsymbol{B}=0.
$$
So combining electron and ion momentum equations, the ideal MHD equation can be written as,
\begin{equation}
\begin{aligned}
& \partial_t \rho+\nabla_{\boldsymbol{x}} \cdot(\rho \boldsymbol{U})=0, \\
& \partial_t(\rho \boldsymbol{U})+\nabla_{\boldsymbol{x}} \cdot(\rho \boldsymbol{U} \boldsymbol{U}+p \mathbb{I})= (\nabla\times\boldsymbol{B}) \times \boldsymbol{B}, \\
& \partial_t \mathscr{E}+\nabla_{\boldsymbol{x}} \cdot((\mathscr{E}+p) \boldsymbol{U})= (\nabla\times\boldsymbol{B}) \cdot\boldsymbol{E},\\
& \partial_t \boldsymbol{B}+\nabla_{\boldsymbol{x}} \times(\boldsymbol{U} \times \boldsymbol{B})=0.\\
\end{aligned}
\label{eq:ideal MHD}
\end{equation}
In this limit, the three-fluid system in Eq.\eqref{threefluid} turns to a neutral and ideal MHD two-fluid system.In summary, as shown in Fig.\ref{fig:asymptotic behavior of BGK-Maxwell}, the kinetic system can be reduced to three fluid system and then Euler-MHD system in different scale.
More specifically, when the collision time $\tau_\alpha$ within a single species is significantly shorter than the characteristic time scale of the system, the kinetic description for that species can be replaced by a fluid description. This transition occurs because frequent collisions rapidly drive the species towards local thermodynamic equilibrium. The interaction between ions and electrons is primarily determined by electromagnetic forces (Lorentz force) and cross-species collisions. The relative strength of these interactions leads to two distinct regimes. When the gyrofrequency $\omega_L$ greatly exceeds the ion-electron collision frequency $\nu_{ie}$ ($\omega_L \gg \nu_{ie}$), the system transitions to an MHD description. In this regime, the magnetic field strongly influences the plasma dynamics. Conversely, if the ion-electron collision frequency significantly surpasses the gyrofrequency ($\nu_{ie} \gg \omega_L$), the system behaves as a single-fluid governed by the Euler equations. In this case, collisional effects dominate over magnetic effects. In the MHD limit, various non-ideal dissipation effects may emerge, depending on the specific physical conditions. Resistivity arises from electron-ion collisions, leading to magnetic field diffusion.Hall Effect becomes significant when ion and electron motions decouple at small scales. Other effects may include ambipolar diffusion or electron inertia, depending on the plasma parameters. In the absence of significant dissipation effects, the system approaches the ideal MHD limit. This regime is characterized by perfect conductivity and frozen-in magnetic field lines.

Now consider the collision effects between charged particles (ions and electrons) and neutral particles. Generally, collisions distribute momentum between the charged and neutral components across the plasma, resulting in the magnetic forces effectively acting on both species. The presence of neutrals makes the magnetic field "heavier" compared to the purely charged particle case, thus influencing the wave speeds. The electrical conductivity, the Hall diffusion coefficient should be adjusted in the partially ionized regime.

In the weakly ionized limit, a new phenomenon called ambipolar diffusion arises due to the velocity separation between neutrals and ions.
The formal derivation of the generalized Ohm's law in the weakly ionized limit, following the approaches in \cite{nakano1986dissipation,o2006explicit}, is provided in \ref{ohms law in weakly ionized limit}. Here the origin of ambipolar diffusion is introduced. Denoting the relative velocity between the ions and neutrals as $\boldsymbol{V}_d \equiv \boldsymbol{U}_i - \boldsymbol{U}_n$, the drag force between the ions and neutrals can be expressed as
$$
\boldsymbol{f}_d = 2\mu_{in}\chi_{in}n_in_n(\boldsymbol{U}_i-\boldsymbol{U}_n).
$$
In the steady steady state, according to the Eq.\eqref{eq:ideal MHD}, the magnetic force will balance the drag force,
$$
2\mu_{in}\chi_{in}n_in_n(\boldsymbol{U}_i-\boldsymbol{U}_n) =(\nabla\times\boldsymbol{B}) \times \boldsymbol{B},
$$
which gives
$$
\boldsymbol{U}_i = \boldsymbol{U}_n + \frac{(\nabla\times\boldsymbol{B}) \times \boldsymbol{B}}{2\mu_{in}\chi_{in}n_in_n}.
$$
Then, the Ohm's law for ions can be written as,
\begin{equation*}
    \boldsymbol{E} + \boldsymbol{U}_i\times\boldsymbol{B} = 0,
\end{equation*}
which gives
\begin{equation*}
    \boldsymbol{E} + \boldsymbol{U}_n\times\boldsymbol{B} = -\frac{(\nabla\times\boldsymbol{B}) \times \boldsymbol{B}\times\boldsymbol{B}}{2\mu_{in}\chi_{in}n_in_n}.
\end{equation*}
In the weakly ionized limit, bulk velocity $\boldsymbol{U}\approx\boldsymbol{U}_n$, therefore, Ohm's law now has a now dissipation term caused by velocity separation between neutrals and charged particles.

\section{UGKWP for PIP}
\label{numerical}

\subsection{General framework}
In the framework of the FVM, the cell averaged conservative variables for species $\alpha$ is $(\boldsymbol{W}_{\alpha})_i = ((\rho_{\alpha})_i, (\rho_{\alpha}\boldsymbol{U}_{\alpha})_i, (\rho_{\alpha}\mathscr{E}_{\alpha})_i)$ on a physical cell $\Omega_i$ are defined as
$$
(\boldsymbol{W}_{\alpha})_i = \frac{1}{|\Omega_i|}\int_{\Omega_i}\boldsymbol{W}_{\alpha}(\boldsymbol{x})\mathrm{d}\boldsymbol{x},
$$
where $|\Omega_i|$ is the volume of cell $\Omega_i$. For a discretized time step $\Delta t=t^{n+1}-t^n$, the evolution of $(\boldsymbol{W}_{\alpha})_i$ is
\begin{equation}
(\boldsymbol{W}_{\alpha})_i^{n+1} = (\boldsymbol{W}_{\alpha})_i^n - \frac{\Delta t}{|\Omega_i|}\sum_{s\in\partial\Omega_i}|l_s|(\mathscr{F}_{\boldsymbol{W}_\alpha})_s + \frac{\Delta t}{\tau_\alpha}(\bar{(\boldsymbol{W}}_{\alpha})_i^n - (\boldsymbol{W}_{\alpha})_i^n)+\Delta t(\boldsymbol{S}_{\alpha})_i^{n+1},
\label{eq:FVM discretization}
\end{equation}
where $l_s\in \partial \Omega_i$ is the cell interface with center $\boldsymbol{x}_s$ and outer unit normal vector $\boldsymbol{n}_{s}$. $|l_s|$ is the area of the cell interface. $(\bar{\boldsymbol{W}}_{\alpha})_i = ((\rho_{\alpha})_i, (\rho_{\alpha}\bar{\boldsymbol{U}}_{\alpha})_i, (\rho_{\alpha}\bar{\mathscr{E}}_{\alpha})_i)$
where $(\bar{\boldsymbol{U}}_{\alpha})_i$ and $(\bar{\mathscr{E}}_{\alpha})_i$ are post-collision velocity and energy in AAP model as Eq.\eqref{eq:aap U and E}. $(\boldsymbol{S}_{\alpha})_i$ is source term due to electromagnetic force. The numerical flux across interface $(\mathscr{F}_{\boldsymbol{W}_\alpha})_s$ can be evaluated from distribution function at the interface,
\begin{equation}
(\mathscr{F}_{\boldsymbol{W}_\alpha})_s = \frac{1}{\Delta t}\int_{t^n}^{t^{n+1}}\boldsymbol{u}\cdot \boldsymbol{n}_sf_\alpha(\boldsymbol{x}_{s},\boldsymbol{u},\boldsymbol{\xi},t)\boldsymbol{\Psi} \mathrm{d}\boldsymbol{\Xi}\mathrm{d}t,
\label{eq:FVM macroscopic flux}
\end{equation}
where $\boldsymbol{\Psi}=(1,\boldsymbol{u},\frac{1}{2}(\boldsymbol{u}^2+\boldsymbol{\xi}^2))$ is the conservative moments of distribution functions with $\boldsymbol{\xi}=(\xi_1,\xi_2,\cdots,\xi_n)$ the internal degree of freedom. $\mathrm{d}\boldsymbol{\Xi}=\mathrm{d}\boldsymbol{u}d\boldsymbol{\xi}$ is the volume element in the phase space. $f_\alpha(\boldsymbol{x}_{s},\boldsymbol{u},\boldsymbol{\xi},t)$ is the distribution function at the cell interface $l_s$.
In the continuum limit where the relaxation time $\tau \rightarrow 0$, the distribution function can be represented analytically using the Chapman-Enskog expansion \cite{xu2001gas}. However, in the transitional and collisionless regimes, the distribution function lacks an analytical representation, necessitating a direct tracking of the distribution function evolution. The evaluation of the distribution function evolution will be presented in Section \ref{microscopic evolution}.

The elementwise equations in Eq.\eqref{eq:FVM discretization} are
\begin{equation}
\begin{aligned}
\left(\rho_\alpha\right)_i^{n+1}= & \left(\rho_\alpha\right)_i^n-\frac{\Delta t}{\left|\Omega_{i}\right|} \sum_{s \in \partial \Omega_{{i}}}\left|l_s\right| (\mathscr{F}_{\rho_\alpha})_s, \\
\left(\boldsymbol{\rho}_\alpha \boldsymbol{U}_\alpha\right)_i^{n+1}= & \left(\boldsymbol{\rho}_\alpha \boldsymbol{U}_\alpha\right)_i^n-\frac{\Delta t}{\left|\Omega_{i}\right|} \sum_{s \in \partial \Omega_{{i}}}\left|l_s\right| (\mathscr{F}_{(\rho\boldsymbol{U})_\alpha})_s\\
& +\frac{\Delta t}{\tau_\alpha}\left(\rho_\alpha^n \bar{\boldsymbol{U}^n}{ }_\alpha-\rho_\alpha^n \boldsymbol{U}_\alpha^n\right)+\frac{\Delta t}{r_{L_i}} n_\alpha^{n+1}\left(\boldsymbol{E}^{n+1}+\boldsymbol{U}_\alpha^{n+1} \times \boldsymbol{B}^{n+1}\right), \\
\left(\boldsymbol{\rho}_\alpha \mathscr{E}_\alpha\right)_i^{n+1}= & \left(\boldsymbol{\rho}_\alpha \mathscr{E}_\alpha\right)_i^n-\frac{\Delta t}{\left|\Omega_{i}\right|} \sum_{s \in \partial \Omega_{i}}\left|l_s\right| (\mathscr{F}_{(\rho\mathscr{E})_\alpha})_s \\
& +\frac{\Delta t}{\tau_\alpha}\left(\rho_\alpha^n \bar{\mathscr{E}^n}{ }_\alpha-\rho_\alpha^n \mathscr{E}_\alpha^n\right)+\frac{\Delta t}{r_{L_i}} n_\alpha^{n+1} \boldsymbol{U}_\alpha^{n+1} \cdot \boldsymbol{E}^{n+1} .
\end{aligned}
\label{eq:fvm fluid elementwise}
\end{equation}

In Eq.\eqref{eq:fvm fluid elementwise}, the source term due to cross-species momentum and energy exchange will be evaluated by the operator splitting method. Lorentz source term is split and coupled with source terms in Maxwell equation to get coupled evolution between fluid species and electromagnetic field. These interaction equations can be solved by the Crank-Nicolson scheme introduced in Section \ref{interaction equation}.

The cell averaged quantities $\boldsymbol{Q}_i$ for electromagnetic variables $\boldsymbol{Q} = \left(E_{x}, E_{y}, E_{z}, B_{x}, B_{y}, B_{z}, \phi, \psi\right)$ in a cell are defined as
$$\boldsymbol{Q}_i =  \frac{1}{|\Omega_i|}\int_{\Omega_i}\boldsymbol{Q}(\boldsymbol{x})\mathrm{d}\boldsymbol{x}, $$
where $(\mathscr{F}_{\boldsymbol{Q}})_s$ is numerical flux across the cell interface $l_s$ which will be presented in section \ref{sec:phm}.
The time evolution formula is
$$
\boldsymbol{Q}_i^{n+1}=\boldsymbol{Q}_i^n+\frac{\Delta t}{\left|\Omega_{i}\right|} \sum_{s \in \partial \Omega_{i}}\left|l_i\right| (\mathscr{F}_{\boldsymbol{Q}})_s+\Delta t (\boldsymbol{S}_{\boldsymbol{Q}})_i^{n+1},
$$
where $(\mathscr{F}_{\boldsymbol{Q}})_s$ is numerical flux across a cell interface, which will be presented in Section \ref{sec:phm}. $\boldsymbol{S}_{\boldsymbol{Q}}$ are sources terms in PHM equations. The componentwise equations are:
$$
\begin{aligned}
\boldsymbol{E}_i^{n+1} & =\boldsymbol{E}_i^n+\frac{\Delta t}{\left|\Omega_{ i}\right|} \sum_{s \in \partial \Omega_{i}}\left|l_s\right| (\mathscr{F}_{\boldsymbol{E}})_s-\frac{\Delta t}{\lambda_D^2 r_{L}}\left(n_i^{n+1} \boldsymbol{U}_i^{n+1}-n_e^{n+1} \boldsymbol{U}_e^{n+1}\right), \\
\boldsymbol{B}_i^{n+1} & =\boldsymbol{B}_i^n+\frac{\Delta t}{\left|\Omega_{ i}\right|} \sum_{s \in \partial \Omega_{i}}\left|l_s\right| (\mathscr{F}_{\boldsymbol{B}})_s, \\
\phi_i^{n+1} & =\phi_i^n+\frac{\Delta t}{\left|\Omega_{ i}\right|} \sum_{s \in \partial \Omega_{{i}}}\left|l_s\right| (\mathscr{F}_{\phi})_s+\frac{\Delta t \chi}{\lambda_D^2 r_{L}}\left(n_i^{n+1}-n_e^{n+1}\right), \\
\psi_i^{n+1} & =\psi_i^n+\frac{\Delta t}{\left|\Omega_{i}\right|} \sum_{s \in \partial \Omega_{i}}\left|l_s\right| (\mathscr{F}_{\psi})_s.
\end{aligned}
$$

General numerical steps are listed as follows:
\begin{enumerate}
    \item \textbf{Coupled free transport and collision process}: Update macroscopic conservative variable $\boldsymbol{W}_\alpha^{n}$ to $\boldsymbol{W}_\alpha^{*}$ considering net flux without force term across the cell interface. The numerical flux is decomposed into wave and particle components according to the UGKWP method as introduced in Section \ref{microscopic evolution} and Section \ref{macroscopic evolution}. Simultaneously, particles' positions are updated from $(\boldsymbol{x}_k^n,\boldsymbol{u}_k^{n})$ to $(\boldsymbol{x}_k^n,\boldsymbol{u}_k^{*})$. Collision process is treated in an annihilation-resampling way as introduced in Section \ref{microscopic evolution}.
    \item \textbf{Electromagnetic field evolution}: Update electromagnetic field $\boldsymbol{E}^{n} \rightarrow \boldsymbol{E}^{*}$, $\boldsymbol{B}^{n} \rightarrow \boldsymbol{B}^{n+1}$, ${\phi}^{n} \rightarrow {\phi}^{*}$ and ${\psi}^{n} \rightarrow {\psi}^{n+1}$ by net flux across cell interface as introduced in Section \ref{sec:phm}.
    \item \textbf{Cross-species collision process}: Update conservative variable $\boldsymbol{W}_\alpha^{*}$ to $\boldsymbol{W}_\alpha^{**}$ considering momentum and energy exchange between different species.
    \item \textbf{Fluid and electromagnetic field interaction process}: Incorporate the interaction between electromagnetic field and charged species $\boldsymbol{W}_\alpha^{**}$ to $\boldsymbol{W}_\alpha^{n+1}$,$\boldsymbol{E}^{*} \rightarrow \boldsymbol{E}^{n+1}$,${\phi}^{*} \rightarrow {\phi}^{n+1}$. Particles' velocities are updated from $\boldsymbol{u}_k^{*}$ to $\boldsymbol{u}_k^{n+1}$ as introduced in Section \ref{interaction equation}.
\end{enumerate}
After four steps, all variables are evolved from $t^n$ to $t^{n+1}$.

The timestep constraint of the current scheme is given by
$$
\Delta t = CFL \frac{\Delta x}{\max(|U+c_s|, c)},
$$
where $c_s$ is the speed of sound, $c$ is the speed of light, $CFL$ is Courant-Friedrichs-Lewy number. Source terms can be treated implicitly, so it will not pose a timestep restriction.

\subsection{Evolution of microscopic distribution function}
\label{microscopic evolution}

In this section, the objective is to employ the particle method to numerically solve the kinetic equation to obtain the particle distribution function at cell interfaces, which will then be used to calculate the numerical flux in Eq.\eqref{eq:FVM macroscopic flux}. The BGK equation without force term can be written as
$$
     \frac{\partial f}{\partial t}+\boldsymbol{u} \cdot \nabla_x f = \frac{g-f}{\tau},
$$
where velocity $\boldsymbol{u} = (u,v,w)$. For simplicity, species subscript $\alpha$ is omitted here. The time-dependent solution at the interface can be written as,
\begin{equation}
    f\left(\boldsymbol{x},\boldsymbol{u}, \boldsymbol{\xi}, t\right)=\frac{1}{\tau} \int_{0}^{t} g(\boldsymbol{x}^{\prime}, \boldsymbol{u}, \boldsymbol{\xi},t^{'}) e^{-\left(t-t^{\prime}\right) / \tau} \mathrm{d} t^{\prime} + e^{-t / \tau} f_{0}\left(\boldsymbol{x}-\boldsymbol{u} t \right),
    \label{eq:BGKsoln}
\end{equation}
where $\boldsymbol{\xi}=(\xi_1,\xi_2,\cdots,\xi_n)$ is the internal degree of freedom, for notation simplicity, $\boldsymbol{\xi}$ is omitted later. $\tau$ is the local mean relaxation time. $f_0$ is the initial gas distribution function at $t=0$, and $g$ is equilibrium distribution along the characteristic line $\boldsymbol{x}^{'} = \boldsymbol{x} - \boldsymbol{u} t^{'}$. Expand the equilibrium distribution function by the Taylor series to the second-order accuracy,
\begin{equation}
g^{'}=g+ g_x \cdot(\boldsymbol{x}^{'}-\boldsymbol{x})+g_t(t^{'}-t),
\label{eq:taylor g}
\end{equation}
where  $g\equiv g(\boldsymbol{x}, \boldsymbol{u},t),g^{'}\equiv g(\boldsymbol{x}^{\prime}, \boldsymbol{u}, t^{'})$.
Substitute them into Eq.\eqref{eq:BGKsoln}, the numerical multiscale evolution solution for simulation particle can be obtained,
\begin{equation}
f(\boldsymbol{x}, \boldsymbol{u},  t)=\left(1-e^{-t / \tau}\right) g^{+}(\boldsymbol{x}, \boldsymbol{u}, t )+e^{-t / \tau} f_{0}\left(\boldsymbol{x} - \boldsymbol{u}t\right),
\label{eq:multiscale BGK soln}
\end{equation}
where,
$$
g^{+}\left( \boldsymbol{x},\boldsymbol{u}, t \right) = g\left( \boldsymbol{x},\boldsymbol{u} ,t\right) + \left( \frac{te^{- t\text{/}\tau}}{1 - e^{- t\text{/}\tau}} - \tau \right)\boldsymbol{u}\cdot g_x\left( \boldsymbol{x},\boldsymbol{u},t \right) +\left( \frac{t}{1 - e^{- t\text{/}\tau}} - \tau \right)g_t\left( \boldsymbol{x},\boldsymbol{u},t \right).
$$
The equation above describes the solution to the BGK equation, where the distribution function $f$ at time $t$ is a combination of the initial distribution function $f_0$ and the Taylor expansion of the equilibrium state $g$. From the particles' perspective, this equation implies that a particle has a probability of $e^{-t/\tau}$ to freely stream during the time interval $[0,t]$, and a probability of $(1 - e^{-t/\tau})$ to collide with other particles. After multiple collisions, the particle distribution will reach the equilibrium state $g^+$.

Typically, the operator splitting technique is employed to decouple the free transport and collision processes for particles, as seen in methods like Particle-in-Cell with Monte Carlo Collisions (PIC/MCC). This approach is suitable for weakly collisional regimes where the mean collision time is not excessively small. However, when the collision time becomes very short, the collision source term becomes numerically stiff, and the mesh size and time step must be strictly constrained. The solution expressed in Eq.\eqref{eq:multiscale BGK soln} suggests that after collisions, particles enter the local equilibrium distribution $g^+$ and the freedom of individual particles degenerates into fluid-like wave dynamics. Therefore, a mean-field description becomes appropriate for modeling those particles. Then after the collision process, the individual particles can be annihilated and resampled at the beginning of the next time step. This concept of microscopic distribution evolution is central to the UGKWP method.

\begin{figure}[H]
    \centering
    \includegraphics[width=0.9\textwidth]{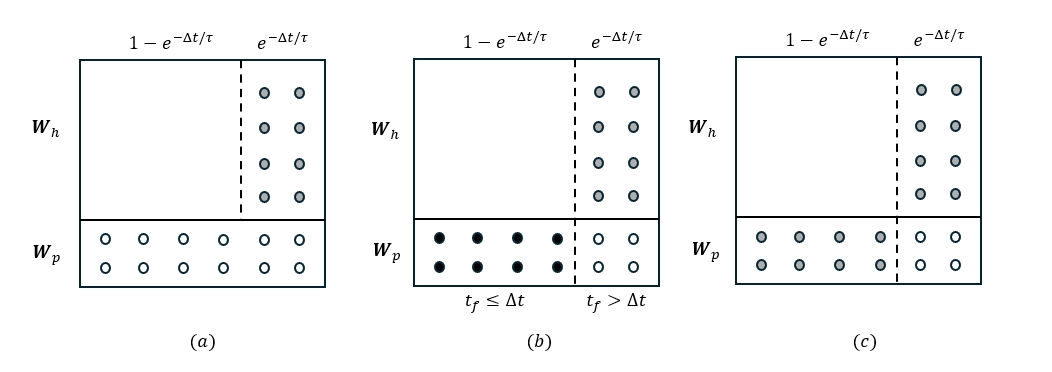}
    \caption{Schematic representation of particle evolution.$\boldsymbol{W}_h$ represents the combination of waves and particles in the equilibrium state.$\boldsymbol{W}_p$ denotes collisionless particles from the previous time step. (a) Initiation of a time step:Grey particles represent collisionless particles sampled from $\boldsymbol{W}_h$. These particles are predicted to remain collisionless during the upcoming time step. (b) Post-streaming phase: Black particles represent those that have undergone collisions, characterized by a free transport time shorter than the simulation time step. These particles are subsequently absorbed into the wave component. (c) Transition to the next time step: The grey particles from stage (b), having remained collisionless, are now incorporated into the $\boldsymbol{W}_p$ population. Simultaneously, a new set of collisionless particles is sampled from $\boldsymbol{W}_h$. }
    \label{fig:algorithm-f-evolution}
\end{figure}

Specifically, in the UGKWP method, the velocity distribution function is represented in a hybrid way. A portion of the distribution is captured analytically through the equilibrium distribution function $g^+$, while the remaining part is represented by stochastic simulation particles $P_k = (m_k, \boldsymbol{x}_k, \boldsymbol{u}_k)$, as illustrated in Figure \ref{fig:algorithm-f-evolution}. Here, $m_k$ denotes the mass of simulation particle $P_k$, which corresponds to a cluster of real gas particles of the same species, and $\boldsymbol{x}_k$ and $\boldsymbol{u}_k$ represent the position and velocity of the simulation particle $P_k$, respectively.

According to Eq.\eqref{eq:multiscale BGK soln}, the cumulative distribution function of the particle's free streaming time $t_{f}$ before the collision is given as
$$
F\left( t_{f} < t \right) = \exp\left( - t\text{/}\tau \right),
$$
from which the free stream time $t_{f}$ can be sampled as $t_{f} = - \tau\ln(\eta)$ with $\eta$ a random varible subject to the uniform distribution $\eta \sim U(0,1)$ . For a time step $\Delta t$ , the particles with $t_{f} \geq \Delta t$ will undergo collisionless free streaming, and the particles with $t_{f} < \Delta t$ will experience collisional interactions. The procedure of updating particles in the UGKWP method is
\begin{enumerate}[Step 1:]
\item At the beginning of the time step, sample the free-streaming time $t_{f,k}$ for each particle $P_k$ from the cumulative distribution function $F(t_f < t) = \exp(-t/\tau)$;

\item During the time step, stream each particle $P_k$ for a time period of $\min(\Delta t, t_{f,k})$. Then identify and retain the collisionless particles, while removing the collisional particles. Calculate the free-transport flux across cell interfaces contributed by the particles and accumulate the total conservative quantities of the particles $\boldsymbol{W}_i^p$;

\item After updating the macroscopic conservative variables, calculate the total conservative quantities of the collisional particles $\boldsymbol{W}_i^h$ from the updated conservative quantities $\boldsymbol{W}_i$ as $\boldsymbol{W}_i^h = \boldsymbol{W}_i - \boldsymbol{W}_i^p$;

\item At the end of the time step, rebuild the velocity distribution. Calculate the analytical distribution $g^{+,c}$ and resample the collisionless particles from the distribution $g^{+,f}$ according to the updated conservative quantities $\boldsymbol{W}_i^h$.
\end{enumerate}

In the procedure outlined above, the algorithm for updating the macroscopic conservative variables will be introduced in the subsequent section. In the distribution rebuilding process, the updated collisional conservative quantities $\boldsymbol{W}_i^h$ are partitioned into two components: the collisional part $g^{+,c}$ and the collisionless part $g^{+,f}$. The collisional part $g^{+,c}$ is given by $(1 - e^{-\Delta t/\tau^{n+1}})g^+$, while the collisionless part $g^{+,f}$ is given by $e^{-\Delta t/\tau^{n+1}}g^+$.
In the Unified Gas-Kinetic Particle (UGKP) method, there is no such separation; instead, all the $\boldsymbol{W}_i^h$ are sampled and represented by particles. However, according to Eq. \eqref{eq:multiscale BGK soln}, from $t^n$ to $t^{n+1}$, $(1 - e^{-\Delta t/\tau^{n+1}})$ of these particles will experience collisions and re-enter the fluid waves. Therefore, these collisional particles can be represented by the fluid waves at all times. Sampling them would consume more computational time, which is undesirable.
In the UGKWP method, only the collisionless part $g^{+,f}$ is sampled. The collisional part $g^{+,c}$ does not have a direct particle representation but is instead represented by the macroscopic conservative variables. Refer to Figure \ref{fig:algorithm-f-evolution} for more intuitive description of the procedure.

The evolution of the microscopic velocity distribution is now solved numerically using particles. In the subsequent section, the method for updating the macroscopic conservative variables based on the velocity distribution will be introduced.

\subsection{Evolution of macroscopic conservative variables}
\label{macroscopic evolution}

The numerical flux of the macroscopic conservative variable in the UGKWP method can be split into the equilibrium flux and free streaming flux according to Eq.\eqref{eq:BGKsoln}. The equilibrium flux is
\begin{equation}
(\mathscr{F}_{\boldsymbol{W}})_s^g = \frac{1}{\Delta t}\int_{t^n}^{t^{n+1}}\boldsymbol{u}\cdot \boldsymbol{n}_s
\left[\frac{1}{\tau} \int_{0}^{t} g(\boldsymbol{x}^{\prime}, \boldsymbol{u}, \boldsymbol{\xi},t^{'}) e^{-\left(t-t^{\prime}\right) / \tau} \mathrm{d} t^{\prime}\right]
\boldsymbol{\Psi} \mathrm{d}\boldsymbol{\Xi}\mathrm{d}t,
\label{eq:macroscopic eq flux}
\end{equation}
and the free streaming flux is
\begin{equation}
(\mathscr{F}_{\boldsymbol{W}})_s^f = \frac{1}{\Delta t}\int_{t^n}^{t^{n+1}}\boldsymbol{u}\cdot \boldsymbol{n}_s
\left[e^{-t / \tau} f_{0}\left(\boldsymbol{x}-\boldsymbol{u} t \right)\right]
\boldsymbol{\Psi} \mathrm{d}\boldsymbol{\Xi}\mathrm{d}t.
\label{eq:macroscopic fr flux}
\end{equation}

The equilibrium flux can be calculated directly from the macroscopic flow field. Assume in the equation \eqref{eq:FVM macroscopic flux}, $\boldsymbol{x}_{s} = \boldsymbol{0}$ and $t^{n} = 0$, the equilibrium $g$ at the cell interface is obtained by conservation constraint
$$
\int_{}^{}{g\boldsymbol{\Psi}}d\boldsymbol{\Xi} = \int_{\boldsymbol{v} \cdot \boldsymbol{n} > 0}^{}{g^{l}\boldsymbol{\Psi}}d\boldsymbol{\Xi} + \int_{\boldsymbol{v} \cdot \boldsymbol{n} < 0}^{}{g^{r}\boldsymbol{\Psi}}d\boldsymbol{\Xi}.
$$
The spatial and time derivatives can be obtained
$$
\int_{}^{}{g_x\boldsymbol{\Psi}}d\boldsymbol{\Xi} = \boldsymbol{W}_x
,\quad \int_{}^{}{g_t\boldsymbol{\Psi}}d\boldsymbol{\Xi} = - \int_{}^{}\boldsymbol{u} \cdot  g_x\boldsymbol{\Psi}d\boldsymbol{\Xi},
$$
where $g^{l}$ and $g^{r}$ are the equilibrium distributions according to the reconstructed left and right side conservative variables at cell interface $\boldsymbol{W}^{l},{\ \boldsymbol{W}}^{r}$, and $\boldsymbol{W}_x$ is the reconstructed spatial derivative of conservative variables at cell interface. The van Leer limiter is used to achieve a second-order accuracy in space reconstruction. Substituting the reconstructed equilibrium distribution into the equilibrium flux, we have
$$
(\mathscr{F}_{\boldsymbol{W}})_s^g = \int_{}^{}\boldsymbol{u} \cdot \boldsymbol{n}_{s}\left( C_{1}g_{0} + C_{2}\boldsymbol{u} \cdot g_{0x} + C_{3}g_{0t} \right)\Psi d\boldsymbol{\Xi},
$$
where the time integration coefficients are
\begin{align*}
&C_{1} = \Delta t - \tau\left( 1 - e^{- \Delta t\text{/}\tau} \right)
,\\
&C_{2} = 2\tau^2(1 - e^{- \Delta t\text{/}\tau}) - \tau\Delta t -\tau \Delta t  e^{- \Delta t\text{/}\tau},\\
&C_{3} = \frac{\Delta t^2}{2} - \tau\Delta t + \tau^2(1-e^{\Delta t/\tau}).
\end{align*}

Next we consider the free stream flux $(\mathscr{F}_{\boldsymbol{W}})_s^f$ . As stated in the last subsection, the initial distribution is represented partially by an analytical distribution $g_{a}^{+,c}$, and partially by particles, and therefore the free stream flux is also calculated partially from the reconstructed analytical distribution as  $(\mathscr{F}_{\boldsymbol{W}})_s^{f,w}$, and partially from particles as  $(\mathscr{F}_{\boldsymbol{W}})_s^{f,p}$. The initial analytical distribution $g_{\alpha}^{+ ,c}$ is reconstructed as
$$
g_{0}^{+ ,c}\left( \boldsymbol{x},\boldsymbol{u} \right) = g_{0}^{+ ,c} + g_{0x}^{+ ,c} \cdot \boldsymbol{x},
$$
which gives
$$
(\mathscr{F}_{\boldsymbol{W}})_s^{f,w} = \int_{}^{}\boldsymbol{u} \cdot \boldsymbol{n}_s\left( C_{4}g_{0}^{+,c} + C_{5}\boldsymbol{u} \cdot g_{0x}^{+,c} \right)\boldsymbol{\Psi}d\boldsymbol{\Xi},
$$
where the time integration coefficients are
\begin{align*}
&C_{4} = \tau\left( 1 - e^{- \Delta t\text{/}\tau} \right) - \Delta te^{- \Delta t\text{/}\tau},
,\\
&C_{5} = \tau\Delta t e^{- \Delta t\text{/}\tau} - \tau^{2}\left( 1 - e^{- \Delta t\text{/}\tau} \right) + \frac{\Delta t^{2}}{2}e^{- \Delta t\text{/}\tau}.
\end{align*}
The net particle flux $(\mathscr{F}_{\boldsymbol{W}})_s^{f,p}$ is calculated as
$$
(\mathscr{F}_{\boldsymbol{W}})_s^{f,p}=  \sum_{k \in P_{\partial\Omega_{i}^{+}}}^{}\boldsymbol{W}_{P_{k}} - \sum_{k \in P_{\partial\Omega_{i}^{-}}}^{}\boldsymbol{W}_{P_{k}},
$$
where $\boldsymbol{W}_{P_{k}} = \left( m_{k},m_{k}\boldsymbol{v}_{k},\frac{1}{2}m_{k}\boldsymbol{v}_{\boldsymbol{k,}}^{2} \right),P_{\partial\Omega_{i}^{-}}$
is the index set of the particles streaming out of cell $\Omega_{i}$
during a time step, and $P_{\partial\Omega_{i}^{+}}$ is the
index set of the particles streaming into cell $\Omega_{i}$. Finally,
the finite volume scheme for conservative variables is
\begin{align}
    \begin{split}
        \boldsymbol{W}_{\boldsymbol{i}}^{\boldsymbol{n} + \boldsymbol{1}} = \boldsymbol{W}_{\boldsymbol{i}}^{\boldsymbol{n}} -& \sum_{s}^{}{\frac{\Delta t}{\left| \Omega_{i} \right|}\left| l_{s} \right|(\mathscr{F}_{\boldsymbol{W}})_s^{g}} - \sum_{s}^{}{\frac{\Delta t}{\left| \Omega_{i} \right|}\left| l_{s} \right|(\mathscr{F}_{\boldsymbol{W}})_s^{f,w}} + \frac{1}{\left| \Omega_{i} \right|}(\mathscr{F}_{\boldsymbol{W}})_s^{f,p} \\ +& \frac{\Delta t}{\tau}(\bar{(\boldsymbol{W}})_i^n - (\boldsymbol{W})_i^n)+\Delta t(\boldsymbol{S})_i^{n+1}.
    \end{split}
\end{align}

The evolution of the macroscopic conservative variables due to species transport and diffusion has been solved by the coupled evolution of the microscopic particle transport and macroscopic fluid dynamics. The evolution of the macroscopic conservative variables from the acceleration induced by the electromagnetic field will be introduced in Section \ref{interaction equation}.

\subsection{Numerical flux of electromagnetic variables}
\label{sec:phm}
LeVeque's finite volume method of wave propagation is used for calculating the evolution of electromagnetic fields here. While other methods, such as the well-known Finite-Difference Time-Domain (FDTD)  method, could potentially be utilized, we have chosen to employ LeVeque's finite volume method to ensure consistency within the overall coding framework, as the fluid evolution is naturally suited to the finite volume scheme. In the case of the FDTD method, additional transformations between surface values and center values may be necessary.

1D numerical flux is illustrated here, for 2D or 3D problems, simply rotating coordinates can be used to get flux in another direction. The general expression for the 1D perfect hyperbolic Maxwell (PHM) system is:
\begin{equation}
    \frac{\partial \boldsymbol{q}}{\partial t}+\boldsymbol{A}_{1} \frac{\partial \boldsymbol{q}}{\partial x}=\boldsymbol{s},
    \label{eq:2D hyperbolic system}
\end{equation}
where
\begin{equation*}
    \boldsymbol{A}_{1}=\left(\begin{array}{cccccccc}
    0 & 0 & 0 & 0 & 0 & 0 & c^{2} \chi & 0 \\
    0 & 0 & 0 & 0 & 0 & c^{2} & 0 & 0 \\
    0 & 0 & 0 & 0 & -c^{2} & 0 & 0 & 0 \\
    0 & 0 & 0 & 0 & 0 & 0 & 0 & \gamma \\
    0 & 0 & -1 & 0 & 0 & 0 & 0 & 0 \\
    0 & 1 & 0 & 0 & 0 & 0 & 0 & 0 \\
    \chi & 0 & 0 & 0 & 0 & 0 & 0 & 0 \\
    0 & 0 & 0 & c^{2} \gamma & 0 & 0 & 0 & 0
    \end{array}\right).
\end{equation*}

The numerical flux across interface $(i-1/2,j)$ is \cite{leveque1997wave}:
\begin{equation}
    \begin{aligned}
    [\mathscr{F}_{\boldsymbol{Q}}]_{i-1 / 2, j}=& \frac{1}{2}\left(\boldsymbol{A}_{1} \boldsymbol{Q}_{i, j}+\boldsymbol{A}_{1} \boldsymbol{Q}_{i-1, j}\right)-\frac{1}{2}\left(\boldsymbol{A}_{1}^{+} \Delta \boldsymbol{Q}_{i-1 / 2}-\boldsymbol{A}_{1}^{-} \Delta \boldsymbol{Q}_{i-1 / 2}\right) \\
    &+\frac{1}{2} \sum_{p} \operatorname{sign}\left(\lambda_{i-1 / 2, j}^{p}\right)\left(1-\frac{\Delta t}{\Delta x} |\lambda_{i-1 / 2, j}^{p}|\right) \mathcal{L}_{1, i-1 / 2, j}^{p} \Phi\left(\theta_{1, i-/ 2, j}^{p}\right),
    \end{aligned}
    \label{eq:PHM flux}
\end{equation}
where $\mathcal{A}_{1}^{+} = R_1\Lambda^+R_1^{-1}$ and $\mathcal{A}_{1}^{-} = R_1\Lambda^-R_1^{-1}$. $R_1$ is the matrix composed of right eigenvectors of $A_1$, and $\Lambda^{+} = diag((\lambda^1)^{+},(\lambda^2)^{+},\cdots,(\lambda^8)^{+})$ with $\lambda^{+} = max(\lambda,0)$ and $\Lambda^{-} = diag((\lambda^1)^{-},(\lambda^2)^{-},\cdots,(\lambda^8)^{-})$ with $\lambda^{-} = min(\lambda,0)$. $\lambda^p$ is the $p$th eigenvalue of $A_1$. Besides, $\Delta \boldsymbol{Q}_{i-1 / 2} = \boldsymbol{Q}_{i+1}-\boldsymbol{Q}_{i}$. The flux slope in Eq.\eqref{eq:PHM flux} is
\begin{equation*}
    \mathcal{L}_{1, i-1 / 2, j}^{p}=\boldsymbol{l}_{1, i-1 / 2, j}^{p} \cdot\left(\boldsymbol{f}_{1, i, j}-\boldsymbol{f}_{1, i-1, j}\right) \boldsymbol{r}_{1, i-1 / 2, j}^{p},
\end{equation*}
where $\boldsymbol{l}$ and $\boldsymbol{r}$ are the left and right eigenvectors corresponding to eigenvalue $\lambda^p$. $\boldsymbol{f}$ is flux function in Eq.\eqref{eq:2D hyperbolic system}.
The limiter function $\Phi(\theta)$ is,
\begin{align*}
    \theta_{1, i-1 / 2, j}^{p} \equiv \frac{\mathcal{L}_{1, I-1 / 2, j}^{p} \cdot \mathcal{L}_{1, i-1 / 2, j}^{p}}{\mathcal{L}_{1, i-1 / 2, j}^{p} \cdot \mathcal{L}_{1, i-1 / 2, j}^{p}},
    \phi(\theta)=\max (0, \min ((1+\theta) / 2,2,2 \theta)),
\end{align*}
where $I=i-1$ if $\lambda_{i-1/2}^p>0$ and $I=i+1$ if $\lambda^p_{i-1/2}<0$. With the limiters, the scheme is second-order accurate in the smooth region and first-order at or near the discontinuity.

\subsection{Electromagnetic field and fluid interaction}
\label{interaction equation}

The first method intuitively separates the interaction between electromagnetic fields and fluid into distinct categories: fluid behavior and individual particle behavior. The collisional particles first relax to a local equilibrium state, and the interaction with the field occurs in a fluid manner. In contrast, the collisionless particles interact with the field in a particle-like way. Consequently, the interaction between the species and the electromagnetic field should be classified into these two distinct categories.

The second method is applied in regimes with strong collisions and strong magnetization, where the Larmor radius $r_L \rightarrow 0$. To overcome the stiffness of the electromagnetic source term, the Crank-Nicolson method is used to implicitly update the moment equation and electromagnetic field. The updated electromagnetic field is then used to update the particle velocities. In this approach, the moments and particles are all implicitly updated. It is worth noting that in weakly collisional and highly rarefied regimes, the moment equation may not be so accurate, so the first method may be preferable.

\subsubsection{Explicit method}
Here is the introduction of the first method. For the fluid part, the acceleration equation is
\begin{equation*}
    \begin{aligned}
    & \frac{\partial (\rho_i^h \boldsymbol{U}^h_i)}{\partial t}=\frac{e n_i^h}{r_L}\left(\boldsymbol{E}+\boldsymbol{U}^h_i \times \boldsymbol{B}\right), \\
    & \frac{\partial (\rho_e^h \boldsymbol{U}^h_e)}{\partial t}=-\frac{e n_e^h}{r_L}\left(\boldsymbol{E}+\boldsymbol{U}^h_e \times \boldsymbol{B}\right).
    \end{aligned}
\end{equation*}
This equation can be discretized explicitly by the forward Euler method. For the particle part, the acceleration of a particle $P$ is given as
\begin{equation}
\frac{\mathrm{d} \boldsymbol{u}_P}{\mathrm{~d} t}=\frac{q_P}{m_Pr_L}\left(\boldsymbol{E}_P+\boldsymbol{u}_P \times \boldsymbol{B}_P\right),
\label{eq:particle acceleration}
\end{equation}
where $\boldsymbol{E}_P, \boldsymbol{B}_P$ is the electric and magnetic field at the particle's position. The $b$-spline shape function of order 1 is chosen to gather the electric and magnetic field to the particle's position.  The zeroth  order of b-spline is the flat-top function $b_0(\xi)$ defined as
$$
b_0(\xi)= \begin{cases}1 & \text { if }|\xi|<1 / 2, \\ 0 & \text { otherwise. }\end{cases}
$$
The subsequent $b$-splines, $b_{\ell}$, are obtained by successive integration via the following generating formula:
\begin{equation}
b_{\ell}(\xi)=\int_{-\infty}^{\infty} \mathrm{d} \xi^{\prime} b_0\left(\xi-\xi^{\prime}\right) b_{\ell-1}\left(\xi^{\prime}\right) .
\label{eqn:shape property}
\end{equation}

\begin{figure}[H]
    \centering
    \includegraphics[width=0.5\textwidth]{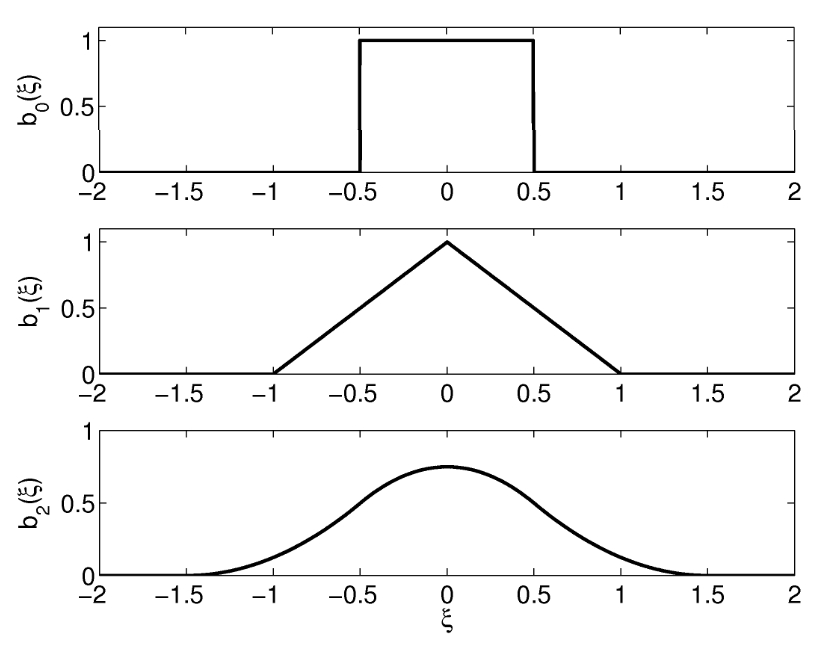}
    \caption{The first three  $b$-spline functions,$b_{\ell}(\xi)$}
    \label{fig:b-spline}
\end{figure}
Based on the $b$-splines, the spatial shape function is chosen as
\begin{equation}
S_{\boldsymbol{x}}\left(\boldsymbol{x}-\boldsymbol{x}_P\right)=\frac{1}{\Delta x_P \Delta y_P \Delta z_P} b_{\ell}\left(\frac{x-x_P}{\Delta x_P}\right) b_{\ell}\left(\frac{y-y_P}{\Delta y_P}\right) b_{\ell}\left(\frac{z-z_P}{\Delta z_P}\right),
\label{eqn:Pic_shaPe}
\end{equation}
where $\Delta x_P, \Delta y_P$, and $\Delta z_P$ are the length of the computational particles in each spatial dimension. The field at the particle position is thus defined as
\begin{equation}
\begin{aligned}
& \boldsymbol{E}_P=\int S_{\boldsymbol{x}}\left(\boldsymbol{x}-\boldsymbol{x}_P\right) \boldsymbol{E}(\boldsymbol{x}) \mathrm{d} \boldsymbol{x}, \\
& \boldsymbol{B}_P=\int S_{\boldsymbol{x}}\left(\boldsymbol{x}-\boldsymbol{x}_P\right) \boldsymbol{B}(\boldsymbol{x}) \mathrm{d} \boldsymbol{x}.
\end{aligned}
\label{eqn:maxshape}
\end{equation}

Boris push \cite{boris1970relativistic} is used to solve the equation \eqref{eq:particle acceleration} numerically. Discretize the Eq.\eqref{eq:particle acceleration} as
$$
\frac{\boldsymbol{u}^{n+1}-\boldsymbol{u}^{*}}{\Delta t}=\frac{q_P}{m_Pr_L}\left[\boldsymbol{E}+\frac{\boldsymbol{u}^{n+1}+\boldsymbol{u}^{*}}{2} \times \boldsymbol{B}\right].
$$
Perform half acceleration to compute $\boldsymbol{v}^{-}$,
$$
\boldsymbol{v}^{-} = \boldsymbol{u}^{*} + \frac{q_P}{m_Pr_L} \boldsymbol{E} \frac{\Delta t}{2}.
$$
Then perform half rotation to compute $\boldsymbol{v}^{'}$.The vector form of the rotation vector is
$$
\boldsymbol{t}=-\hat{b} \tan \left(\frac{\theta}{2}\right) \equiv\frac{q_P}{m_Pr_L}\boldsymbol{B} \frac{\Delta t}{2},
$$
and
$$
\boldsymbol{v}^{\prime}=\boldsymbol{v}+\boldsymbol{v} \times \boldsymbol{t}.
$$
Then perform second half rotation to compute $\boldsymbol{v}^{+}$
$$
\boldsymbol{v}^{+}=\boldsymbol{v}^{-}+\boldsymbol{v}^{\prime} \times \boldsymbol{s},
$$
where
$$
\boldsymbol{s}=\frac{2 \boldsymbol{t}}{1+t^2}.
$$
Finally, perform second-half acceleration to compute $\boldsymbol{v}^{n+1}$.
To enhance temporal accuracy, the leap-frog method can also be employed. In the leap-frog method, velocity updates are performed at half-time steps (time level $n+1/2$), while position updates occur at full-time steps. This staggered approach allows for second-order accuracy in time while requiring only first-order computations at each step. Refer to \cite{birdsall2018plasma,brieda2019plasma} for more details.

Once the fluid part and the particle part have been updated, then the electric field can be updated according to the equation
\begin{equation*}
    \begin{aligned}
    & \frac{\partial \boldsymbol{E}}{\partial t}=-\frac{1}{\lambda_D^2 r_L}\left(\boldsymbol{J}_i-\boldsymbol{J}_e\right),\\
    &\frac{1}{\chi} \frac{\partial \phi}{\partial t}=\frac{{n_i}-{n_e}}{{\lambda_D}^2 {r_{L}}}.
    \end{aligned}
\end{equation*}
This equation is also discretized explicitly with current density being the average as time step $n$ and $n+1$.

\subsubsection{Implicit method}

The moment equation of interaction with the electromagnetic field is
\begin{equation*}
    \begin{aligned}
    & \frac{\partial (\rho_i \boldsymbol{U}_i)}{\partial t}=\frac{e n_i}{r_L}\left(\boldsymbol{E}+\boldsymbol{U}_i \times \boldsymbol{B}\right), \\
    & \frac{\partial (\rho_e \boldsymbol{U}_e)}{\partial t}=-\frac{e n_e}{r_L}\left(\boldsymbol{E}+\boldsymbol{U}_e \times \boldsymbol{B}\right), \\
    & \frac{\partial \boldsymbol{E}}{\partial t}=-\frac{e}{\lambda_D^2 r_L}\left(\boldsymbol{U}_i-\boldsymbol{U}_e\right),\\
    &\frac{1}{\chi} \frac{\partial \phi}{\partial t}=\frac{{n_i}-{n_e}}{{\lambda_D}^2 {r_{L}}}.
    \end{aligned}
\end{equation*}
The above equations can be discretized by the Crank-Nicolson scheme,
\begin{equation*}
    \begin{aligned}
        &\boldsymbol{U}_i^{n+1} - \boldsymbol{U}_i^{**} = \frac{e\Delta t}{m_ir_L}(\frac{\boldsymbol{E}^{n+1} + \boldsymbol{E}^{*}}{2} + \frac{\boldsymbol{U}_i^{n+1} + \boldsymbol{U}_i^{*}}{2}\times\boldsymbol{B}^{n+1}), \\
        &\boldsymbol{U}_e^{n+1} - \boldsymbol{U}_e^{**} = \frac{e\Delta t}{m_er_L}(\frac{\boldsymbol{E}^{n+1} + \boldsymbol{E}^{*}}{2} + \frac{\boldsymbol{U}_e^{n+1} + \boldsymbol{U}_e^{*}}{2}\times\boldsymbol{B}^{n+1}),\\
        &\boldsymbol{E}^{n+1} - \boldsymbol{E}^{*} = -\frac{e\Delta t}{\lambda_D^2r_L}(n_i\frac{\boldsymbol{U}_i^{n+1} + \boldsymbol{U}_i^{*}}{2}- n_e\frac{\boldsymbol{U}_e^{n+1} + \boldsymbol{U}_e^{*}}{2}),\\
        &\boldsymbol{\phi}^{n+1} - \boldsymbol{\phi}^{*} = \frac{\chi\Delta t}{\lambda_D^2r_L}(n_i^{*} - n_e^{*}),
    \end{aligned}
\end{equation*}
which forms a linear system $\boldsymbol{A}\boldsymbol{x}=\boldsymbol{b}$, with
\begin{equation*}
    \boldsymbol{b}=(U_{ix}^{**}, U_{iy}^{**},U_{iy}^{**},U_{ex}^{**}, U_{ey}^{**},U_{ey}^{**},E_x^{*},E_y^{*},E_z^{*})^{T},
\end{equation*}
\begin{equation*}
    \boldsymbol{x}=(U_{ix}^{n+1}, U_{iy}^{n+1},U_{iy}^{n+1},U_{ex}^{n+1}, U_{ey}^{n+1},U_{ey}^{n+1},E_x^{n+1},E_y^{n+1},E_z^{n+1})^{T},
\end{equation*}
and
\begin{equation*}
    \boldsymbol{A}=\left(\begin{array}{ccccccccc}
    1 & -\frac{\alpha B^{n+1}_z}{2} & \frac{\alpha B^{n+1}_y}{2} & 0& 0 & 0 & -\frac{\alpha}{2} & 0 & 0 \\
    \frac{\alpha B^{n+1}_z}{2} & 1 & -\frac{\alpha B^{n+1}_x}{2} & 0& 0 & 0 & 0 & -\frac{\alpha}{2} & 0 \\
    -\frac{\alpha B^{n+1}_y}{2} & \frac{\alpha B^{n+1}_x}{2} & 1 & 0& 0 & 0 & 0 & 0 & -\frac{\alpha}{2} \\
    0 & 0 & 0 & 1 & -\frac{\beta B^{n+1}_z}{2} & \frac{\beta B^{n+1}_y}{2} &  -\frac{\beta}{2} & 0 & 0\\
    0 & 0 & 0 &  \frac{\beta B^{n+1}_z}{2} & 1 &  -\frac{\beta B^{n+1}_x}{2} & 0 &  -\frac{\beta}{2} & 0\\
    0 & 0 & 0 &  -\frac{\beta B_y}{2} &  \frac{\beta B_x}{2} & 1 & 0 & 0 &  -\frac{\beta}{2}\\
    -\frac{\gamma n_i}{2} & 0 & 0 & \frac{\gamma n_e}{2} & 0 & 0 & 1 & 0 & 0\\
    0 & -\frac{\gamma n_i}{2} & 0 & 0 & \frac{\gamma n_e}{2} & 0 & 0 & 1 & 0 \\
    0 & 0 & -\frac{\gamma n_i}{2} & 0 & 0 & \frac{\gamma n_e}{2} & 0 & 0 & 1
    \end{array}\right),
\end{equation*}
where $\alpha = \frac{e\Delta t}{m_ir_L}, \beta = -\frac{e\Delta t}{m_er_L}, \gamma =-\frac{e\Delta t}{\lambda_D^2r_L} $,
$\boldsymbol{B}^{n+1}$ is obtained at the last step. This system can be solved by the Gaussian Elimination method with partial pivoting.

After this step, the electric field $\boldsymbol{E}$ and magnetic field $\boldsymbol{B}$ are all updated to the next time step $n+1$. Then, the particle velocities and positions can be updated using the same method as stated in the previous section, but with the field quantities evaluated at the $n+1$ time step. Compared to the implicit method in PIC \cite{markidis2010multi}, here, the iterative procedure is in the moment equation rather than the particle equation.

\section{Numerical results}
\label{results}
\subsection{Landau damping}

In this section, the Landau damping case is used to test the algorithm's capability to capture the kinetic behavior of the plasma. The results obtained using the UGKWP-PIP method are compared to those from the PIC method. The implementation of the collisional PIC method is illustrated in \ref{collisional PIC}. The results show that the UGKWP-PIP method can effectively capture the kinetic phenomenon of Landau damping, and demonstrates a faster computational speed compared to the PIC method in the strongly collisional plasma regime.

Landau damping is on the scale of plasma frequency $\omega_{pe}$. In this scale, ions are assumed to form an immobile positive charge background and electrostatic force is the dominant. Therefore, the system can be modeled by BGK-Vlasov-Poisson (BGK-VP) equations as follows,
\begin{equation}
    \begin{aligned}
    & \frac{\partial f_e}{\partial t}+\boldsymbol{u}_e \cdot \nabla_x f_e+\frac{e \boldsymbol{E}}{m_e} \cdot \nabla_u f_e= \frac{g_e - f_e}{\tau_e}, \\
    & \boldsymbol{E} = -\nabla \phi, \quad \Laplace \phi = -\frac{e}{\epsilon_0}(n_i - \int  f_e(\boldsymbol{v}) d^3v) .
    \end{aligned}
\end{equation}
The initial distribution for Landau damping is
$$f_{0}(x, u)=\frac{1}{\sqrt{2 \pi}}(1+\alpha \cos (k x)) \mathrm{e}^{-\frac{u^{2}}{2}}.$$
Electrostatic field energy is defined as $$|E|_{L2}=\frac{\epsilon_0}{2}\int \boldsymbol{E}^2dx$$

For linear Landau damping, $\alpha=0.01$ and $k=0.3$, for nonlinear Landau damping is $\alpha=0.5$ and $k=0.5$. In this simulation, cells $N_{grid}=128$, particles per cell $N_{pc}=1000$, domain length $L=2\pi/k$.

\begin{figure}[H]
\begin{subfigure}[b]{0.48\textwidth}
\centering
    \includegraphics[width=0.95\textwidth]{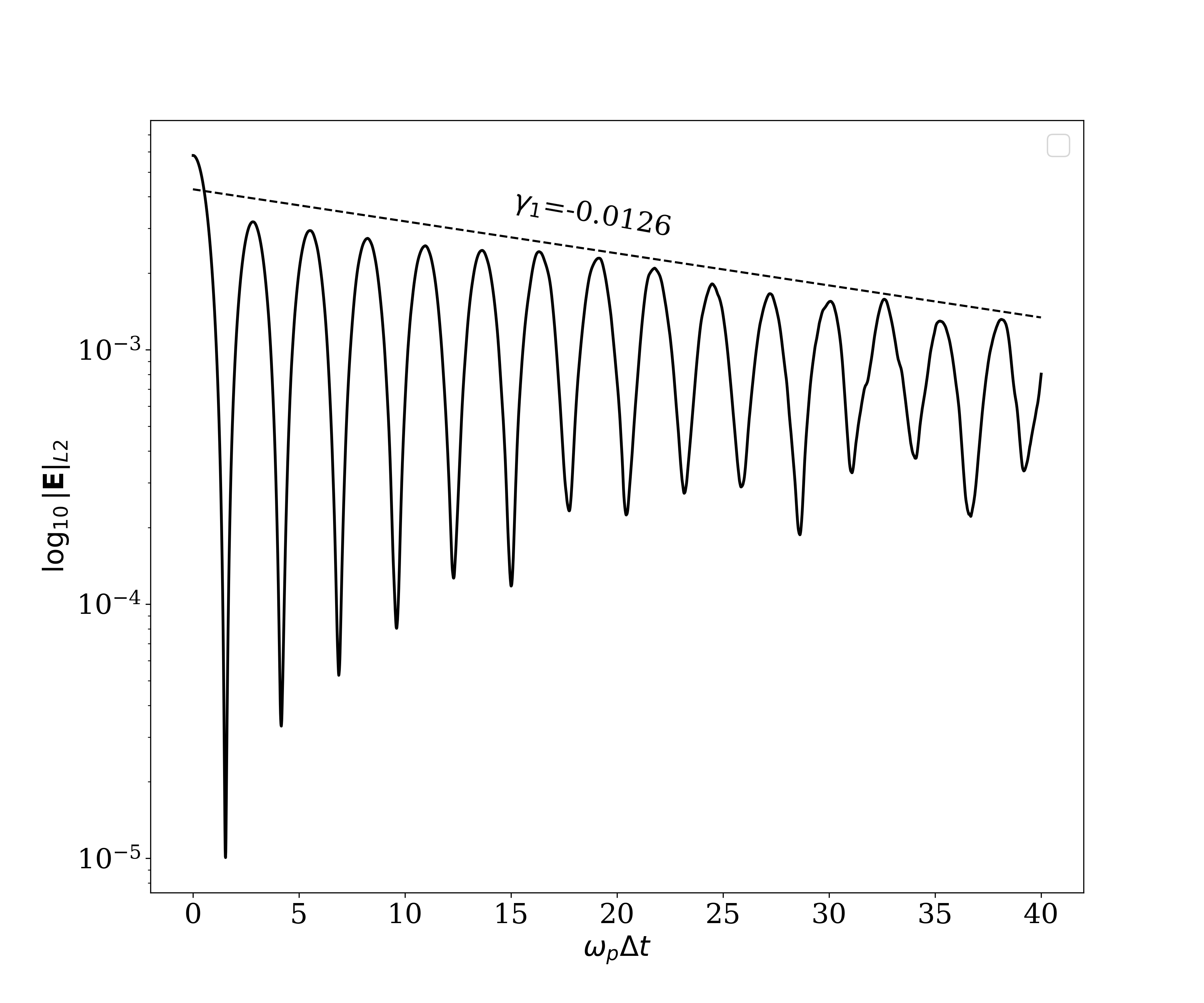}
\caption{}
\label{fig:LLD-kninf-UGKWP}
\end{subfigure}
\begin{subfigure}[b]{0.48\textwidth}
\centering
    \includegraphics[width=0.95\textwidth]{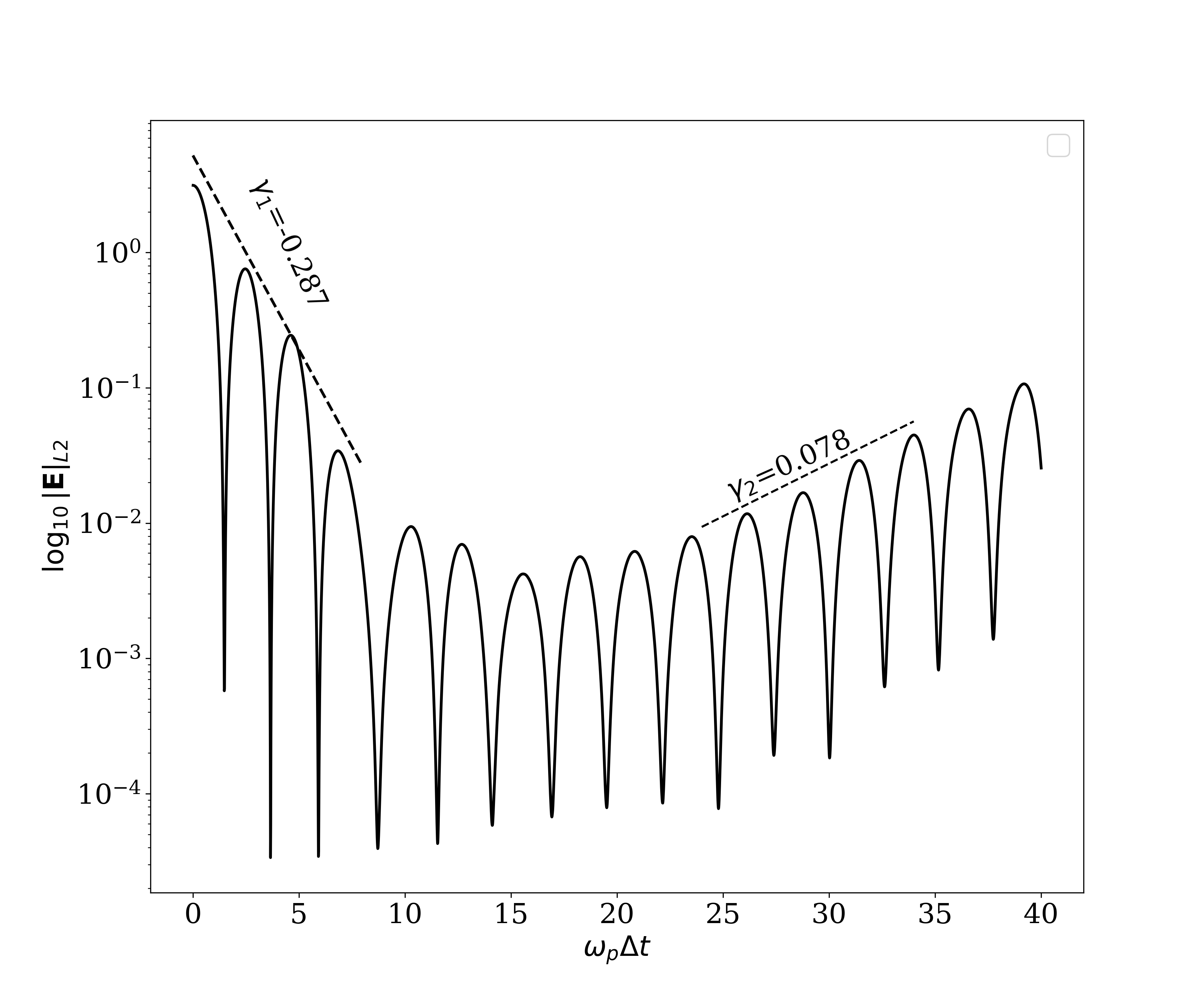}
\caption{}
\label{fig:NLD-kninf-UGKWP}
\end{subfigure}
\caption{(a)Linear Landau damping and (b)nonlinear Landau damping at $Kn=\infty$. The time evolution of electrostatic energy predicted by UGKWP-PIP is the same as the theoretical results.}
\label{fig:LDtheory}
\end{figure}

Figure \ref{fig:LDtheory} shows the results of both linear Landau damping and nonlinear Landau damping in the collisionless limit, where the Knudsen number $Kn=\infty$. The reference damping rates shown are given by the theoretical predictions for this collisionless regime.
In this collisionless limit, the UGKWP-PIP method automatically simplifies to the standard PIC method and accurately reproduces the theoretically predicted Landau damping rates.

\begin{figure}[H]
\begin{subfigure}[b]{0.48\textwidth}
\centering
    \includegraphics[width=0.95\textwidth]{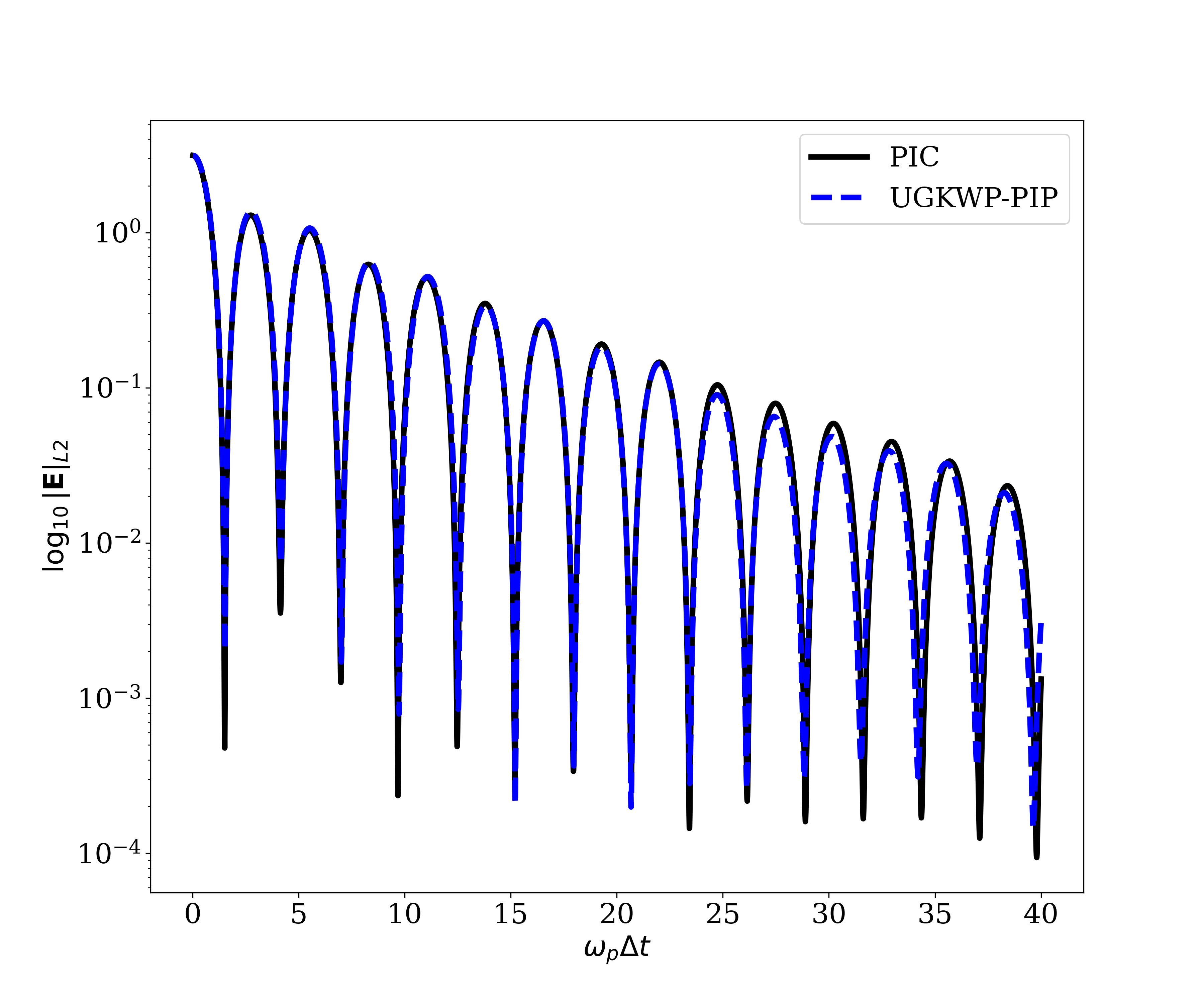}
\caption{}
\label{fig:NLD-kn1.0-UGKWP-PICMCC-mesh128}
\end{subfigure}
\begin{subfigure}[b]{0.48\textwidth}
\centering
    \includegraphics[width=0.95\textwidth]{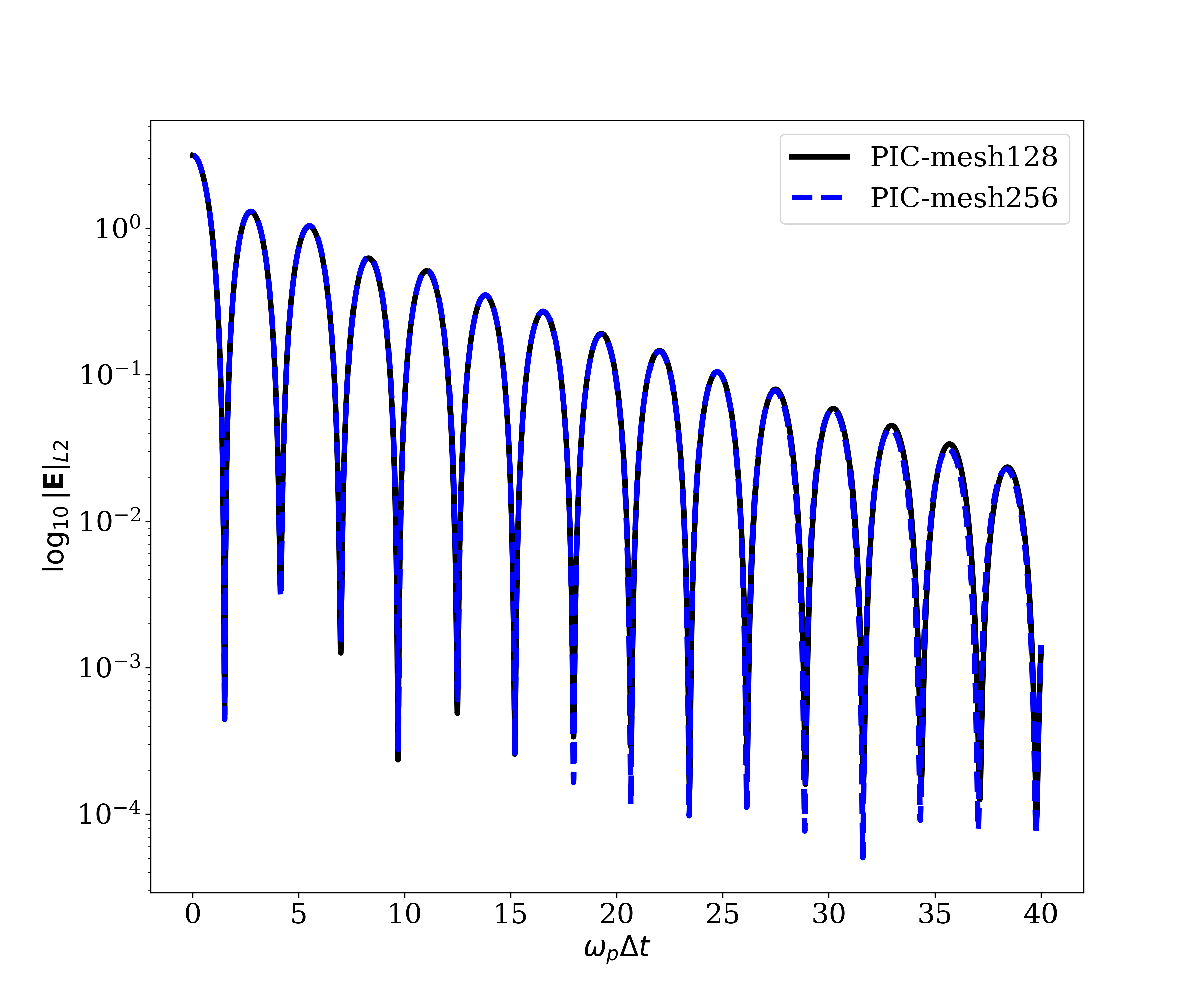}
\caption{}
\label{fig:NLD-kn1.0-PICMCC-different-mesh}
\end{subfigure}
\caption{Nonlinear Landau damping at $Kn=1$. The time evolution of electrostatic energy predicted by UGKWP-PIP and PIC. (a) Comparison of evolution between UGKWP-PIP and PIC when the mesh size is 128. Two methods predict the same results. (b) Evolution of PIC at different mesh sizes: as mesh gets finer, PIC gives the same results.}
\label{fig:NLDkn1}
\end{figure}

Figure \ref{fig:NLD-kn1.0-UGKWP-PICMCC-mesh128} shows the results of nonlinear Landau damping in the weakly collisional regime where the Knudsen number is $Kn=1$. The results obtained using the UGKWP-PIP method and the standard PIC method agree very well in this regime.
To further verify the accuracy of these results, Figure \ref{fig:NLD-kn1.0-PICMCC-different-mesh} presents the PIC simulation results using different mesh sizes. When the mesh size is doubled from 128 to 256, the results remain nearly identical. This suggests that the results shown are the accurate and mesh-converged solutions for nonlinear Landau damping in the weakly collisional $Kn=1$ regime.

\begin{figure}[H]
\begin{subfigure}[b]{0.48\textwidth}
\centering
    \includegraphics[width=0.95\textwidth]{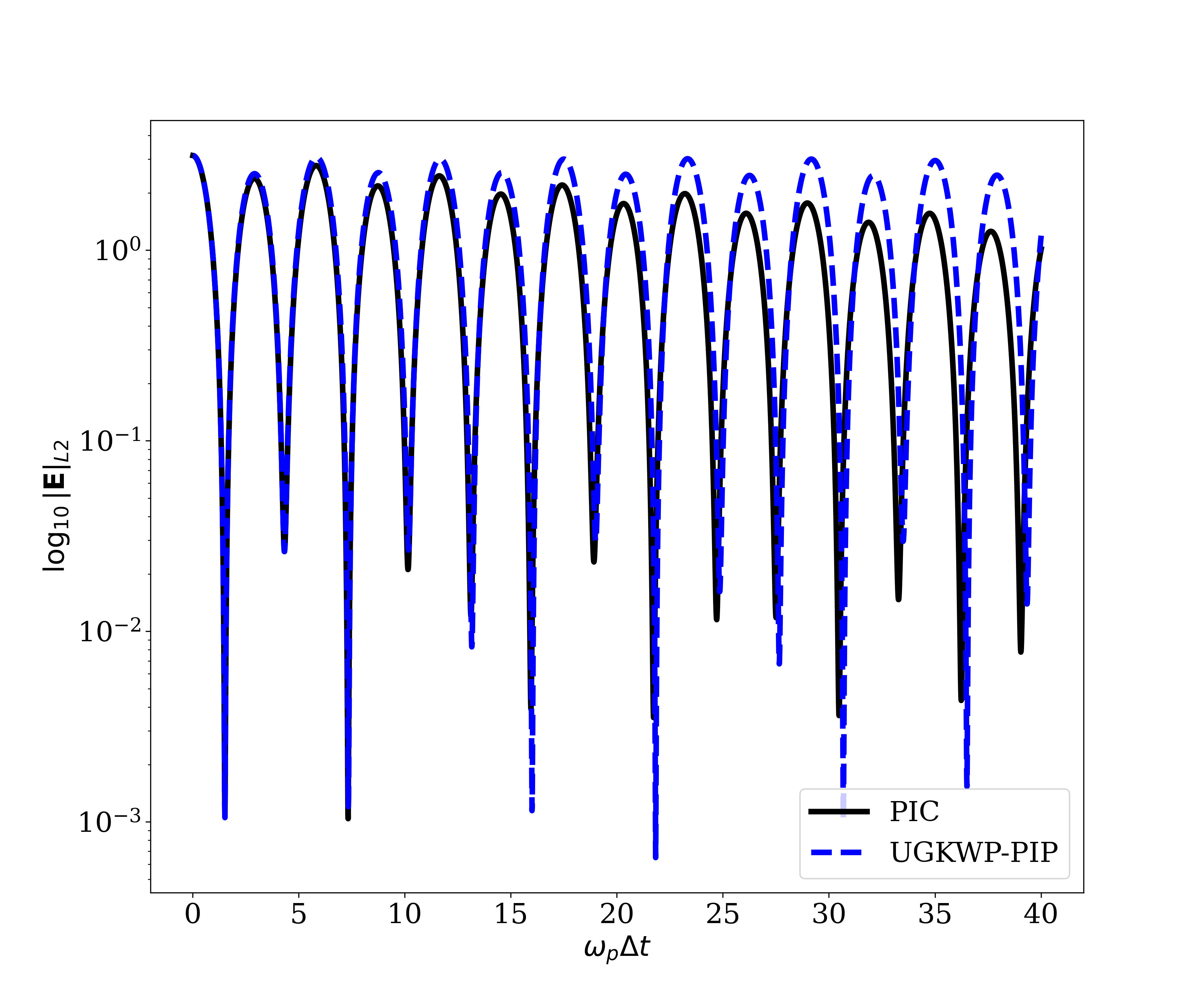}
\caption{}
\label{fig:NLD-Kn0.001-UGKWP-vs-PICMCC-mesh128}
\end{subfigure}
\begin{subfigure}[b]{0.48\textwidth}
\centering
    \includegraphics[width=0.95\textwidth]{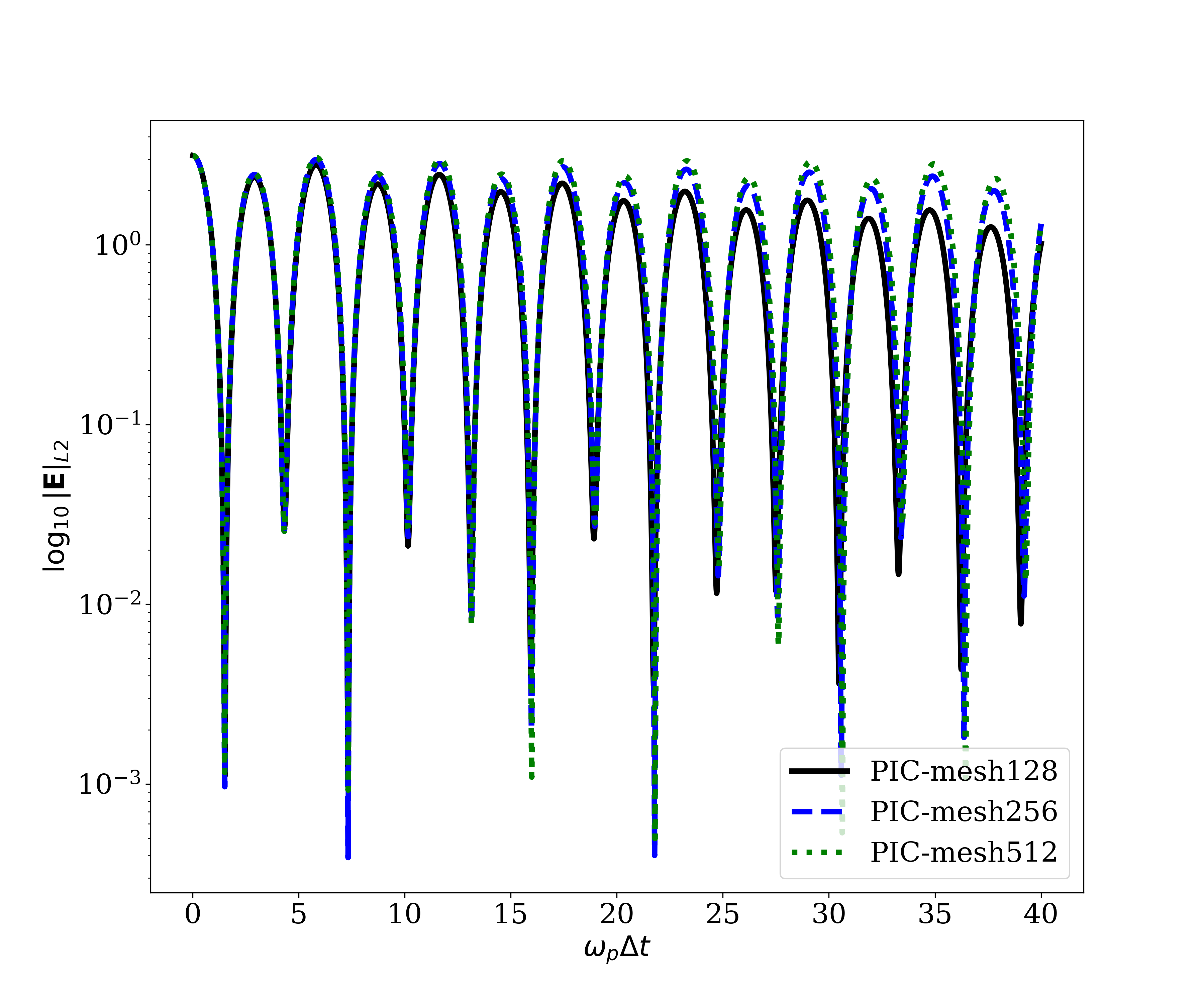}
\caption{}
\label{fig:NLD-Kn0.001-PICMCC-different-mesh}
\end{subfigure}
\caption{Nonlinear Landau damping at $Kn=0.001$. The time evolution of electrostatic energy predicted by UGKWP-PIP and PIC. (a) Comparison of evolution between UGKWP-PIP and PIC when the mesh size is 128. PIC fails to accurately capture the evolution due to coarse mesh. (b) Evolution of PIC at different mesh sizes: with the mesh refinement, PIC converges to the right solution, the same as UGKWP-PIP predicts in (a).}
\end{figure}

Figure \ref{fig:NLD-Kn0.001-UGKWP-vs-PICMCC-mesh128} presents the results of nonlinear Landau damping in the strongly collisional regime where the Knudsen number is $Kn=0.001$. In this regime, the results obtained using the UGKWP-PIP method differ from those predicted by the standard PIC method.
To further investigate the true solution, Figure \ref{fig:NLD-Kn0.001-PICMCC-different-mesh} examines the PIC simulation results under the refinement of mesh sizes to 256 and 512. These results show that as the mesh size is increased, i.e. the cell size becomes smaller, the PIC solution gradually converges towards to the result predicted by the UGKWP-PIP method.
This is reasonable, as the PIC method employs an operator splitting approach, which imposes strict restrictions on the cell size and time step. In this strongly collisional regime, the PIC method requires significantly more computational time compared to the UGKWP-PIP approach. When the mesh size is increased to 512, the PIC method is measured to consume 15 times more computational time than that of the UGKWP-PIP method. This demonstrates the unified-preserving property of the current method.

\begin{figure}[H]
\begin{subfigure}[b]{0.48\textwidth}
\centering
    \includegraphics[width=0.95\textwidth]{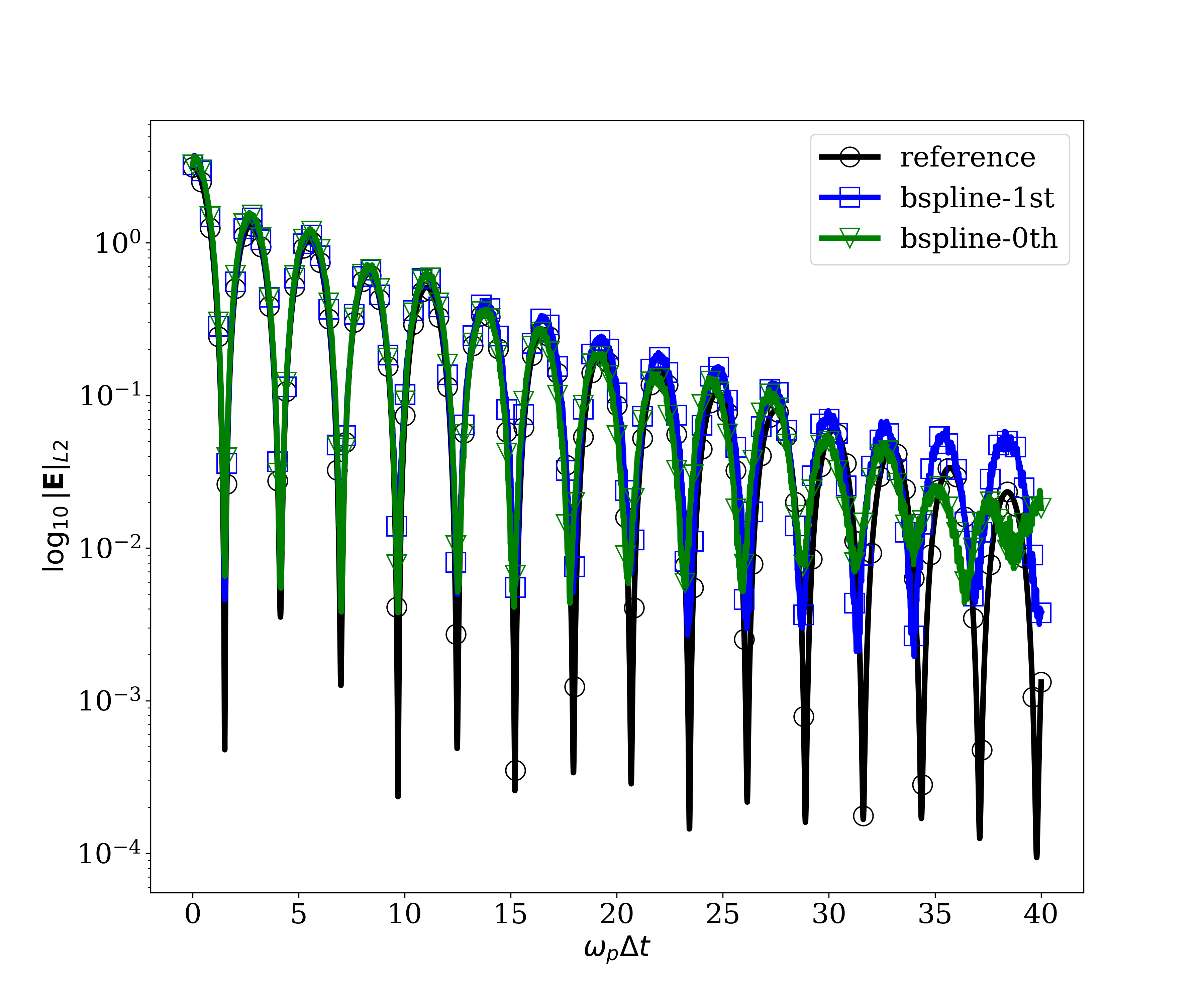}
\caption{}
\label{fig:bspline-kn1-NS}
\end{subfigure}
\begin{subfigure}[b]{0.48\textwidth}
\centering
    \includegraphics[width=0.95\textwidth]{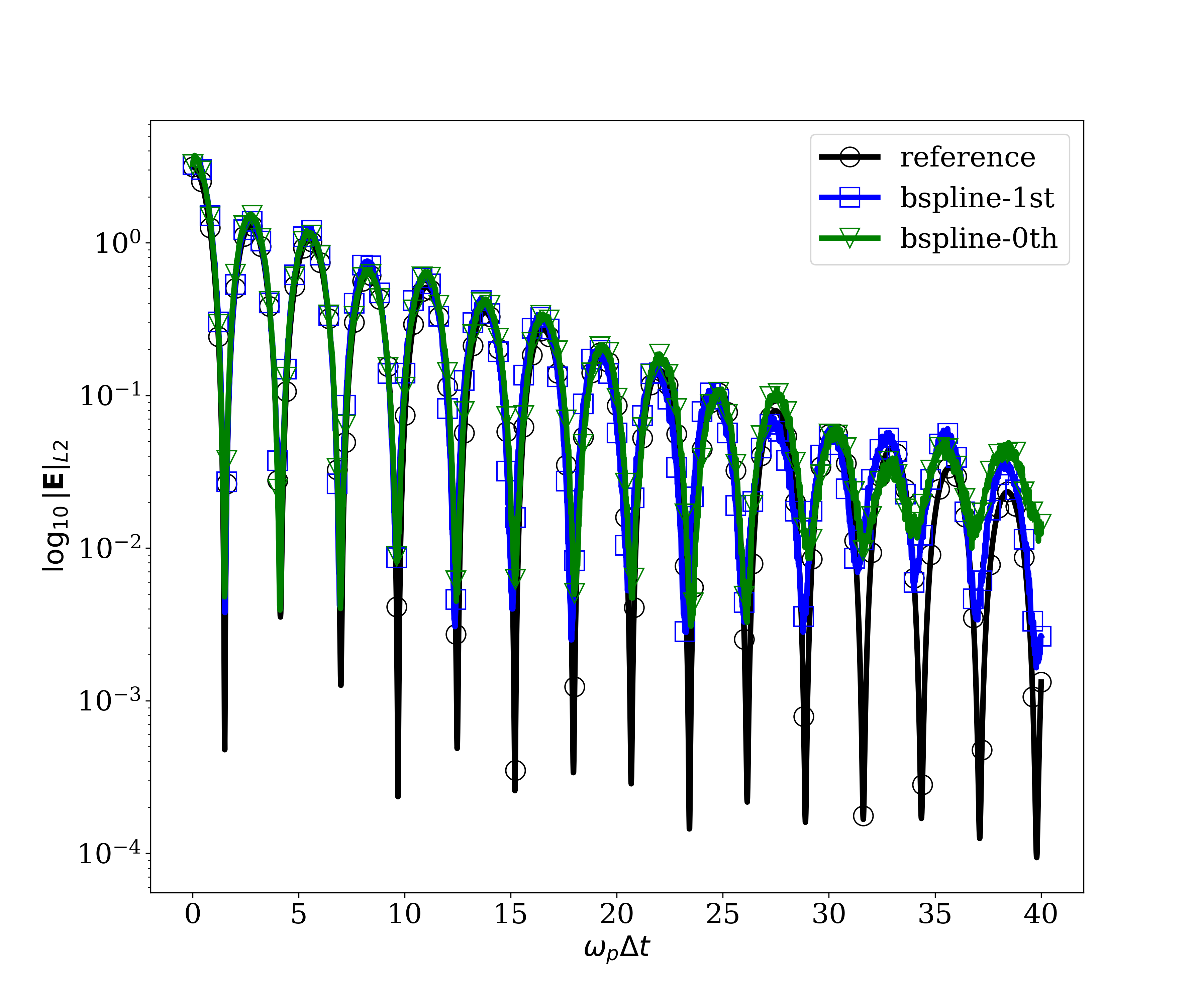}
\caption{}
\label{fig:bspline-kn1-QS}
\end{subfigure}
\caption{Effects of order of b-spline shape function on nonlinear Landau damping at $Kn=1$. (a) Noise start: in this case, both the 0th and 1st b-spline shape function deviates from the reference solution a lot, but the 1st b-spline one preserves the information better while the 0th one decays faster. (b) quiet start: in this case, 1st b-spline shape function is better than the 0th one in both the accuracy and the information-preserving capability. }
\end{figure}

Figures \ref{fig:bspline-kn1-NS} and \ref{fig:bspline-kn1-QS} test the effects of the shape function used in the numerical scheme. Figure \ref{fig:bspline-kn1-NS} employs a "noise start" method (see more details in \ref{collisional PIC}), while Figure \ref{fig:bspline-kn1-QS} utilizes a "quiet start" method.
In both cases, the results obtained using the 1st-order b-spline shape function are shown to be superior to those from the 0th-order b-spline shape function, which is equivalent to using no shape function at all. This indicates that the higher-order b-spline shape function can better capture the kinetic effects in the weakly collisional regime.

\subsection{Multiscale Brio-Wu tests}

In this section, the Brio-Wu cases are first tested for different values of the Knudsen number, proving the multiscale capability of the current numerical scheme. Next, the Brio-Wu cases are evaluated at the ion inertial scale and the electron inertial scale. Finally, the Brio-Wu cases are tested in a weakly ionized plasma to examine the dissipation effects caused by collisions between the plasma and neutral particles.

The initial condition for the test is

$$
\left[\begin{array}{c}
\rho_e \\
u_e \\
v_e \\
w_e \\
p_e \\
\rho_i \\
u_i \\
v_i \\
w_i \\
p_i \\
B_x, \\
B_y, \\
B_z, \\
E_x, \\
E_y, \\
E_z
\end{array}\right]_l=\left[\begin{array}{c}
1.0 \frac{m_e}{m_i} \\
0 \\
0 \\
0 \\
0.5 \\
1.0 \\
0 \\
0 \\
0 \\
0.5 \\
0.75 \\
1.0 \\
0 \\
0 \\
0 \\
0
\end{array}\right]
\quad
\left[\begin{array}{c}
\rho_e \\
u_e \\
v_e \\
w_e \\
p_e \\
\rho_i \\
u_i \\
v_i \\
w_i \\
p_i \\
B_x, \\
B_y, \\
B_z \\
E_x, \\
E_y \\
E_z
\end{array}\right]_r=\left[\begin{array}{c}
0.125 \frac{m_e}{m_i} \\
0 \\
0 \\
0 \\
0.05\\
0.125 \\
0 \\
0 \\
0 \\
0.05 \\
0.75 \\
-1.0 \\
0 \\
0 \\
0 \\
0
\end{array}\right]
$$

\begin{figure}[H]
\centering
\begin{subfigure}[b]{0.48\textwidth}
    \centering
    \includegraphics[width=0.98\textwidth]{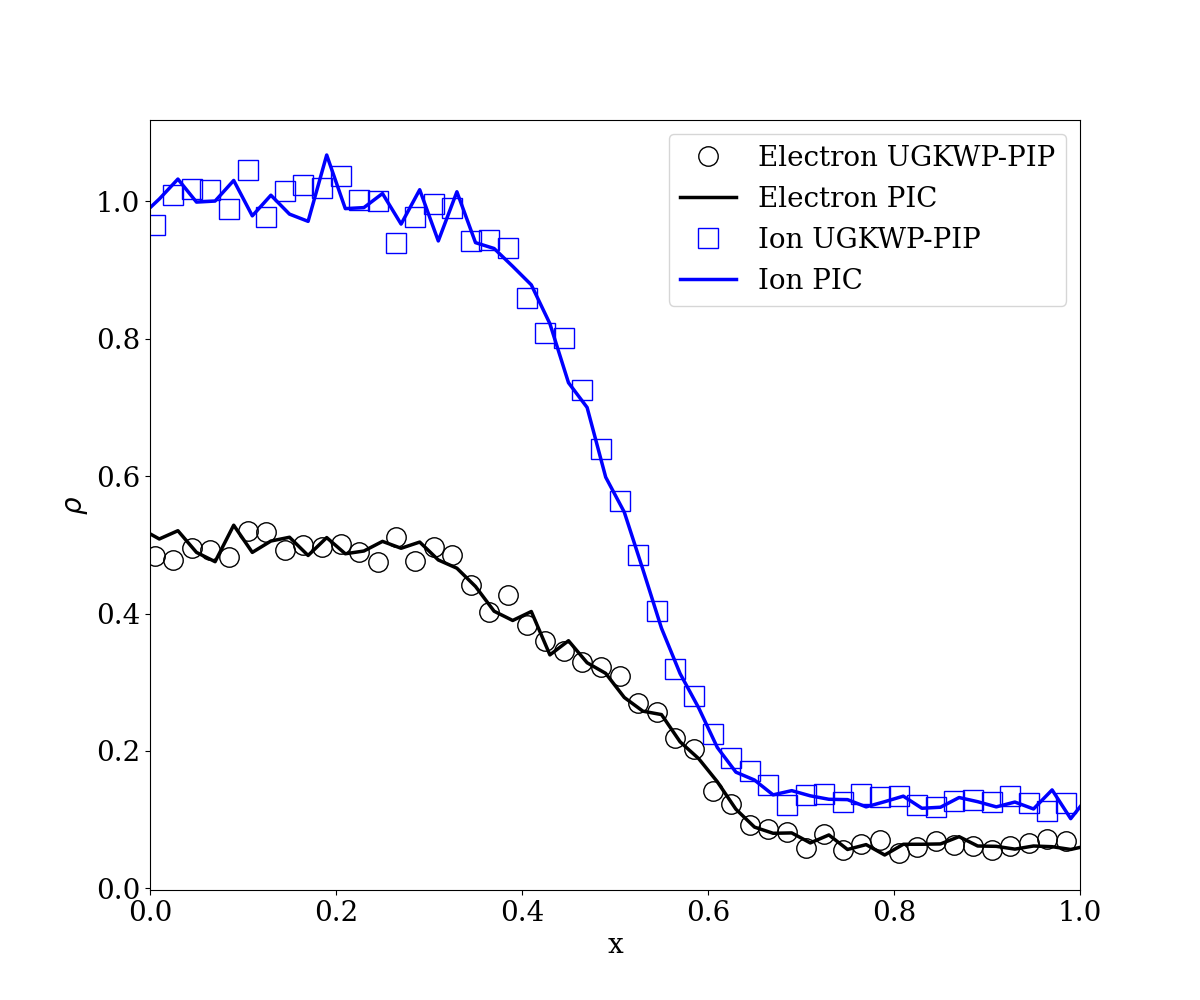}
    \caption{}
\end{subfigure}
\begin{subfigure}[b]{0.48\textwidth}
    \centering
    \includegraphics[width=0.98\textwidth]{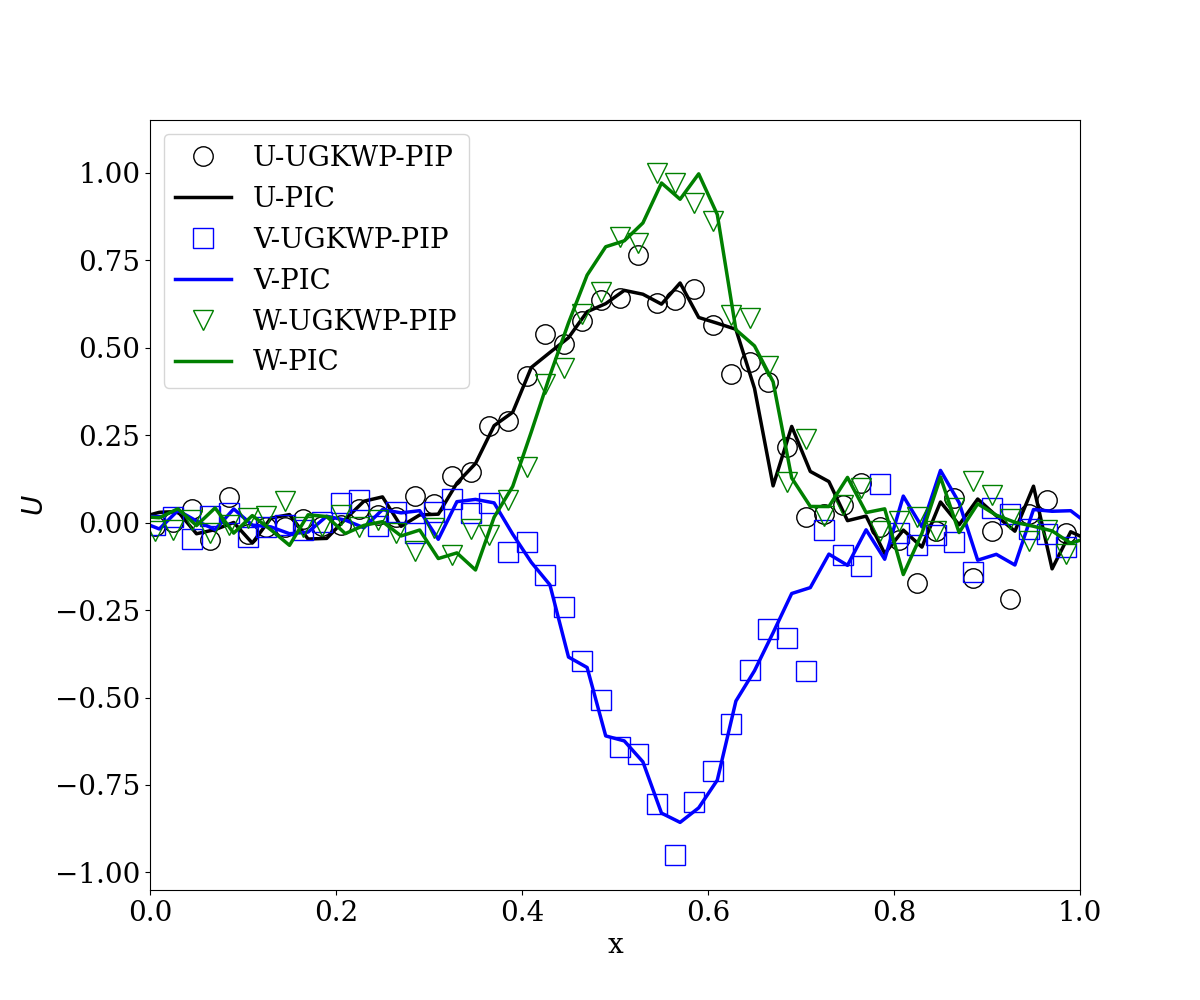}
    \caption{}
\end{subfigure}
\vskip\baselineskip
\begin{subfigure}[b]{0.48\textwidth}
    \centering
    \includegraphics[width=0.98\textwidth]{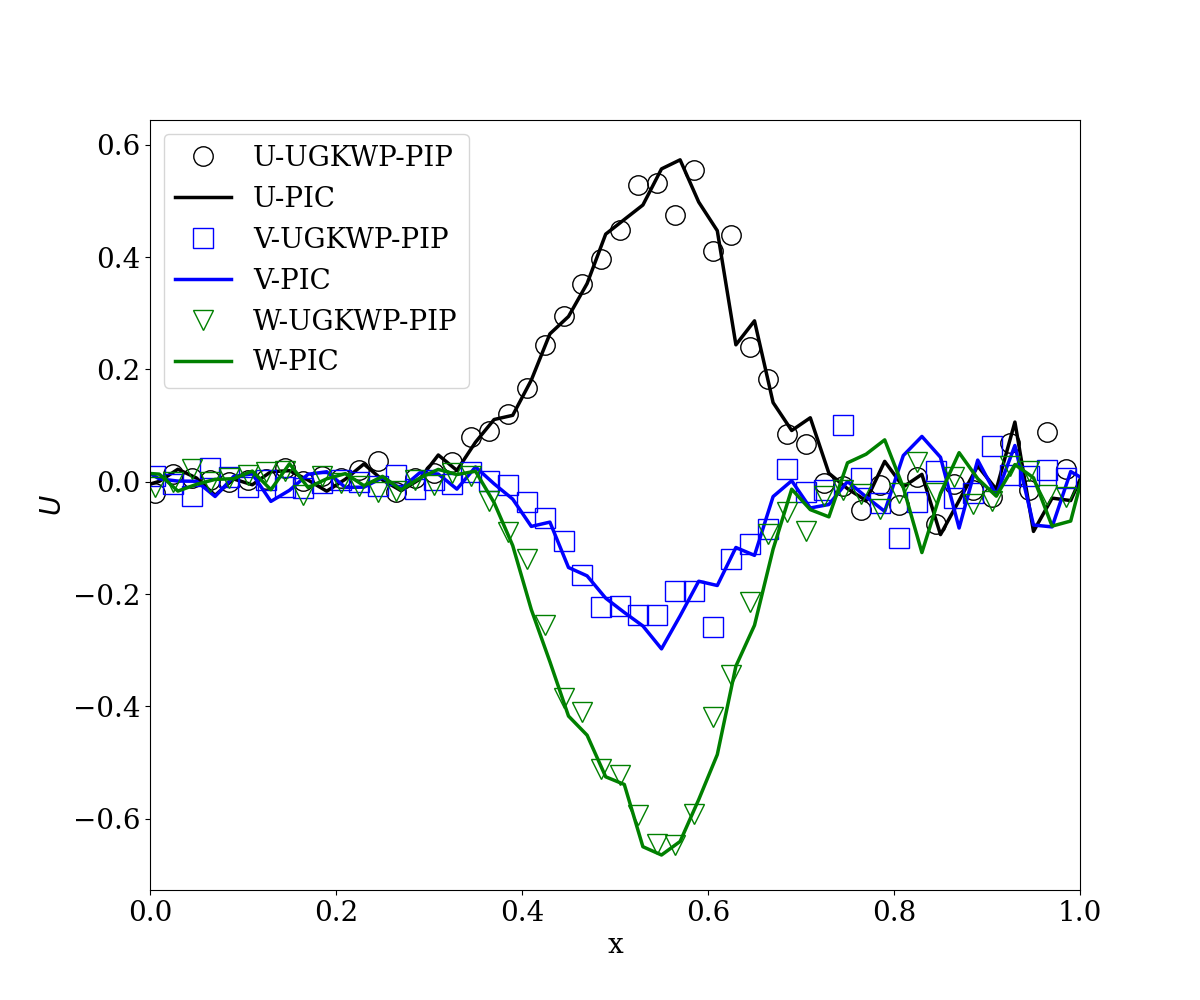}
    \caption{}
\end{subfigure}
\begin{subfigure}[b]{0.48\textwidth}
    \centering
    \includegraphics[width=0.98\textwidth]{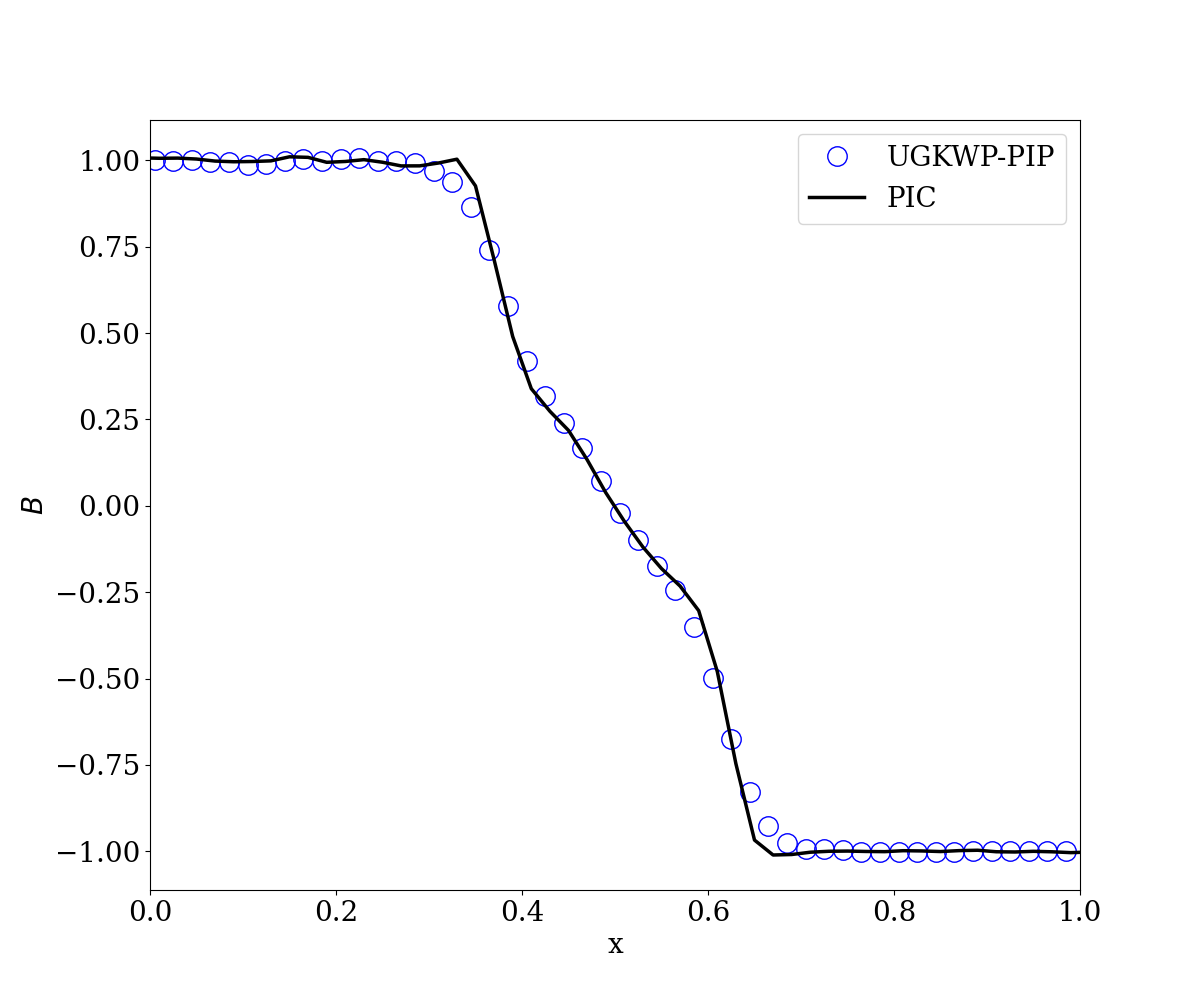}
    \caption{}
\end{subfigure}
\caption{(a) Density, (b) electron velocity, (c) ion velocity (d) $B_y$  profile at the fully ionized limit, $Kn=1.0, r_L=0.1$. }
\label{fig:briokn1}
\end{figure}

\begin{figure}[H]
\centering
\begin{subfigure}[b]{0.48\textwidth}
    \centering
    \includegraphics[width=0.98\textwidth]{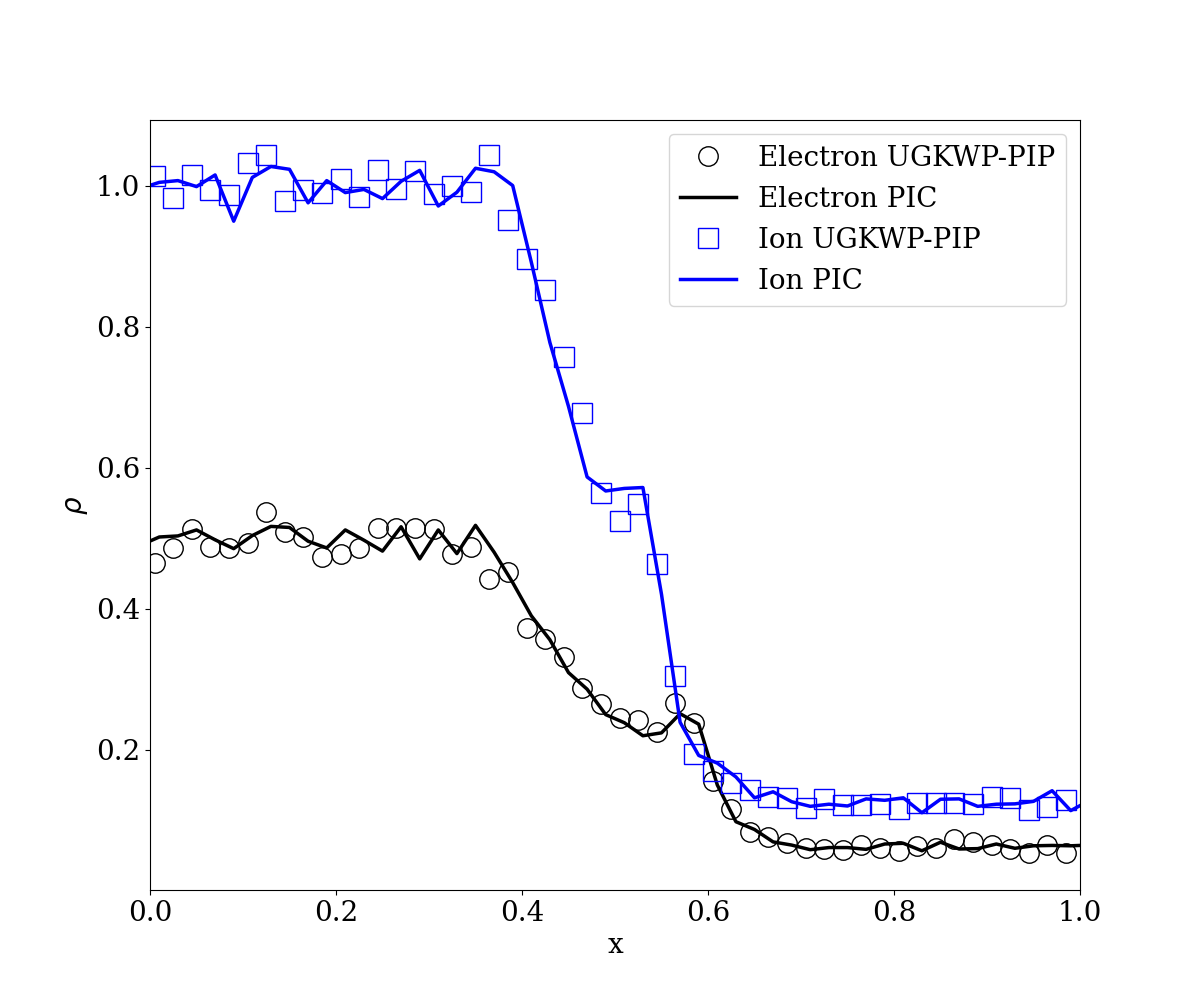}
    \caption{}
\end{subfigure}
\begin{subfigure}[b]{0.48\textwidth}
    \centering
    \includegraphics[width=0.98\textwidth]{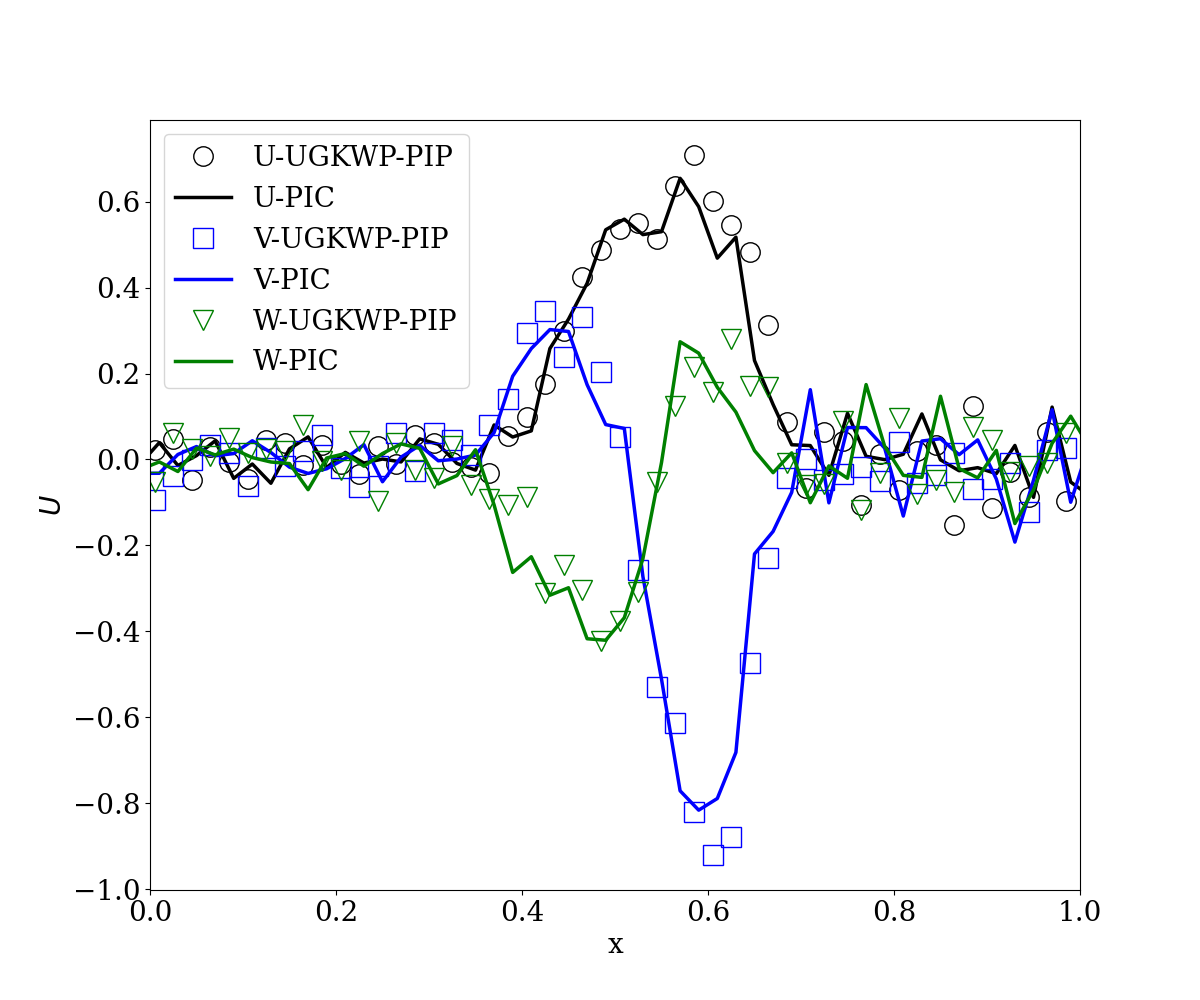}
    \caption{}
\end{subfigure}
\vskip\baselineskip
\begin{subfigure}[b]{0.48\textwidth}
    \centering
    \includegraphics[width=0.98\textwidth]{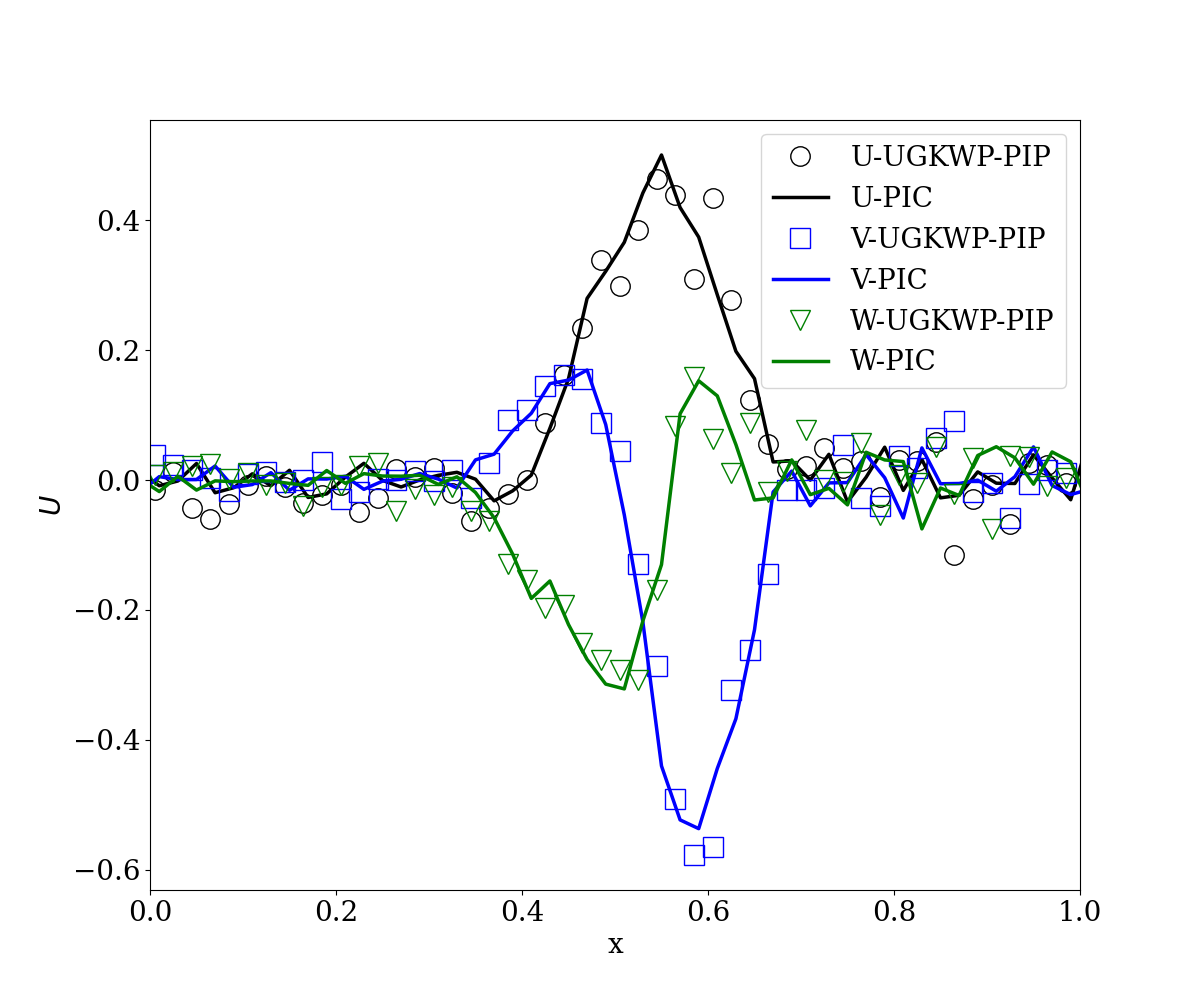}
    \caption{}
\end{subfigure}
\begin{subfigure}[b]{0.48\textwidth}
    \centering
    \includegraphics[width=0.98\textwidth]{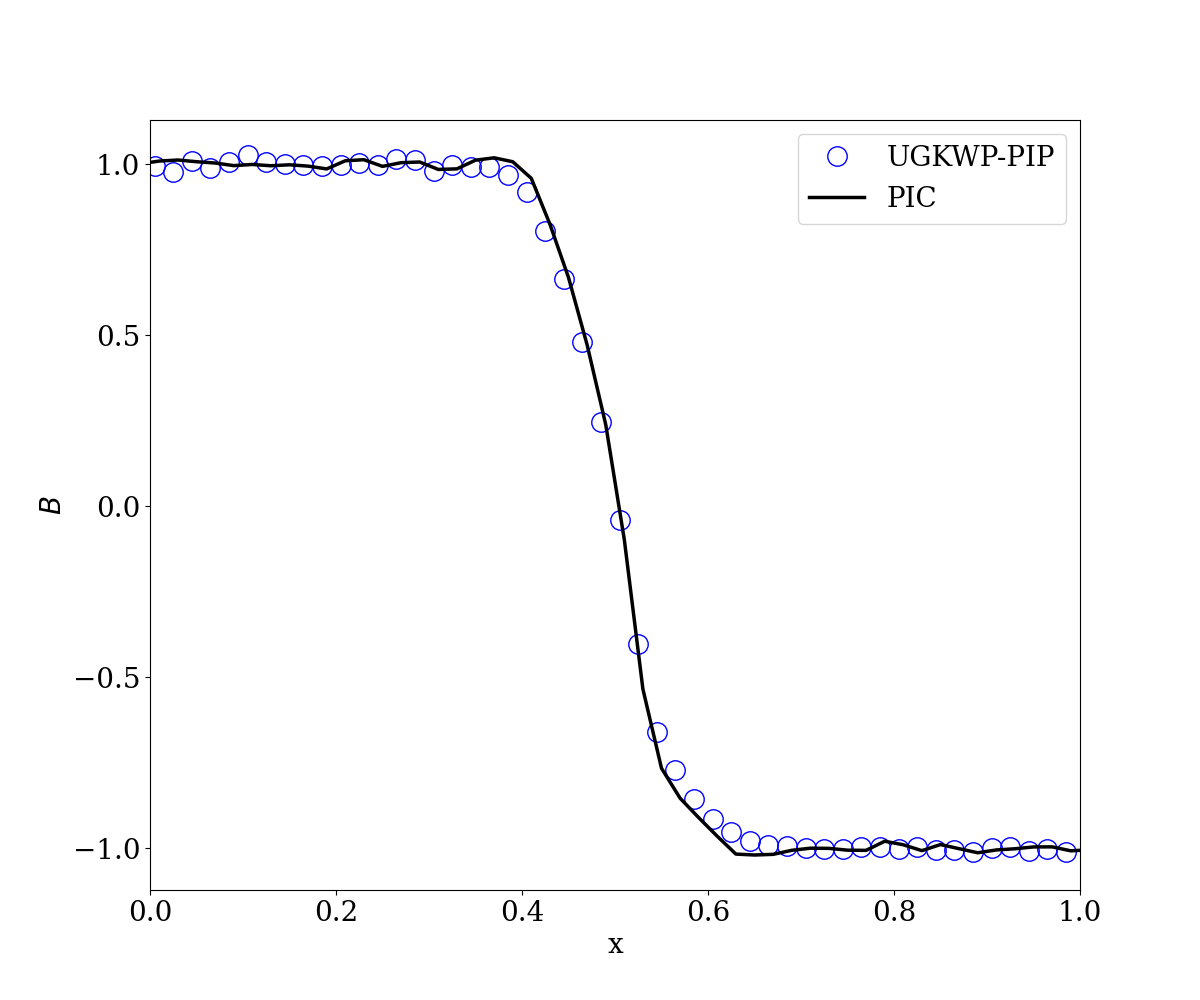}
    \caption{}
\end{subfigure}
\caption{(a) Density, (b) electron velocity, (c) ion velocity (d) $B_y$ profile at the fully ionized limit, $Kn=0.01, r_L=0.01$. }
\label{fig:briokn0.01}
\end{figure}

\begin{figure}[H]
\centering
\begin{subfigure}[b]{0.48\textwidth}
    \centering
    \includegraphics[width=0.98\textwidth]{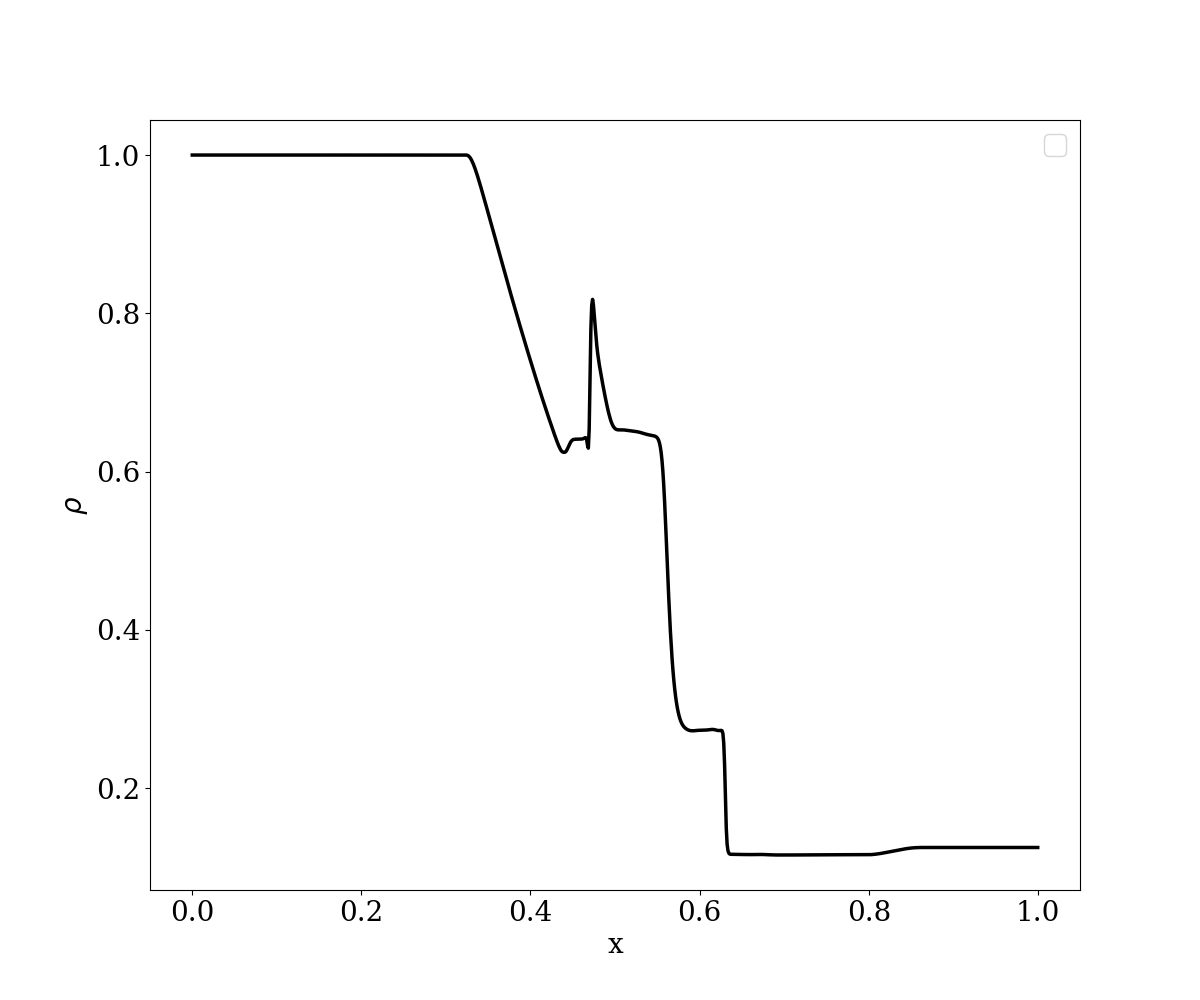}
    \caption{}
\end{subfigure}
\begin{subfigure}[b]{0.48\textwidth}
    \centering
    \includegraphics[width=0.98\textwidth]{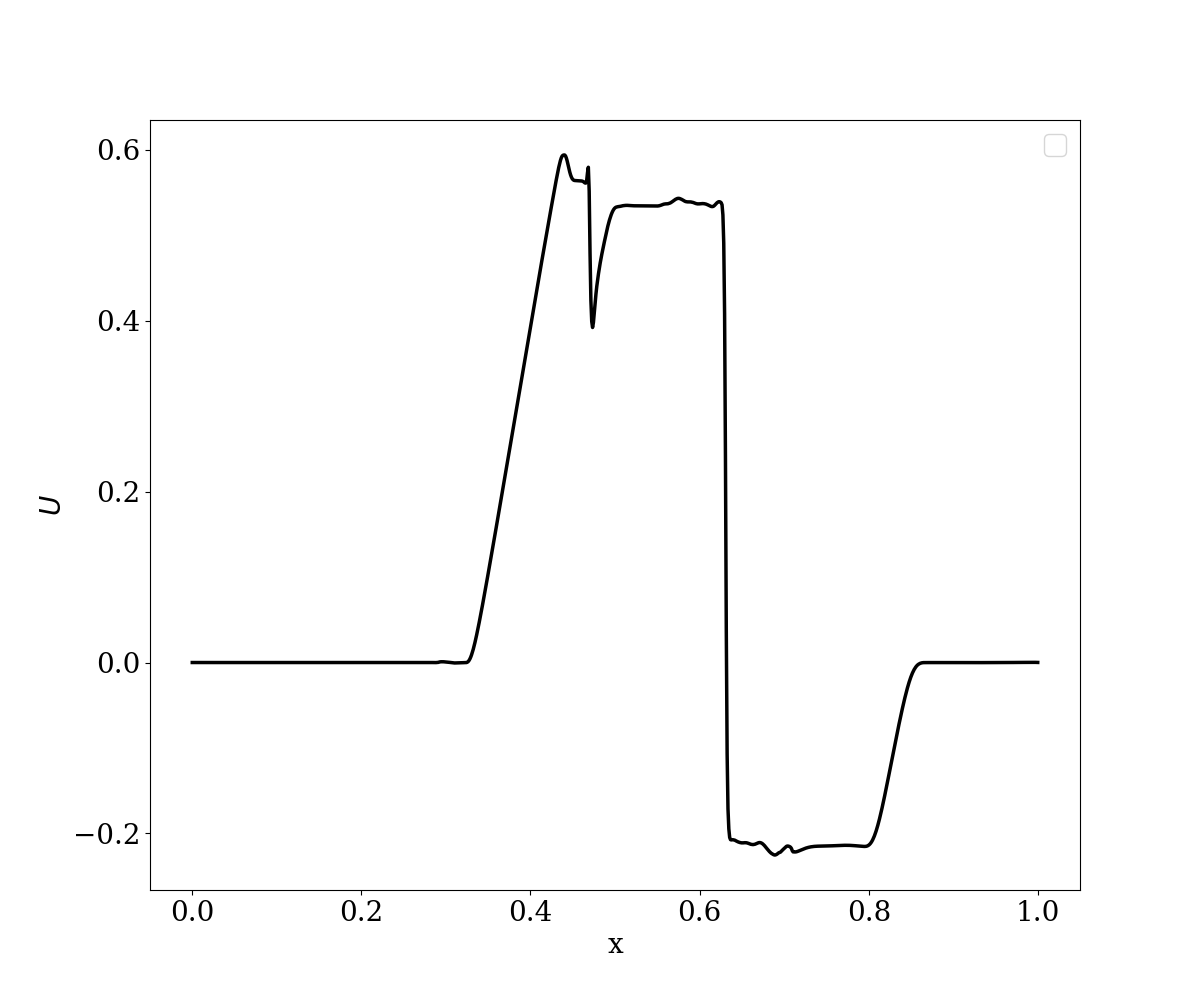}
    \caption{}
\end{subfigure}
\vskip\baselineskip
\begin{subfigure}[b]{0.48\textwidth}
    \centering
    \includegraphics[width=0.98\textwidth]{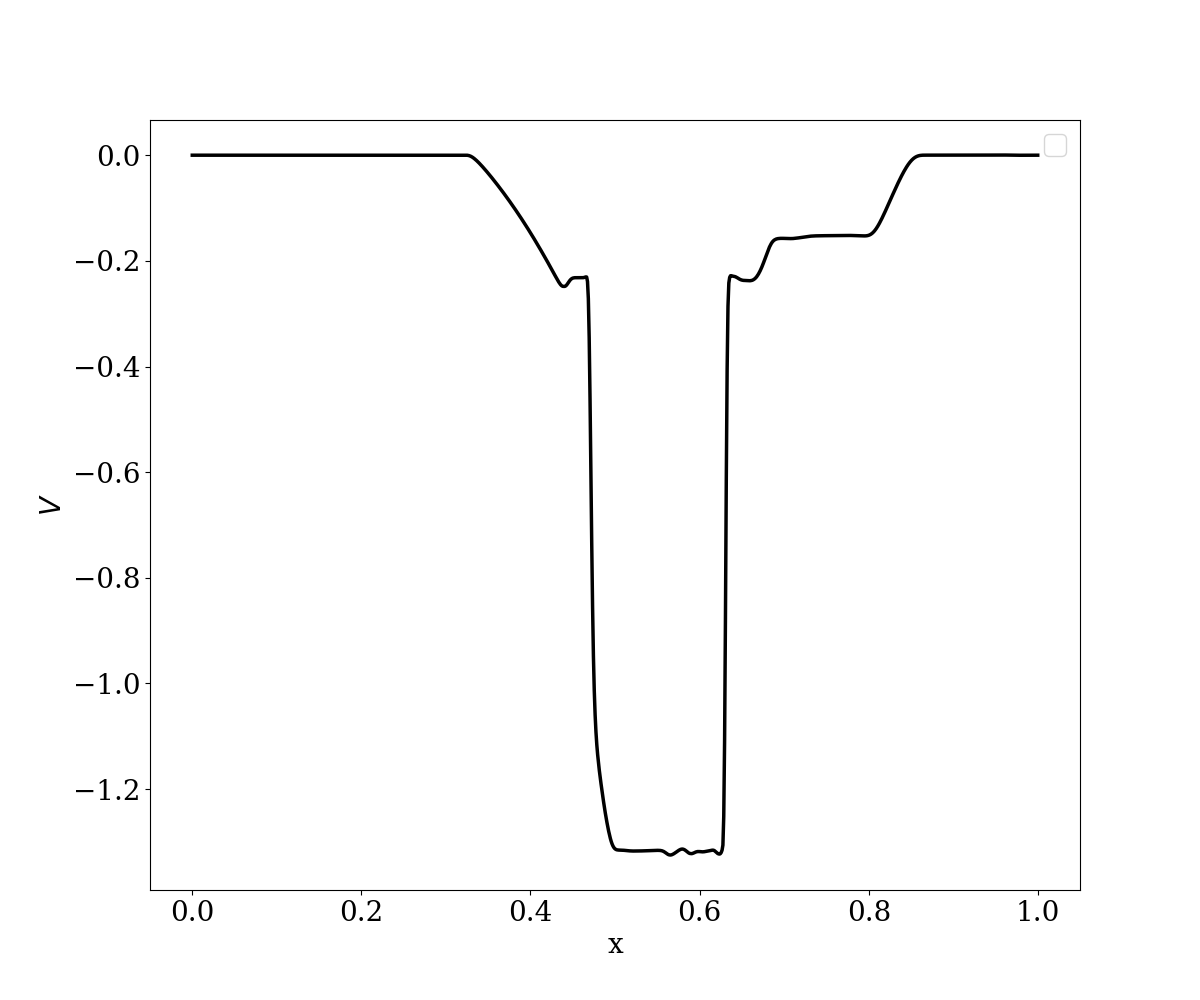}
    \caption{}
\end{subfigure}
\begin{subfigure}[b]{0.48\textwidth}
    \centering
    \includegraphics[width=0.98\textwidth]{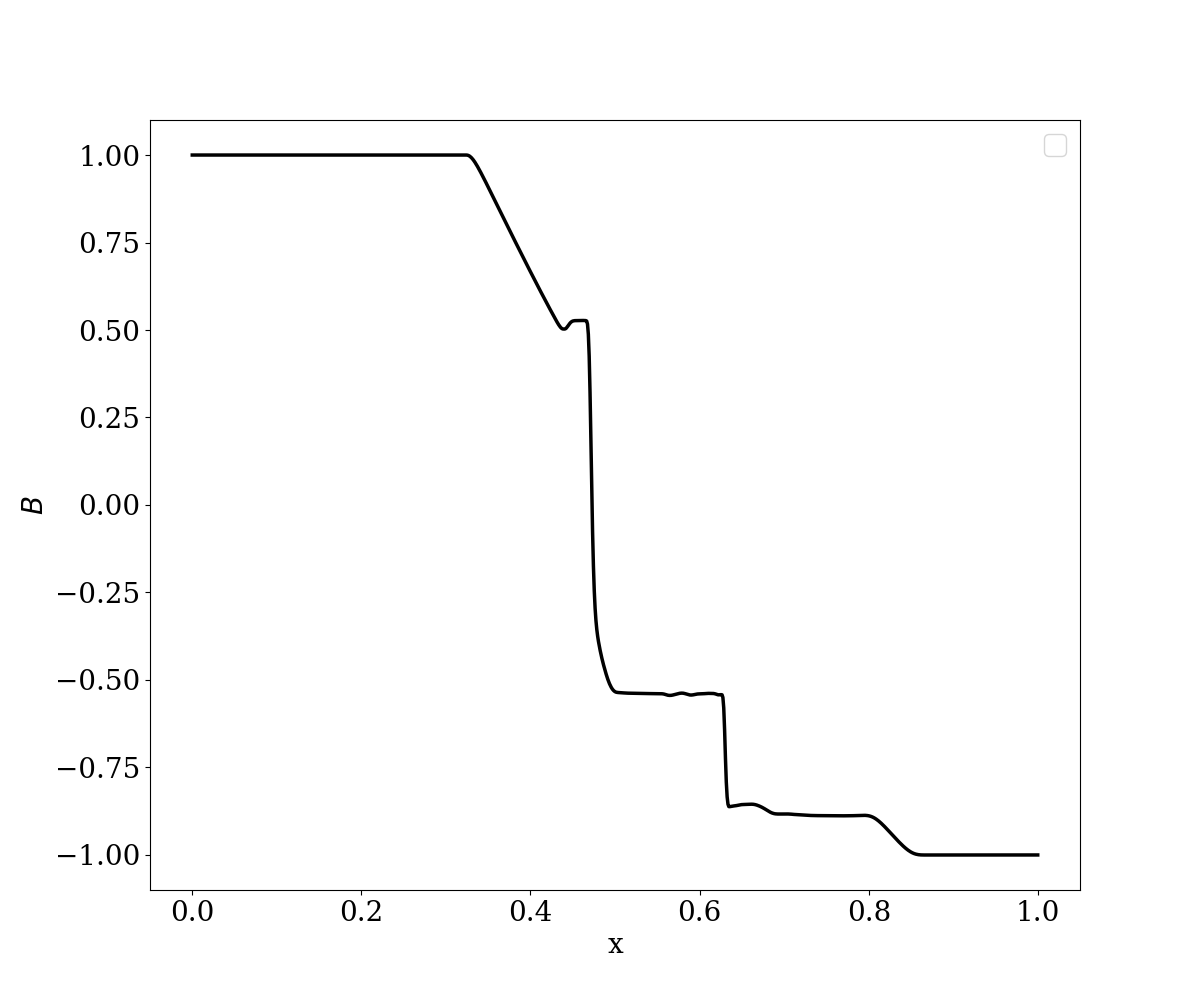}
    \caption{}
\end{subfigure}
\caption{(a) Density, (b) x-velocity, (c) y-velocity (d) $B_y$ profile at fully ionized limit, $Kn=0, r_L=0.0001$ from the current UGKWP. }
\label{fig:briokn0}
\end{figure}

Figure \ref{fig:briokn1} to \ref{fig:briokn0} shows Brio Wu shock tube results across a range of Knudsen numbers, from $Kn=1$ (rarefied regime) to $Kn=0$ (continuum regime). In the rarefied regime, the solution of the UGKWP-PIP agrees well with the PIC method. In this rarefied regime, the increased viscosity due to enhanced free molecular transport causes the shock wave to disappear, resulting in a smooth transition from the left to right states. As the Knudsen number decreases towards the continuum regime ($Kn=0$), the discontinuity across the shock wave begins to re-emerge due to the diminishing viscosity effects. In the continuum regime, the results converge to the ideal MHD solution.

\begin{figure}[H]
\centering
\begin{subfigure}[b]{0.48\textwidth}
    \centering
    \includegraphics[width=0.98\textwidth]{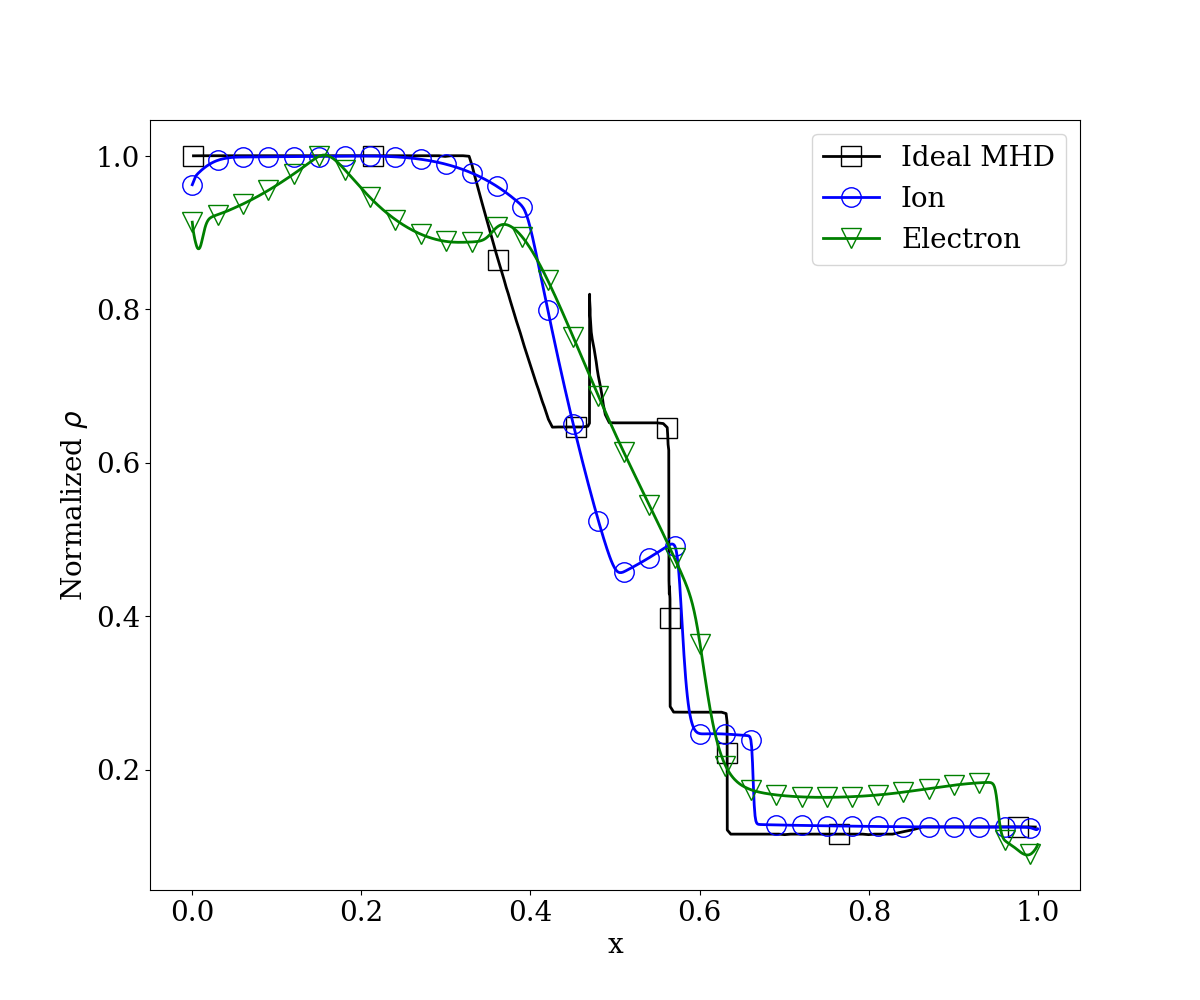}
    \caption{}
\end{subfigure}
\begin{subfigure}[b]{0.48\textwidth}
    \centering
    \includegraphics[width=0.98\textwidth]{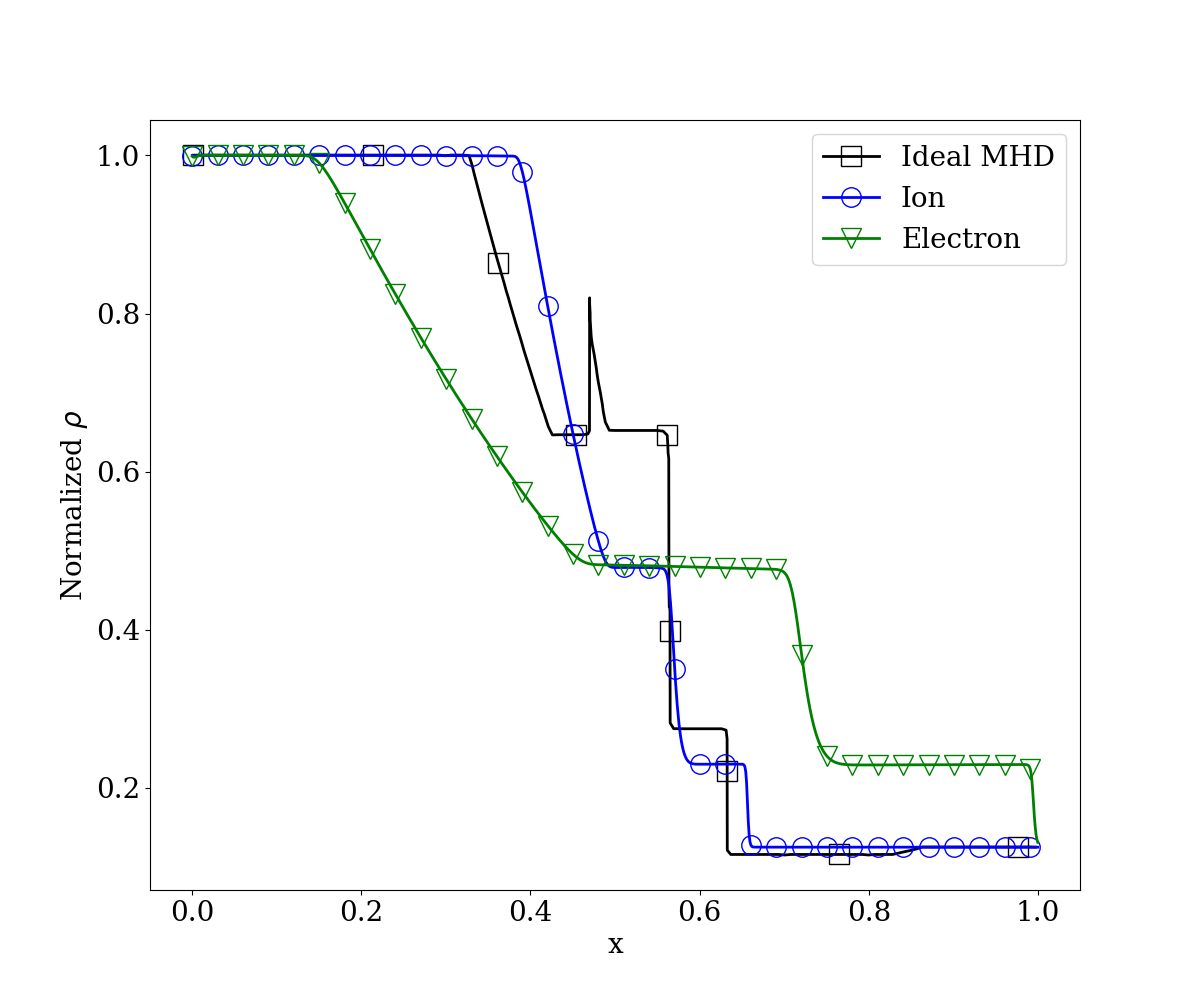}
    \caption{}
\end{subfigure}
\caption{Normalized density profile at fully ionized limit, $Kn=0$, (a) $r_L=1$, (b) $r_L=10$. As $r_L$ gets larger, the system goes from the MHD scale to the Hall-MHD scale, i.e. ion inertia scale, at $r_L=1$, where ions are demagnetized and come to Euler riemann solution and electrons are still partially frozen on the magnetic line. Finally, after electrons are demagnetized, the system goes to the electron inertia scale at $r_L=10$.}
\label{fig:ioninertia}
\end{figure}

Figure \ref{fig:ioninertia} shows the results across different Larmor radius. According to Eq.\eqref{eq:skin length}, as the Larmor radius increases, the ion inertial length $d_i$ also increases. When the ion inertial length becomes comparable to the characteristic length scale, $d_i \sim \mathcal{O}(1)$, the magnetic field lines are no longer frozen into the ion motion, and the wave structure of the ions transitions from a MHD structure to an Euler structure. Similarly, when the electron inertial length $d_e$ becomes comparable to the characteristic length scale, $d_e \sim \mathcal{O}(1)$, the electrons also experience a similar transition in their wave structures.

\begin{figure}[H]
\centering
\begin{subfigure}[b]{0.31\textwidth}
    \centering
    \includegraphics[width=0.98\textwidth]{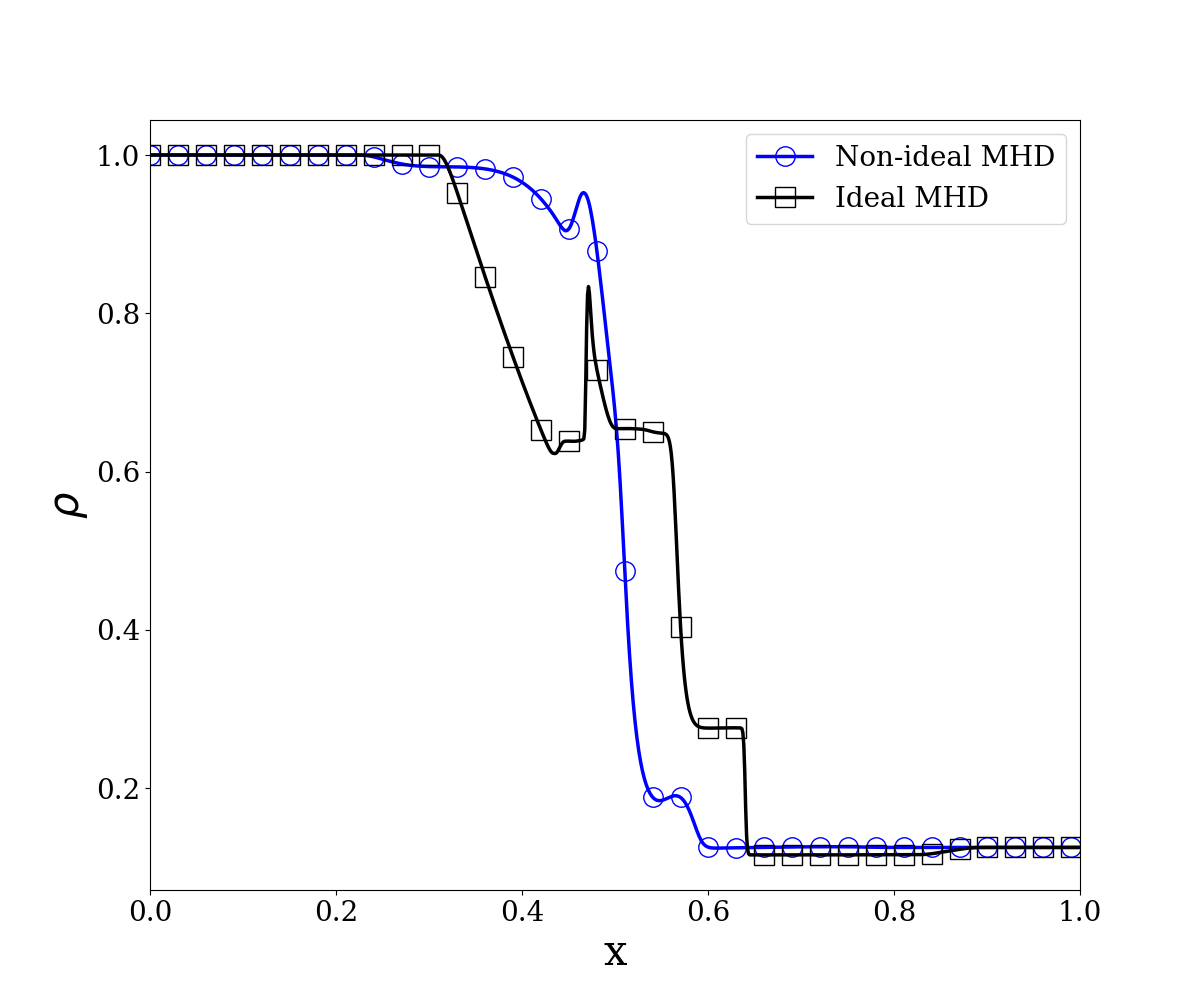}
    \caption{}
\end{subfigure}
\begin{subfigure}[b]{0.31\textwidth}
    \centering
    \includegraphics[width=0.98\textwidth]{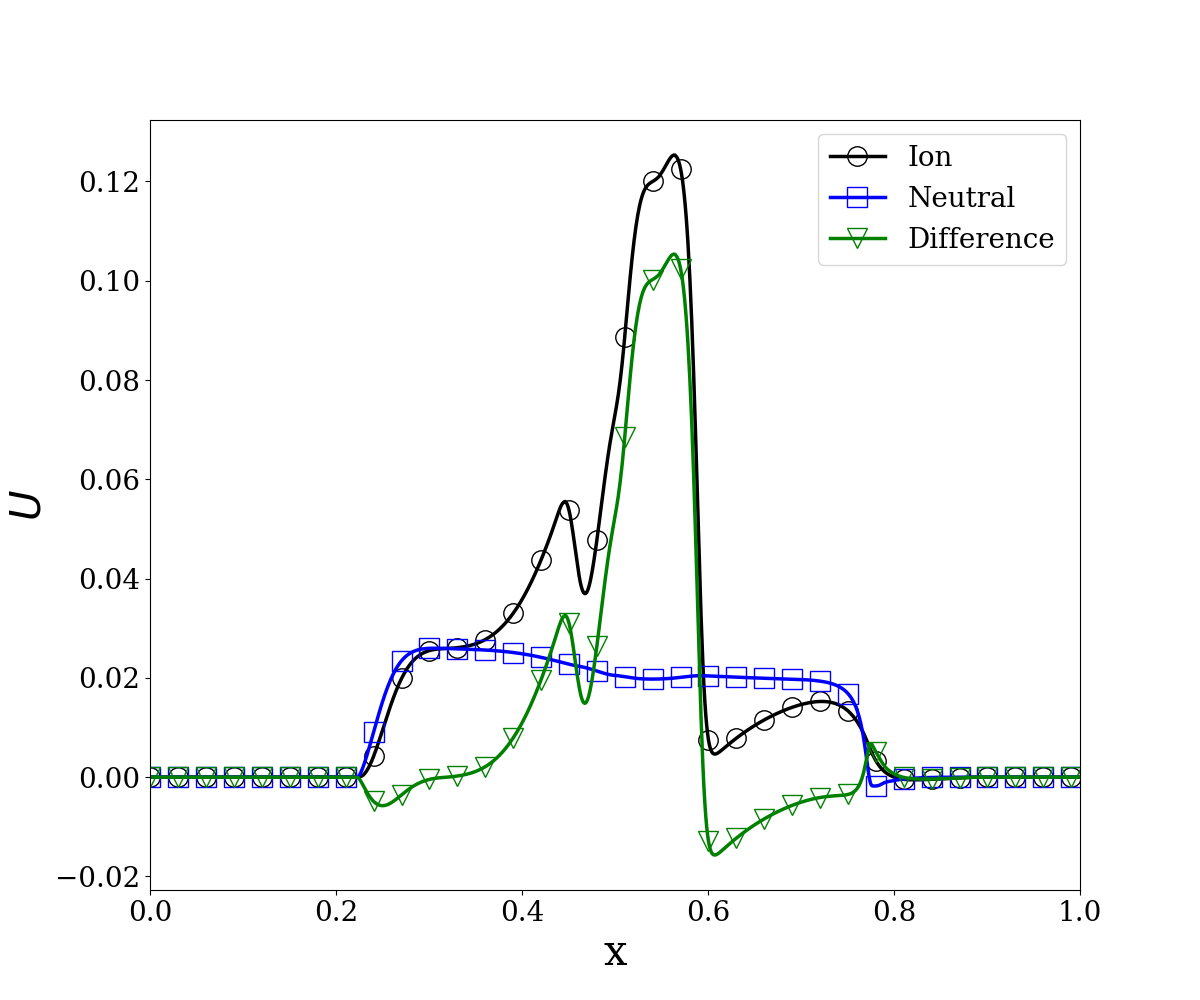}
    \caption{}
\end{subfigure}
\begin{subfigure}[b]{0.31\textwidth}
    \centering
    \includegraphics[width=0.98\textwidth]{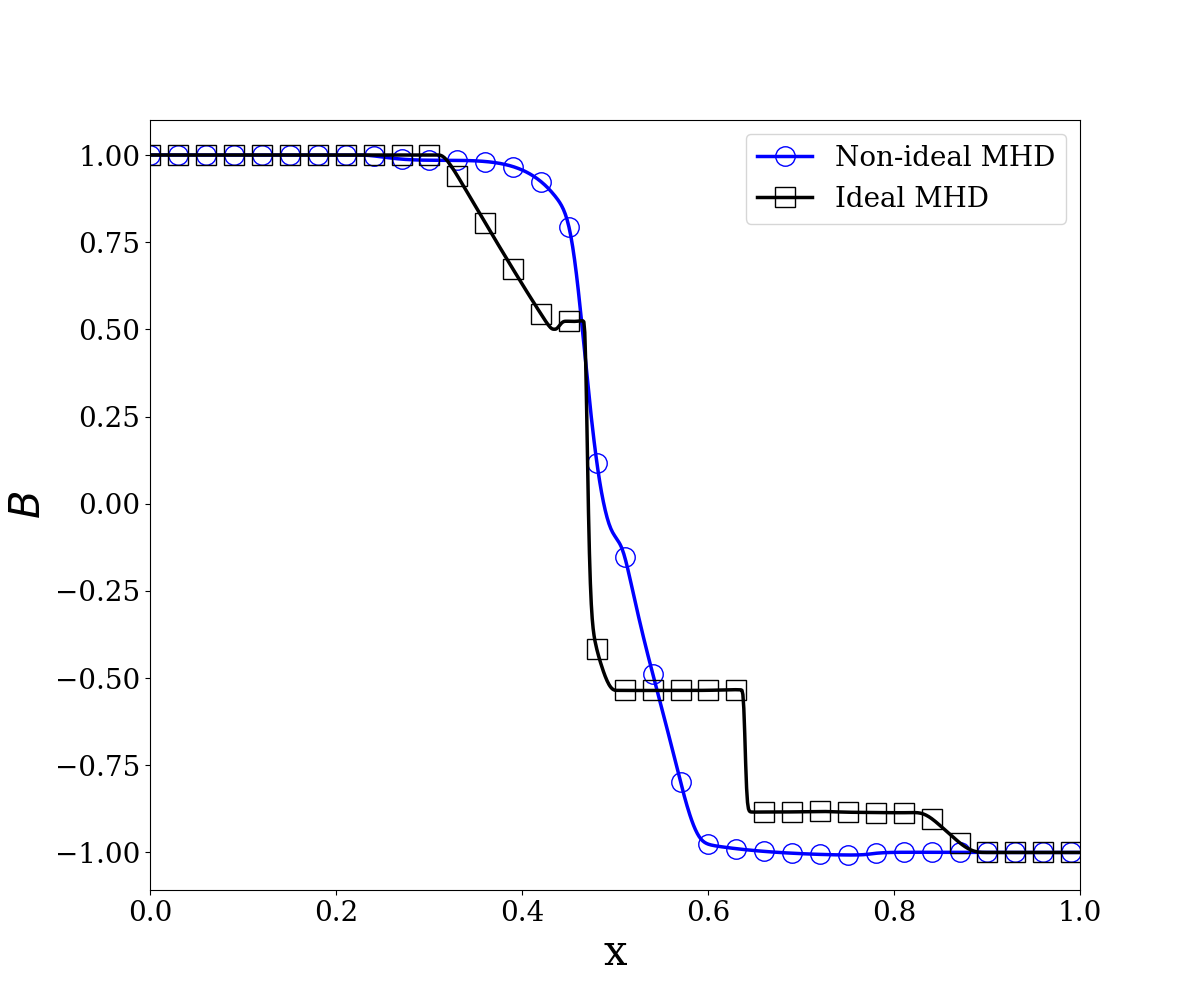}
    \caption{}
\end{subfigure}
\caption{(a) Density, (b) ion velocity, (c) magnetic field profile in the presence of neutral particles at $Kn=0,r_L=0.0001$. Due to the collision between charged species and neutral species, the magnetic force lines become  "heavier" and the shock speed becomes smaller. The discontinuity disappears in the magnetic field due to the dissipation provided by the cross-species collision.}
\label{fig:ambipolarkn0}
\end{figure}

Figure \ref{fig:ambipolarkn0} shows the results from a simulation using the Brio-Wu setup for ions and electrons but in the presence of a neutral gas. The neutral gas density $n_n = 10.0$ is much larger than the ion density $n_i=1.0$, as is typical in a weakly ionized plasma. The interaction coefficients between the ions and neutrals, and the electrons and neutrals, are fixed at $\chi_{in} = 2$ and $\chi_{en}=2$ in the momentum equation respectively. The Larmor radius is $r_L=0.0001$.
According to Eq.\eqref{eq:diffusivitycoef}, in this setup, the resistive diffusivity $r_O=0.0001$, the Hall diffusivity $r_H=0.0$, and the ambipolar diffusivity $r_A=2.5\times 10^{8}$. Therefore, the ambipolar diffusion effect is the dominant dissipation mechanism in this weakly ionized plasma setting.

From Figure \ref{fig:ambipolarkn0}(a), we can see that the right-going magnetosonic shock speed is slowed down under the influence of the neutral gas. This is reasonable as the magnetic forces now act on both the neutral and charged particle species. Figure \ref{fig:ambipolarkn0}(b) clearly shows the velocity gap between the ions and neutrals, which will give rise to a drag force between the two species. Finally, Figure \ref{fig:ambipolarkn0}(c) demonstrates that the discontinuity in the magnetic field is smoothed out due to the dissipation provided by the ambipolar diffusion mechanism.

\subsection{Orszag-Tang vortex}

In this section, the Orszag-Tang vortex problem was tested to explore how neutrals influence MHD shocks. This problem was originally designed to study the MHD turbulence \cite{orszag1979small}. It was intensively studied later and gradually became a benchmark problem of 2D MHD codes to test the capability to handle the formation of MHD shocks and shock-shock interactions  \cite{dahlburg1989evolution,zachary1994higher,jiang1999high}. The computational domain is $[0,2\pi]\times[0,2\pi]$, and the initial conditions are:
\begin{center}

\begin{tabular}{ |p{3cm}||p{3cm}|p{3cm}|  }
 \hline
 \multicolumn{3}{|c|}{Initial condition} \\
 \hline
 Item & Ions & Electrons \\
 \hline
 m& 1.0 & 0.04  \\
 n& $\gamma^2 $ & $\gamma^2 $  \\
 p & $\gamma $& $\gamma$\\
 $V_x$ &$-\sin(y) $& $-\sin(y)$\\
 $V_y$ &$\sin(x) $ & $\sin(x)$\\
 $B_y$ & $\sin(2x) $ & $\sin(2x)$   \\
 \hline
\end{tabular},
\end{center}
where $\gamma=5/3$ is specific heat capacity.

\begin{figure}[H]
\centering
\begin{subfigure}[b]{0.48\textwidth}
    \centering
    \includegraphics[width=0.98\textwidth]{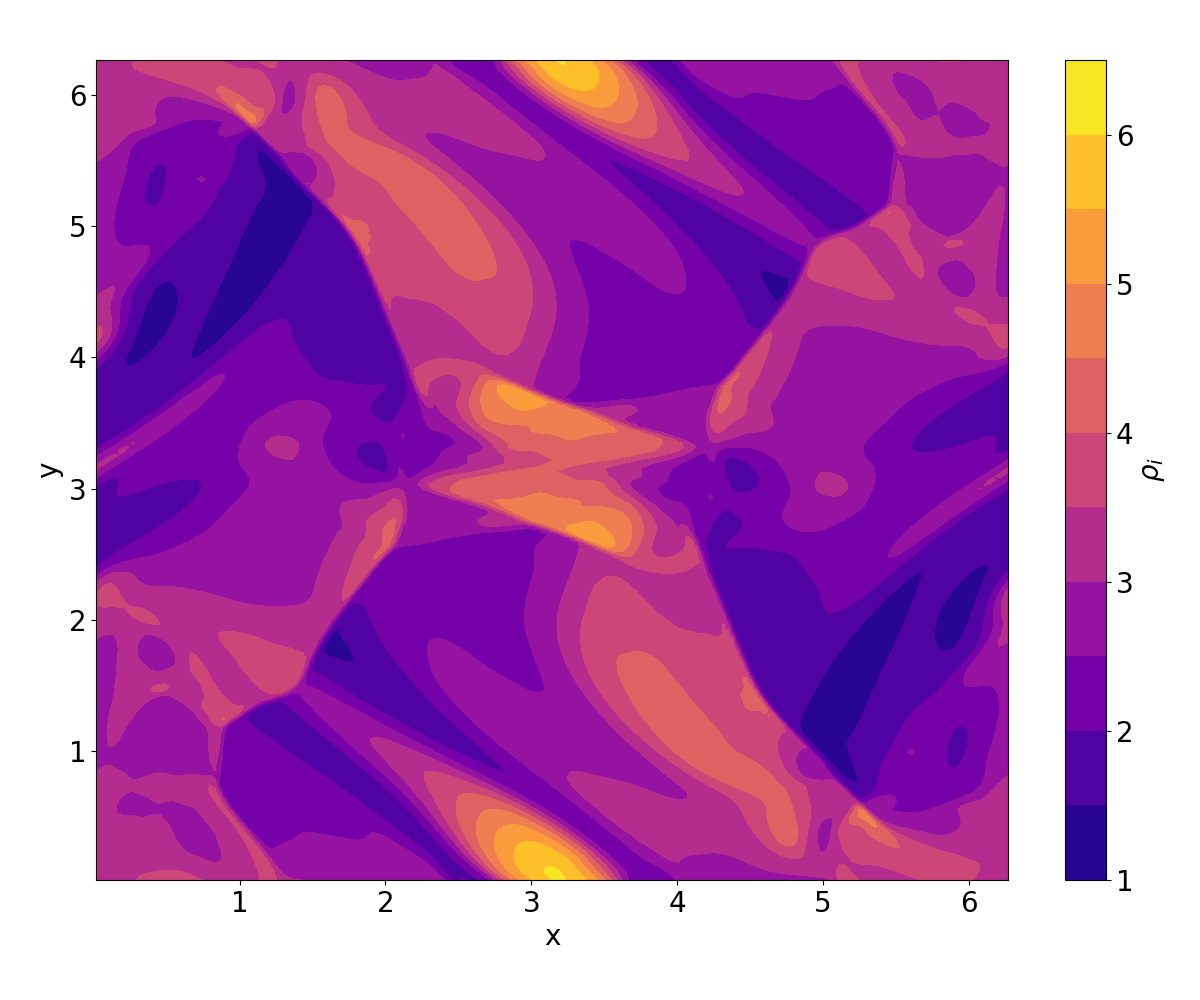}
    \caption{}
\end{subfigure}
\begin{subfigure}[b]{0.48\textwidth}
    \centering
    \includegraphics[width=0.98\textwidth]{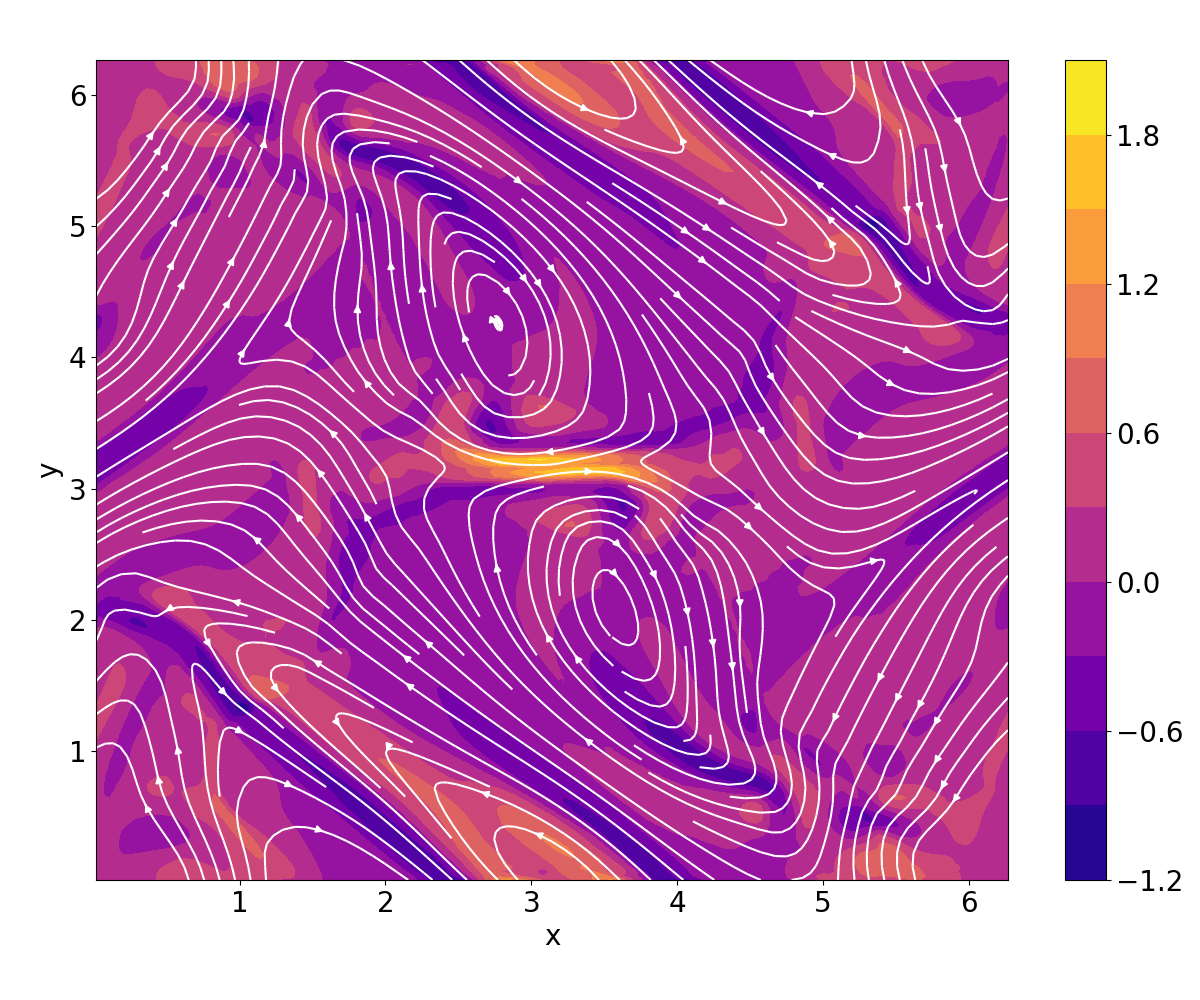}
    \caption{}
\end{subfigure}
\caption{Orszag-tang ovetex at $Kn=0, t=3$.(a) Density of ion, (b) magnetic lines (while solid line) and the contours of the out-of-plane current density $J_z$.}
\label{fig:Orszag-Kn0-r0.1}
\end{figure}

\begin{figure}
    \centering
    \includegraphics[width=0.6\textwidth]{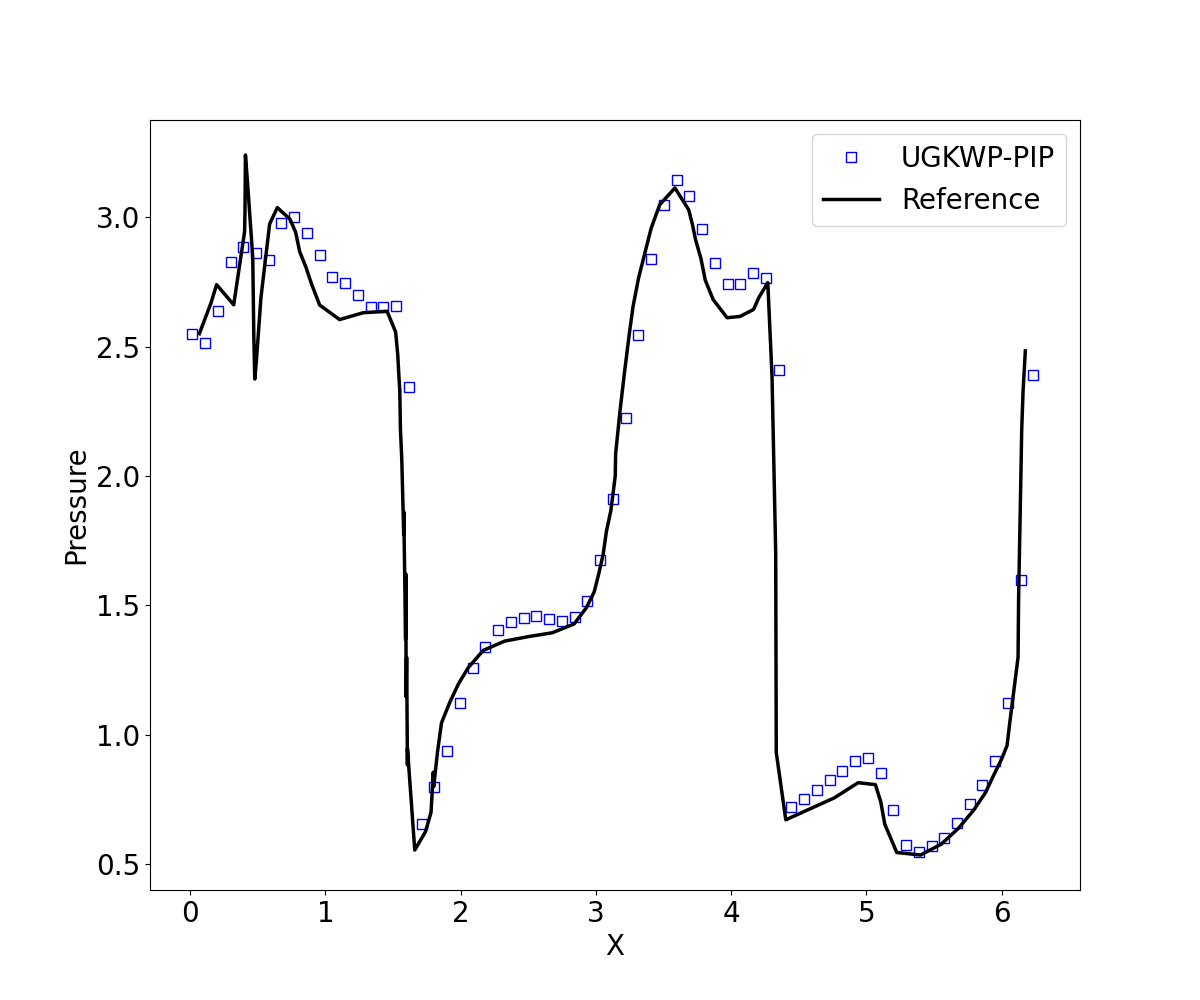}
    \caption{Orszag-tang ovetex at $Kn=0, t=3$, pressure comparison over $y=0.625\pi$.}
    \label{fig:Orszag-Kn0-r0.1-p}
\end{figure}

Figure \ref{fig:Orszag-Kn0-r0.1} shows the results at $t=3$ in the ideal MHD limit. The pressure profile matches well with the reference solution, as shown in Figure \ref{fig:Orszag-Kn0-r0.1-p}. This proves the algorithm's capability to capture shock in the ideal MHD limit.

\begin{figure}[H]
\centering
\begin{subfigure}[b]{0.48\textwidth}
    \centering
    \includegraphics[width=0.98\textwidth]{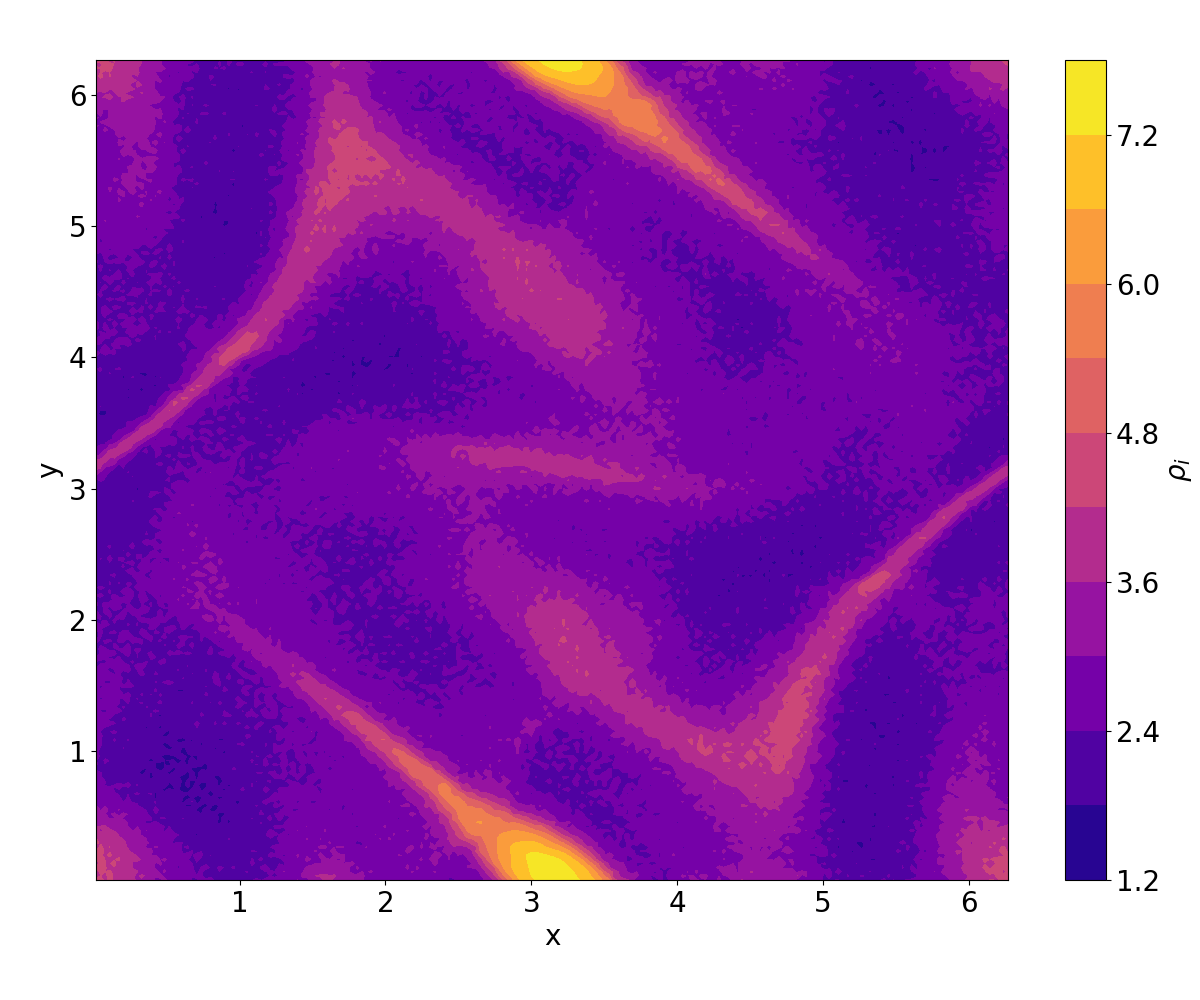}
    \caption{}
\end{subfigure}
\begin{subfigure}[b]{0.48\textwidth}
    \centering
    \includegraphics[width=0.98\textwidth]{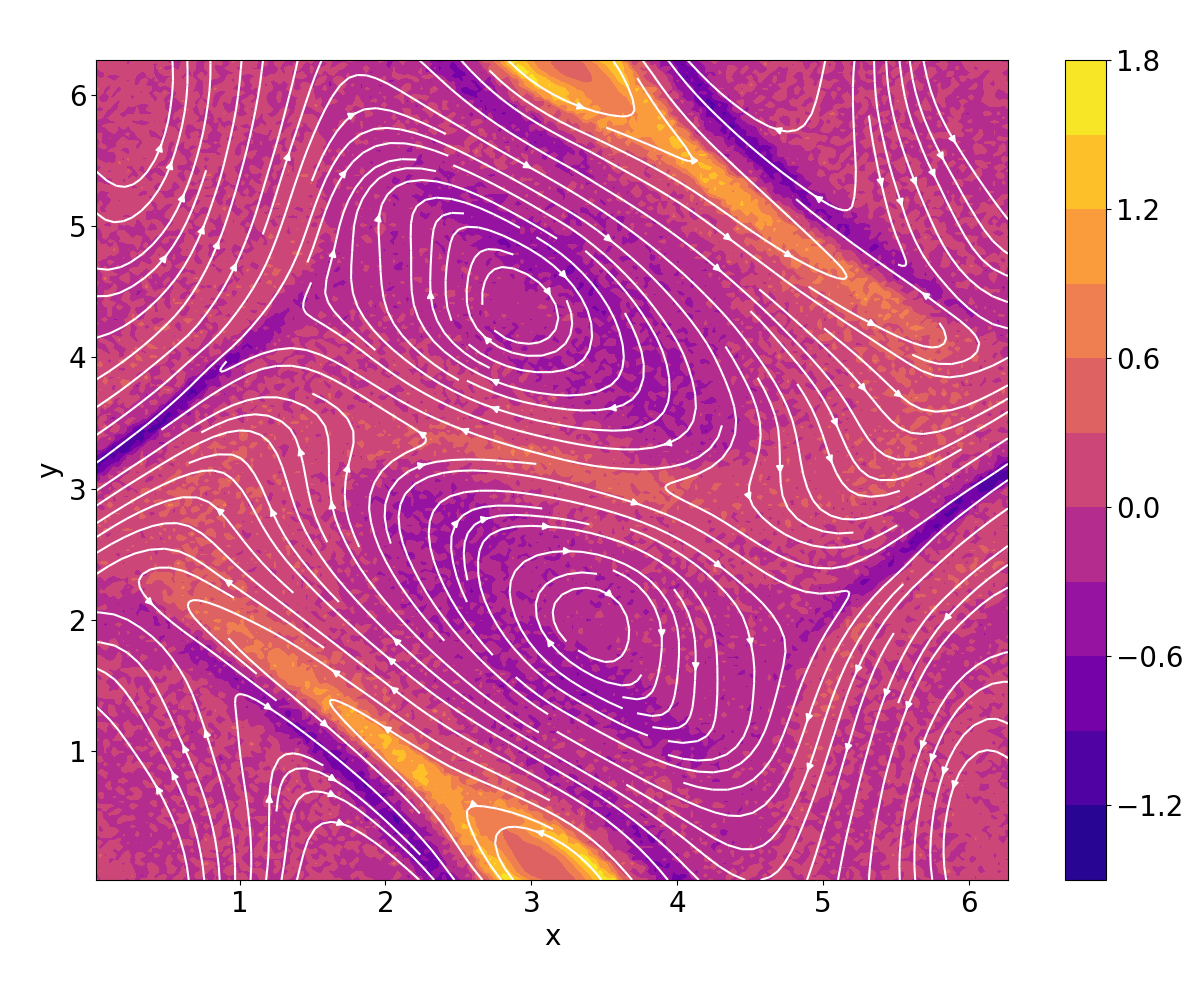}
    \caption{}
\end{subfigure}
\caption{Orszag-tang ovetex at $Kn=0.001, t=3$.(a) Density of ion, (b) magnetic lines (white solid line), and the contours of the out-of-plane current density $J_z$.}
\label{fig:Orszag-Kn0.001-r0.1}
\end{figure}

Figure \ref{fig:Orszag-Kn0.001-r0.1} shows the results at $Kn=0.001$. It can be seen that due to the increased viscosity caused by rarefied effects, the shock is smoothed out. The magnetic field line is similar to that of the ideal MHD case in Figure \ref{fig:Orszag-Kn0-r0.1}, but the current density in the center is less intensive, which may indicate that the magnetic field gradient is not as sharp as in the ideal MHD case.

\begin{figure}[H]
\centering
\begin{subfigure}[b]{0.48\textwidth}
    \centering
    \includegraphics[width=0.98\textwidth]{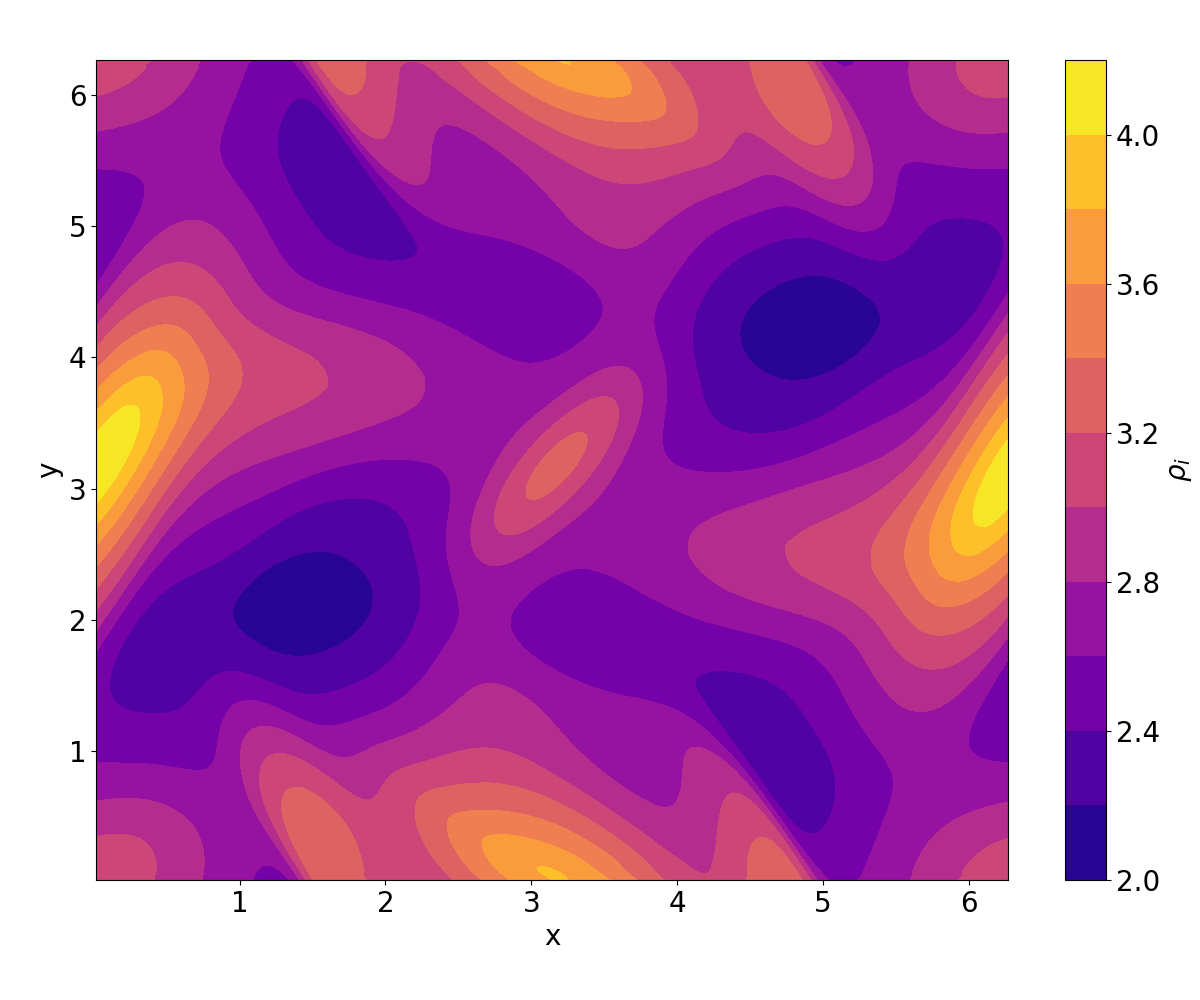}
    \caption{}
\end{subfigure}
\begin{subfigure}[b]{0.48\textwidth}
    \centering
    \includegraphics[width=0.98\textwidth]{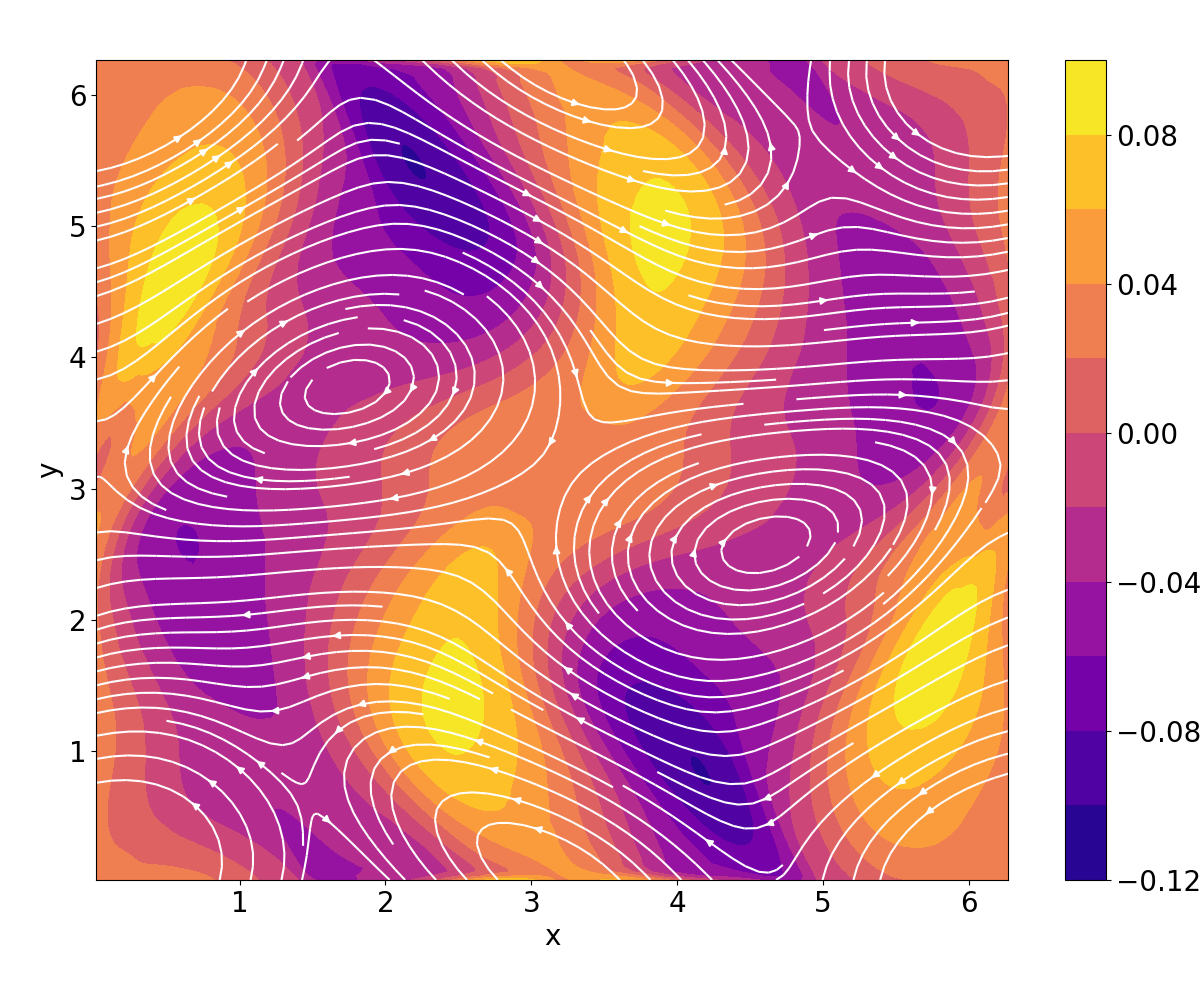}
    \caption{}
\end{subfigure}
\caption{Orszag-tang ovetex with neutral species background at $Kn=0, t=3$.(a) Density of ion, (b) magnetic lines (white solid line), and the contours of the out-of-plane current density $J_z$.}
\label{fig:Orszag-Kn0-neutral}
\end{figure}

Figure \ref{fig:Orszag-Kn0-neutral} illustrates the simulation results at $Kn=0$ and $t=3$ in a neutral species background with a uniform number density of $n_n=10.0$. The interaction coefficients between ions and neutrals $\chi_{in}$ and electrons and neutrals $\chi_{en}$ are both set to 10 in the momentum equation.
Compared to the ideal-MHD limit, the flow field exhibits smoother characteristics and a slower evolution. This deceleration may be attributed to the magnetic force now acting on a larger number of particles in the system, including the neutral species. As a result, the overall dynamics of the flow field are more gradual.

\subsection{Magnetic Reconnection}

The simulation uses the same initial conditions as the Geospace Environment Modeling (GEM) challenge problem. The initial condition is the so-called "Harris-Sheet" current sheet solution. The initial magnetic field is given by
$$
\boldsymbol{B}(y)=B_0 \tanh (y / \lambda) \boldsymbol{e}_{\boldsymbol{x}},
$$
And a corresponding current sheet is carried by the electrons
$$
\boldsymbol{J}_e=-\frac{B_0}{\lambda} \operatorname{sech}^2(y / \lambda) \boldsymbol{e}_{\boldsymbol{z}} .
$$
The initial number densities of electrons and ions are
$$
n_e=n_i=1 / 5+\operatorname{sech}^2(y / \lambda) .
$$
The electron and ion pressures are set to be
$$
P_i=5 P_e=\frac{5 B_0}{12} n(y),
$$
where $B_0=0.1, m_i=25 m_e$ and $\lambda=0.5$ is the thickness of the sheet. The electromagnetic correction potentials are set $\phi=\psi=0$ initially. The computational domain is $\left[-L_x / 2, L_x / 2\right] \times\left[-L_y / 2, L_y / 2\right]$ with $L_x=8 \pi, L_y=4 \pi$, which is divided into $200 \times 100$ cells.

Periodic boundaries are applied at $x= \pm L_x / 2$ and conducting wall boundaries at $y= \pm L_y / 2$. For electric field, $\boldsymbol{n}\times(\boldsymbol{E}_2-\boldsymbol{E}_1)=0$, i.e. transverse field being the same across the boundary. Here 1 stands for inner domain, 2 stands for outer domain. Across the boundar $y=\pm L_y/2$, we have $E_{1x}=E_{2x}=0, E_{1z}=E_{2z}=0$ since there is no electric field inside conductor. Assuming there is no surface current, then Amphere's laws give that $\partial_y B_z = \partial_y B_x = 0$. Similarly, $\boldsymbol{n}\cdot(\boldsymbol{B}_2-\boldsymbol{B}_1)=0$, we have $B_{1y}=B_{2y}=0$. Assuming there's no surface charge, then we have $\partial_y E_y=0$ across the boundary. For the fluid variable, we use wall boundary, where the surface velocity is 0 and the temperature and number density are the same as the neighboring cell to surface.

To initiate reconnection, the magnetic field is perturbed with $\delta \boldsymbol{B}=\boldsymbol{e}_{\boldsymbol{z}} \times \nabla_{\boldsymbol{x}} \psi$, where
$$
\psi(x, y)=0.1 B_0 \cos \left(2 \pi x / L_x\right) \cos \left(\pi y / L_y\right) .
$$
s.t.
\begin{align*}
    \delta \boldsymbol{B} =& -0.1 B_0 2\pi/L_x \sin \left(2 \pi x / L_x\right) \cos \left(\pi y / L_y\right) \boldsymbol{e}_y \\
    &+ 0.1 B_0 \pi/L_y \cos \left(2 \pi x / L_x\right) \sin \left(\pi y / L_y\right) \boldsymbol{e}_x
\end{align*}

The reconnected flux is defined by
$$
\phi(t)=\frac{1}{2 L_x} \int_{-L_x / 2}^{L_x / 2}\left|B_y(x, 0, t)\right| d x
$$

\begin{figure}[H]
\centering
\begin{subfigure}[b]{0.48\textwidth}
    \centering
    \includegraphics[width=0.98\textwidth]{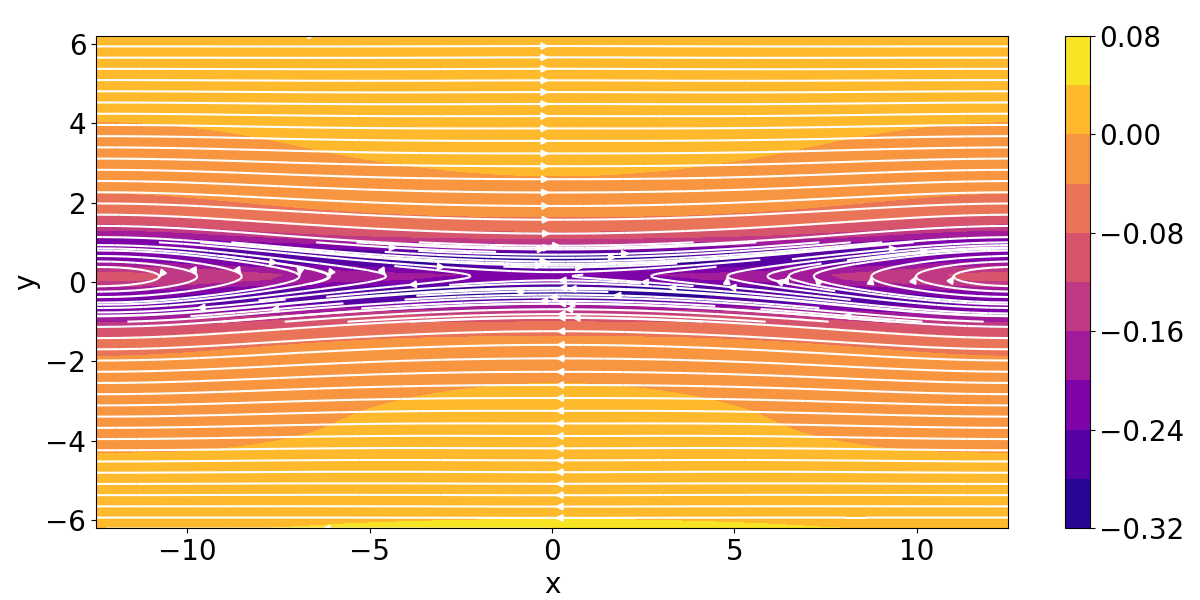}
    \caption{}
\end{subfigure}
\begin{subfigure}[b]{0.48\textwidth}
    \centering
    \includegraphics[width=0.98\textwidth]{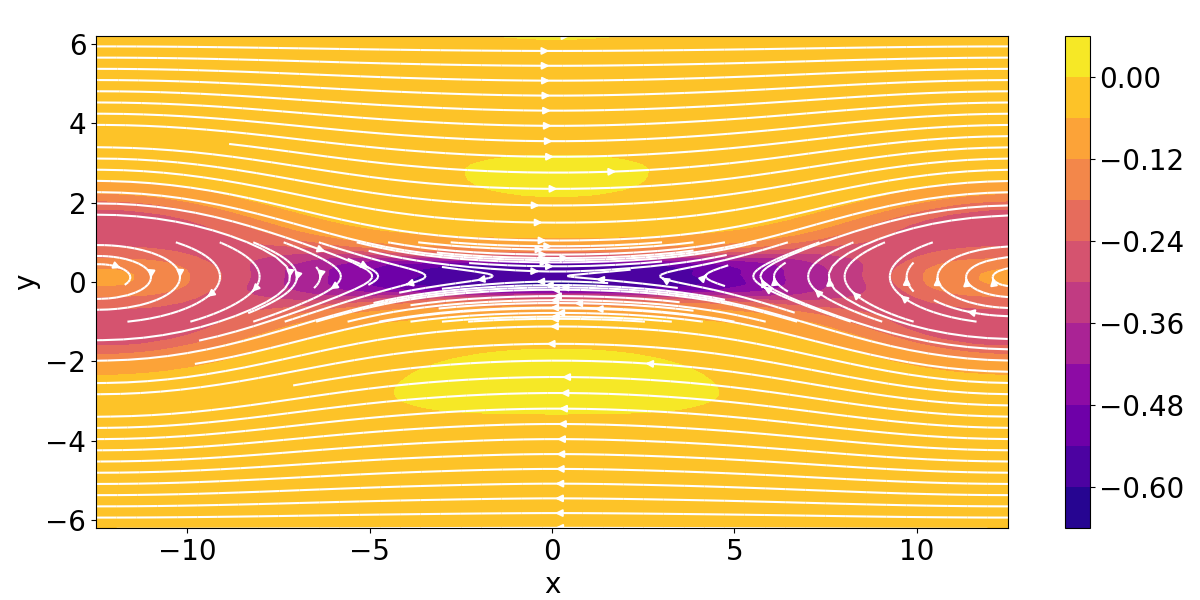}
    \caption{}
\end{subfigure}
\vskip\baselineskip
\begin{subfigure}[b]{0.48\textwidth}
    \centering
    \includegraphics[width=0.98\textwidth]{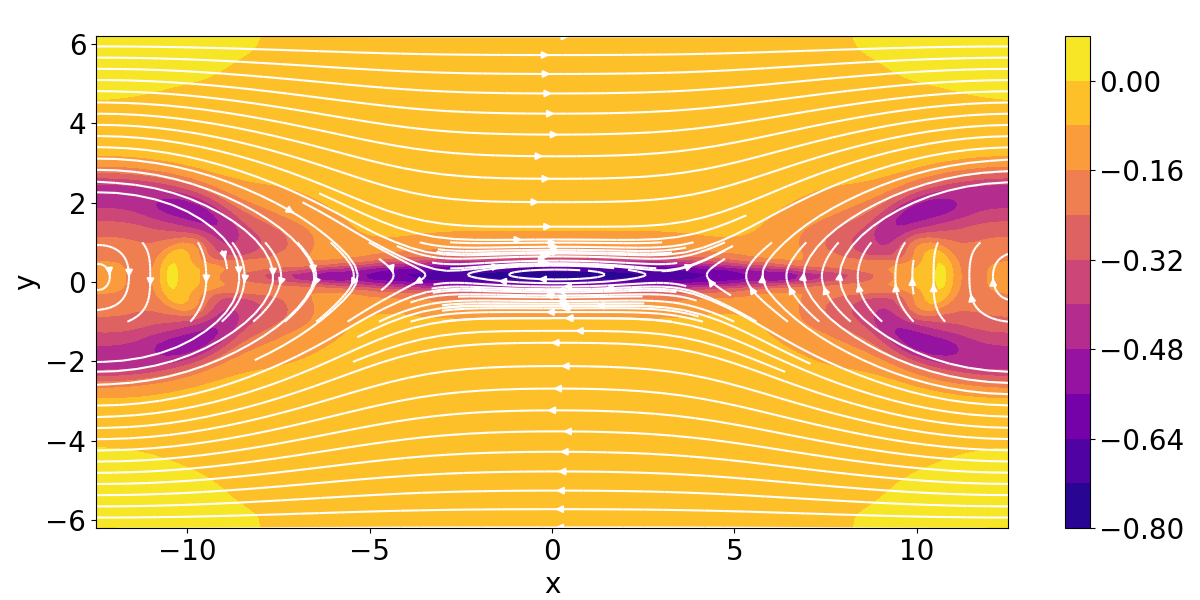}
    \caption{}
\end{subfigure}
\begin{subfigure}[b]{0.48\textwidth}
    \centering
    \includegraphics[width=0.98\textwidth]{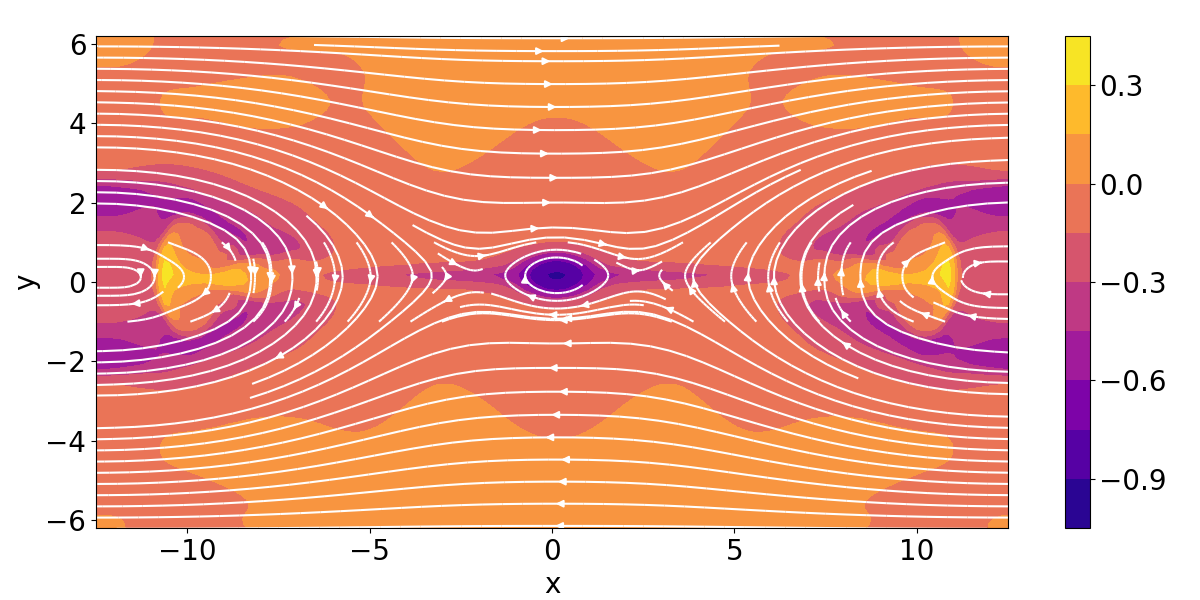}
    \caption{}
\end{subfigure}
\caption{Magnetic reconnection at $r_L=1, Kn=0, Nx=200$. Time evolution of magnetic lines (while lines) and the contours of out-of-plane current density $J_z$ at (a)$\omega_{pi}\Delta t=10$, (b)$\omega_{pi}\Delta t=20$, (c)$\omega_{pi}\Delta t=30$, (d)$\omega_{pi}\Delta t=10$.}
\label{fig:GEM-Kn0-r1-B-streamplot}
\end{figure}

Figure \ref{fig:GEM-Kn0-r1-B-streamplot} shows the time evolution of magnetic field lines (white lines) and the contours of out-of-plane current density $J_z$ at different times. As time progresses, the magnetic field lines in the center of the domain break and rejoin, and finally, an X-point reconnection pattern is formed.

\begin{figure}
    \centering
    \includegraphics[width=0.6\textwidth]{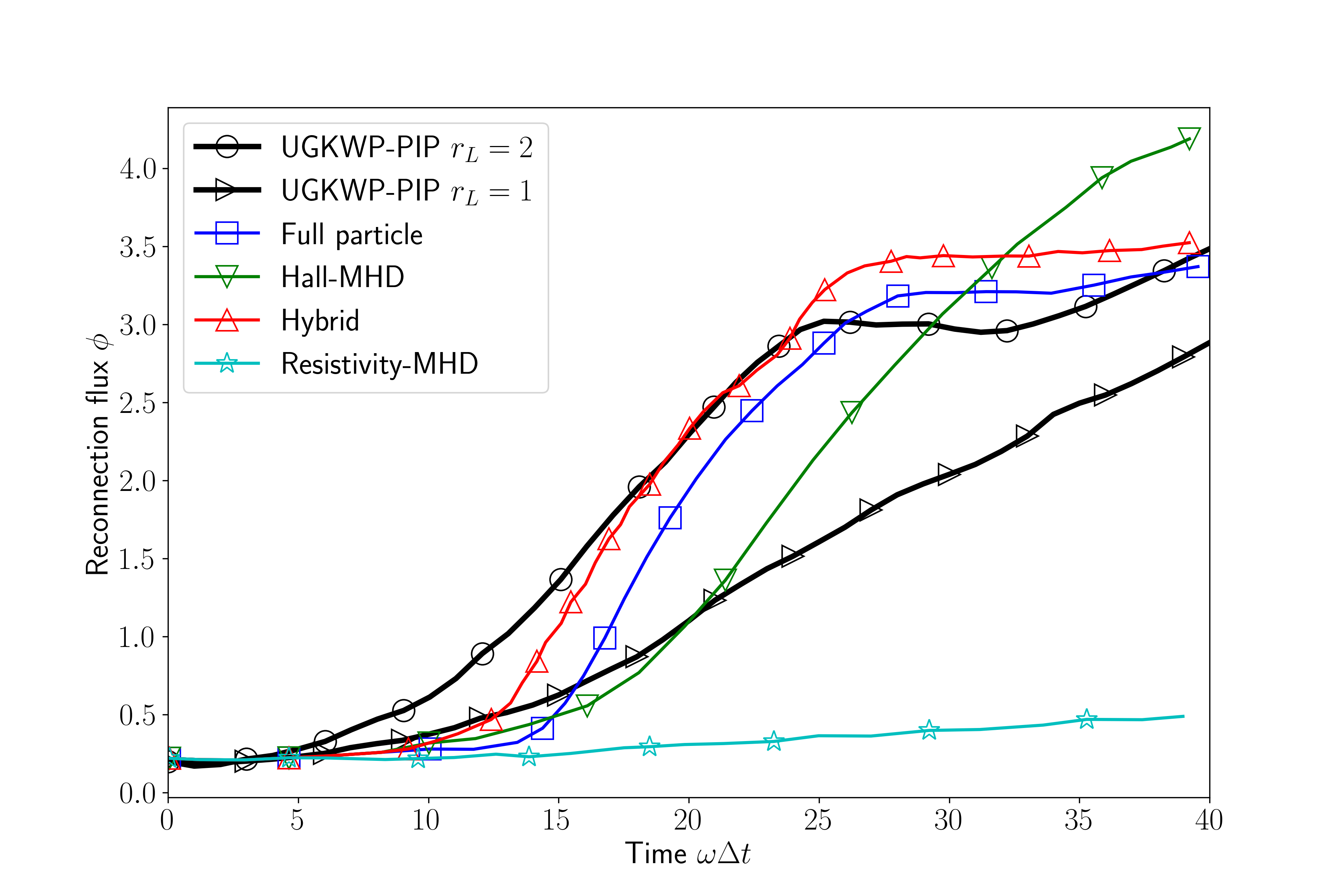}
    \caption{Magnetic reconnection flux at $Kn=0$: The two black lines represent results, and it can be observed that our algorithm closely matches the reconnection velocity of Hall-MHD and Particle methods, which achieves faster reconnection rates than normal resistivity MHD model
    \label{fig:GEM-Kn0-r1-reconnection-flux}
}
\end{figure}

\begin{figure}[H]
\centering
\begin{subfigure}[b]{0.48\textwidth}
    \centering
    \includegraphics[width=0.98\textwidth]{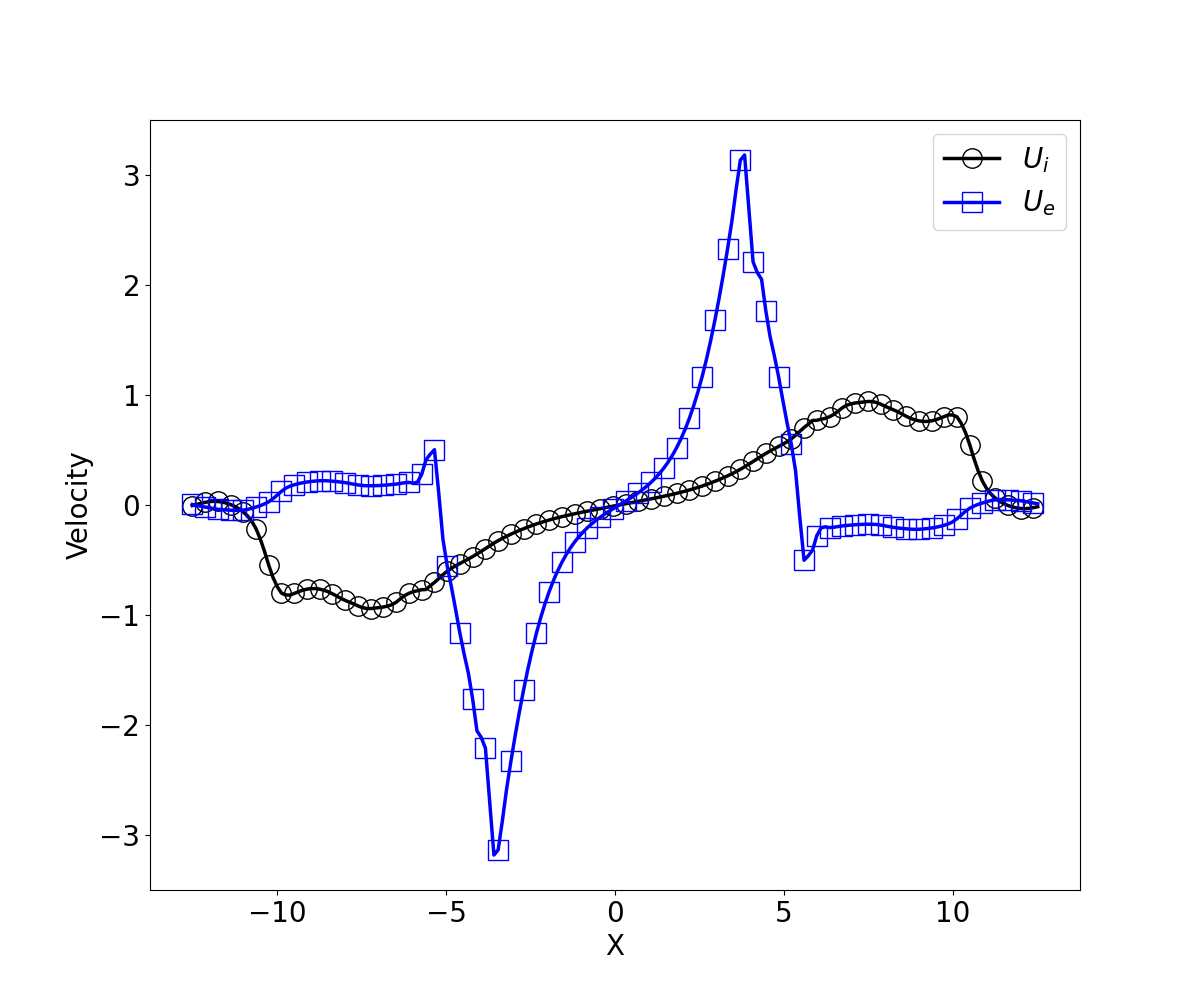}
    \caption{}
\end{subfigure}
\begin{subfigure}[b]{0.48\textwidth}
    \centering
    \includegraphics[width=0.98\textwidth]{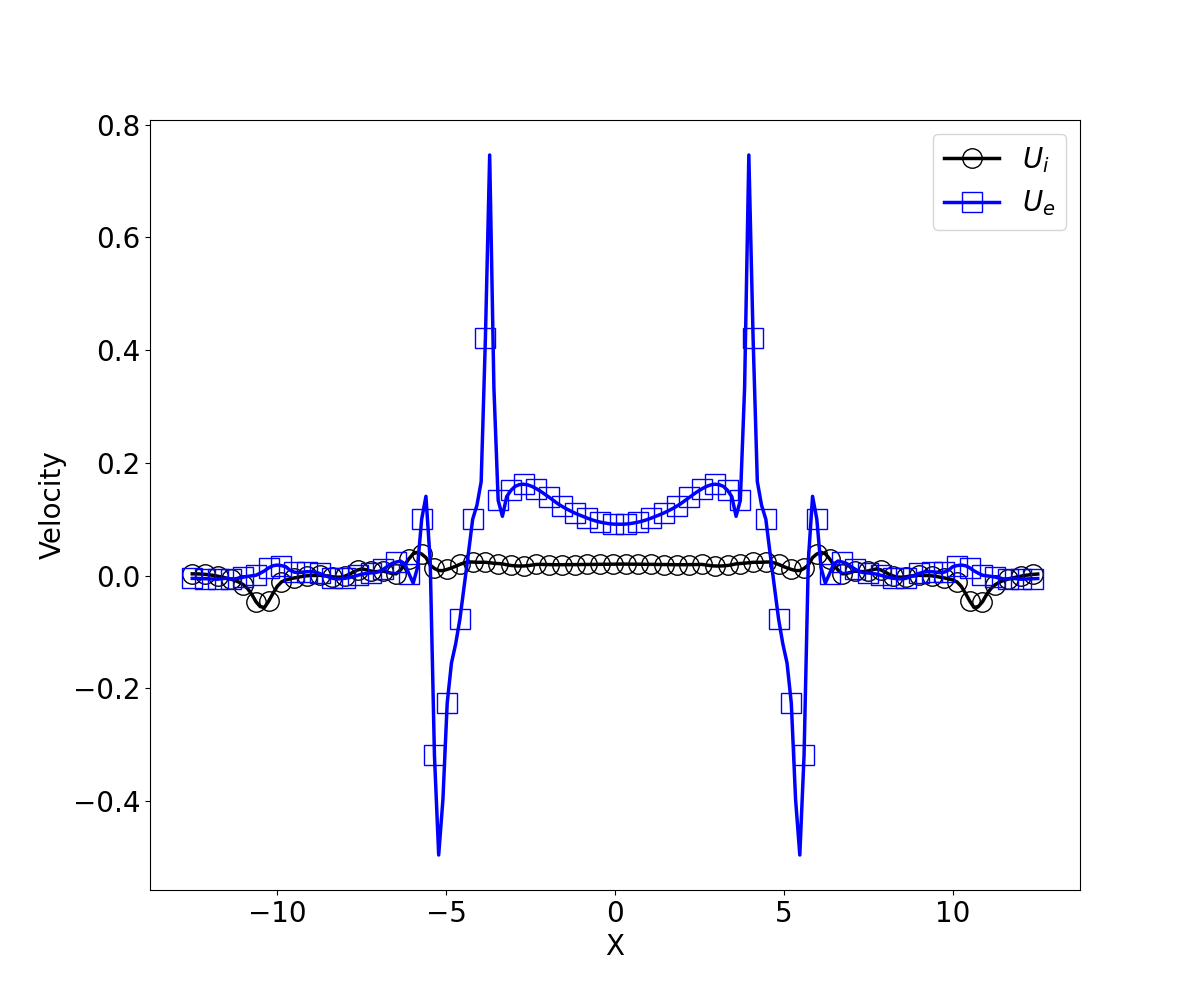}
    \caption{}
\end{subfigure}
\caption{Magnetic reconnection at $r_L=1, Kn=0, Nx=200$. (a) x-component velocity $U$, (b) y-component velocity of electrons and ions at $y=0$. The velocity is separated due to different magnetization levels of ions and electrons. The electrons move at a higher speed than ions in both x and y components.}
\label{fig:GEM-Kn0-r1-GEM-Kn0-r1-velocity_compare}
\end{figure}

Figure \ref{fig:GEM-Kn0-r1-reconnection-flux} shows the reconnection flux evolution over time. When the normalized Larmor radius $r_L=2$, the reconnection flux is comparable to that of Hall-MHD, realizing a fast reconnection rate.
Figure \ref{fig:GEM-Kn0-r1-GEM-Kn0-r1-velocity_compare}(a) shows the x-component velocity of ions and electrons at time $\omega_{pi}\Delta t = 30$, where a clear velocity separation between ions and electrons is observed. This is very similar to Hall-MHD, since in Hall-MHD, the ions are demagnetized in the diffusion regime, which is on the ion inertial scale, while electrons are still frozen to the magnetic field lines. Compared to resistive-MHD where the reconnection electric field is provided by resistivity, the Hall effect provides a more efficient electric field for reconnection. This test shows that the model can reproduce the Hall-MHD effect.

\begin{figure}[H]
\centering
\begin{subfigure}[b]{0.48\textwidth}
    \centering
    \includegraphics[width=0.98\textwidth]{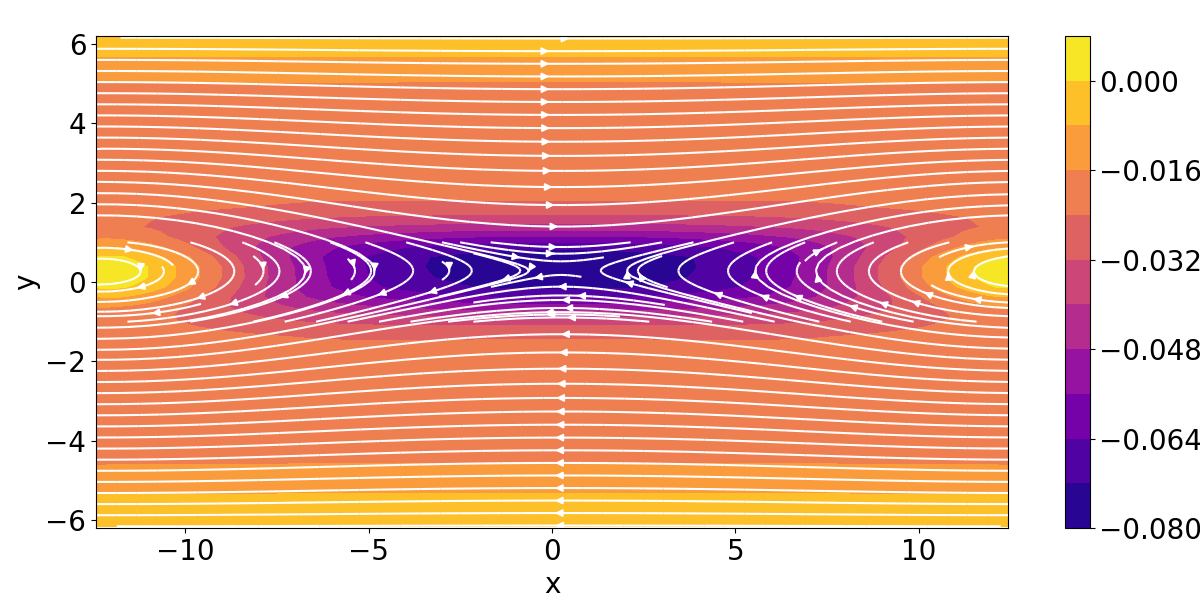}
    \caption{}
\end{subfigure}
\begin{subfigure}[b]{0.48\textwidth}
    \centering
    \includegraphics[width=0.98\textwidth]{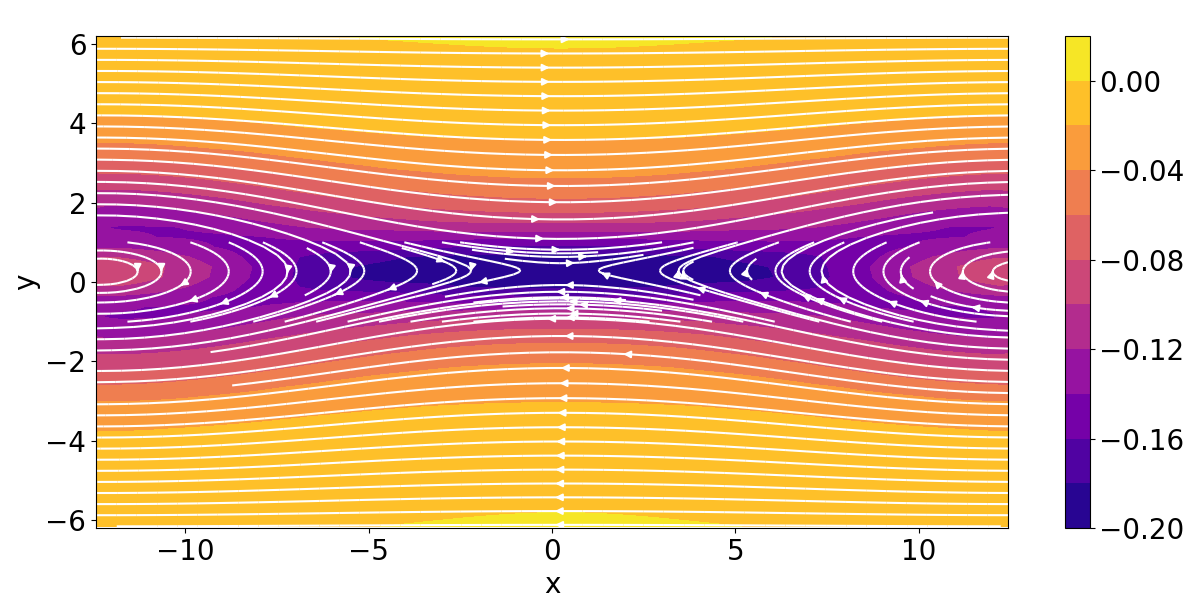}
    \caption{}
\end{subfigure}
\caption{(Magnetic lines (white solid line) and contours of current density $J_z$ of (a) strong coupled PIP $\chi=1.0$ (b) weakly coupled $Kn=0, \chi=0.1$}
\label{fig:chicompareB}
\end{figure}

Figure \ref{fig:chicompareB} shows the magnetic field lines and current density $J_z$ contour at $\omega_{pi}\Delta t =30$ when a neutral particle background is introduced. The neutral density is set to $n=1.0$, and the interaction coefficient $\chi$ of every species in the system is set uniformly to (a) $1.0$ (representing strongly coupled) and (b) $0.1$ (representing weakly coupled) in the momentum equation.

\begin{figure}
    \centering
    \includegraphics[width=0.6\textwidth]{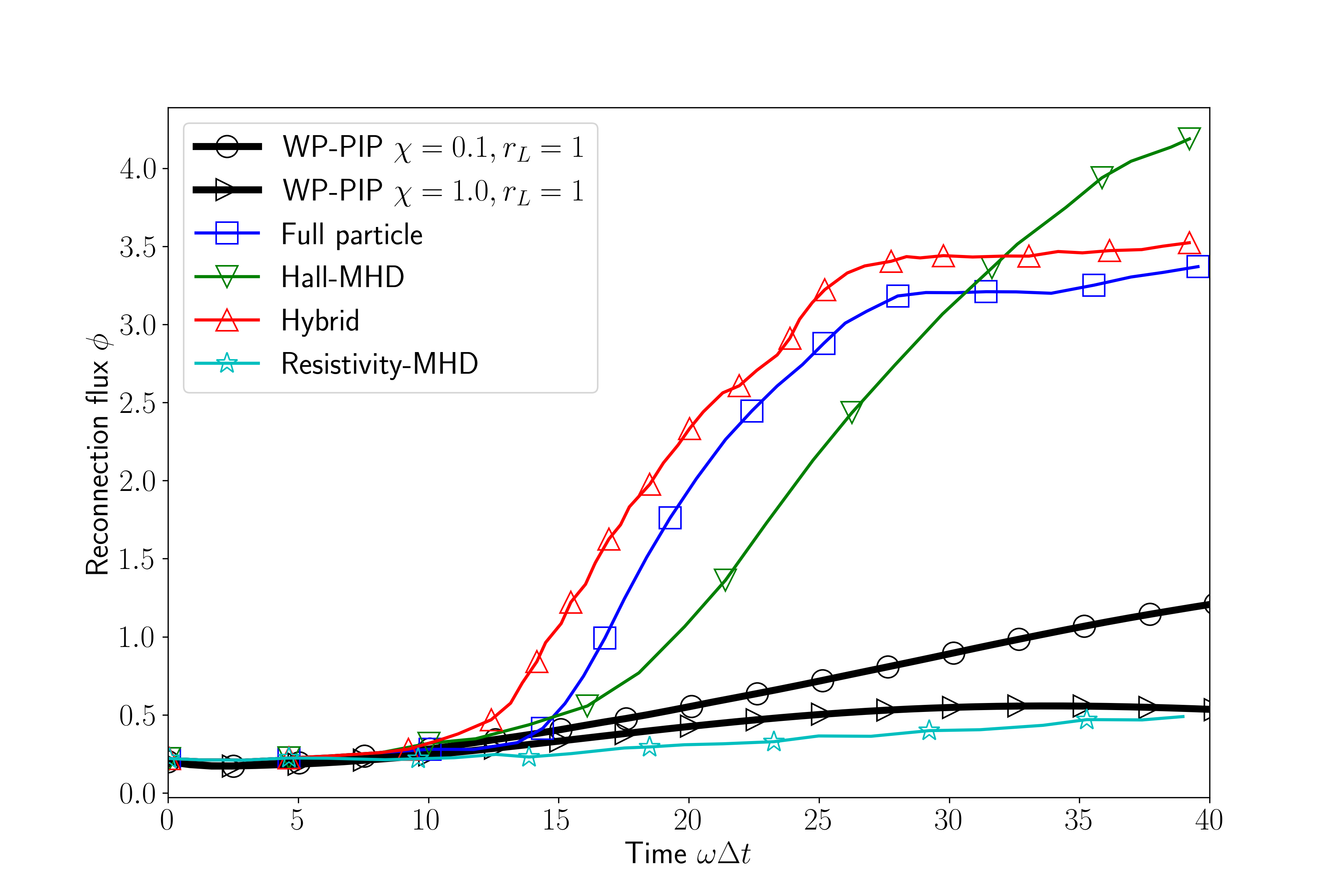}
    \caption{Magnetic reconnection flux at $Kn=0$ with uniform neutral background. The case of strong coupling between ions, electrons, and neutrals, the momentum is redistributed evenly among the different species, and the velocity separation between them disappears. Therefore the reconnection process is driven primarily by resistive diffusion. Conversely, in the case of weak coupling between the species, the velocity separation is smaller compared to the pure collisionless case. Here, the reconnection is supported partially by the Hall effect and partially by resistive diffusion.}
    \label{fig:chicompareflux}
\end{figure}

\begin{figure}[H]
\centering
\begin{subfigure}[b]{0.48\textwidth}
    \centering
    \includegraphics[width=0.98\textwidth]{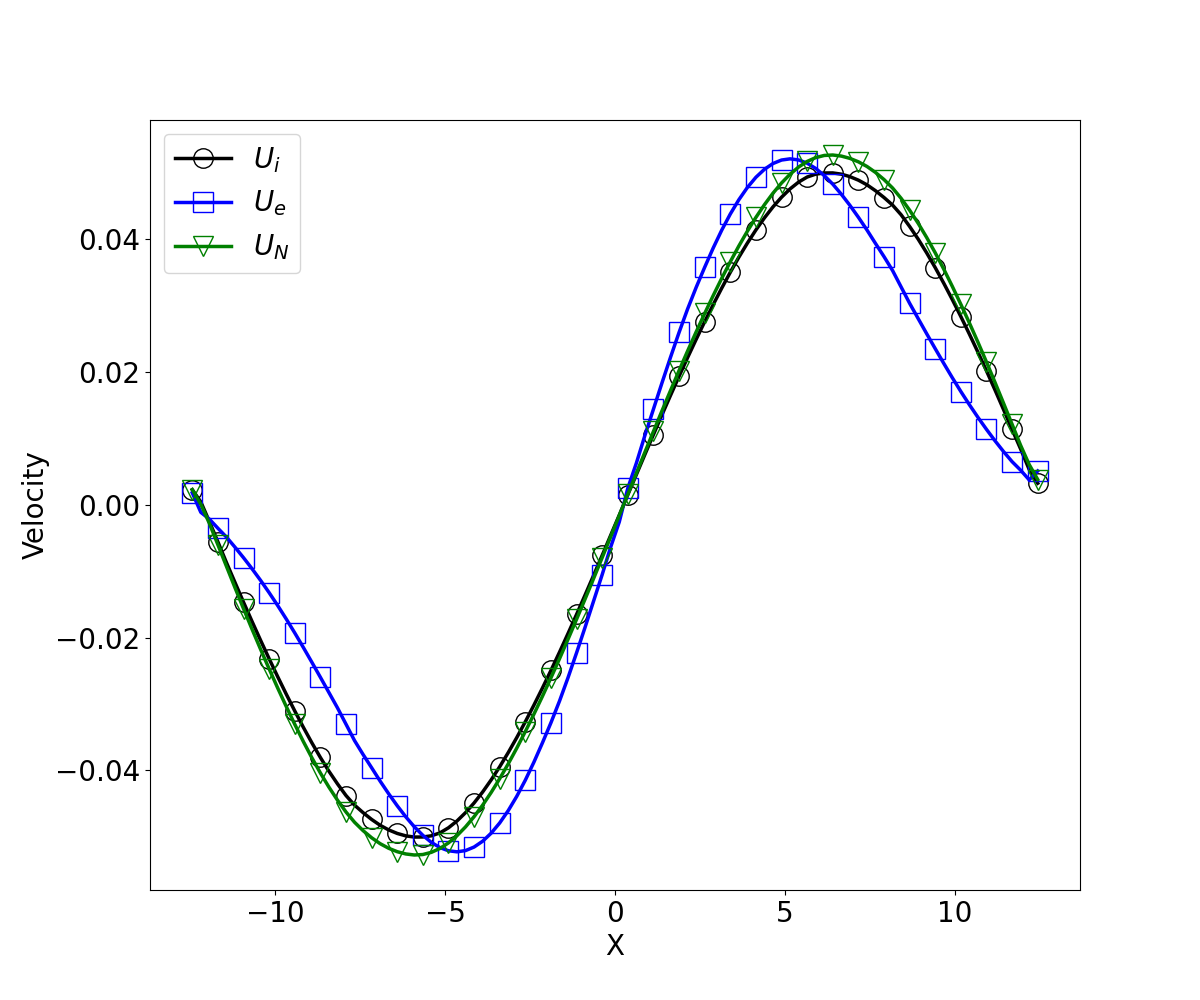}
    \caption{}
\end{subfigure}
\begin{subfigure}[b]{0.48\textwidth}
    \centering
    \includegraphics[width=0.98\textwidth]{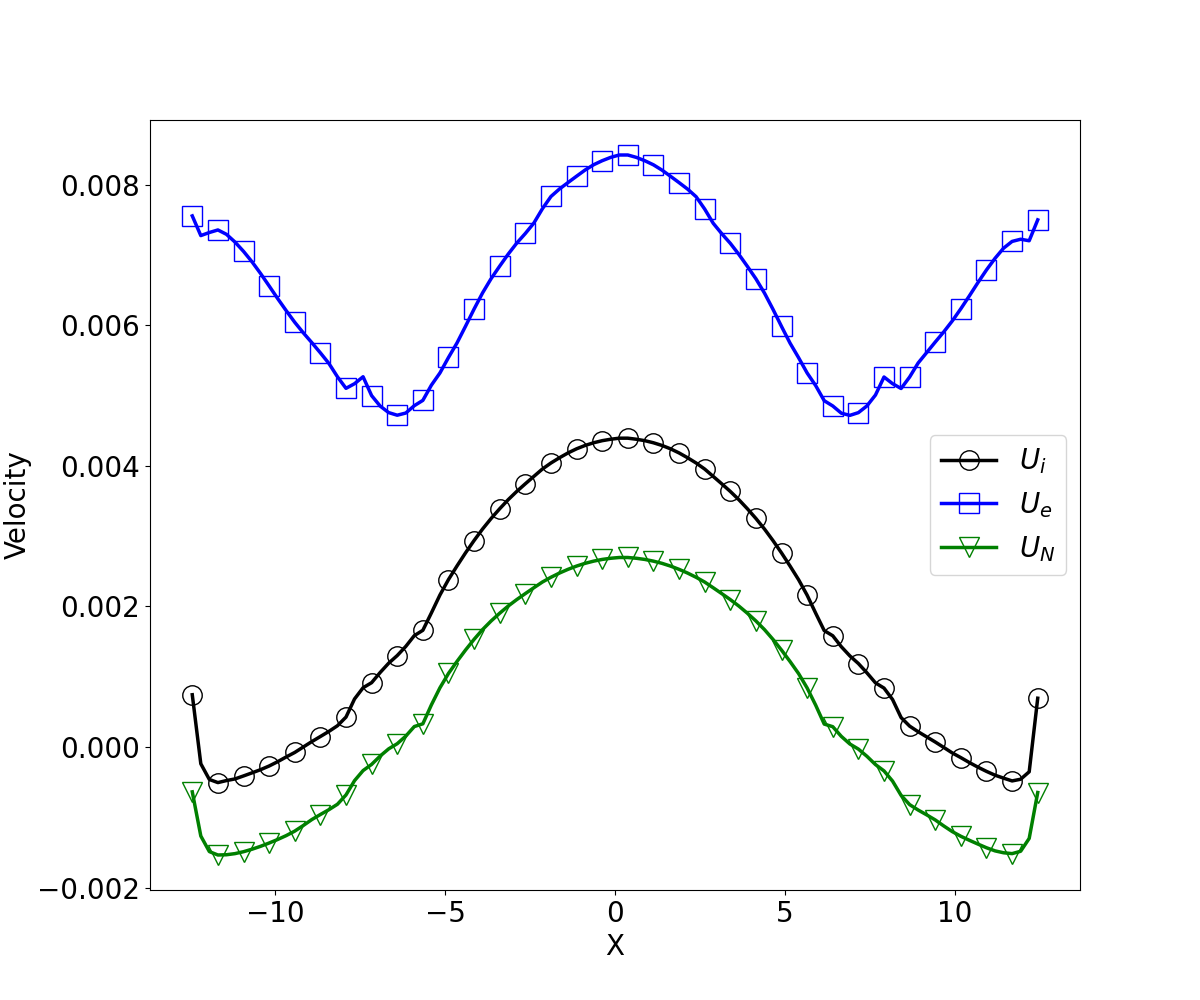}
    \caption{}
\end{subfigure}
\vskip\baselineskip
\begin{subfigure}[b]{0.48\textwidth}
    \centering
    \includegraphics[width=0.98\textwidth]{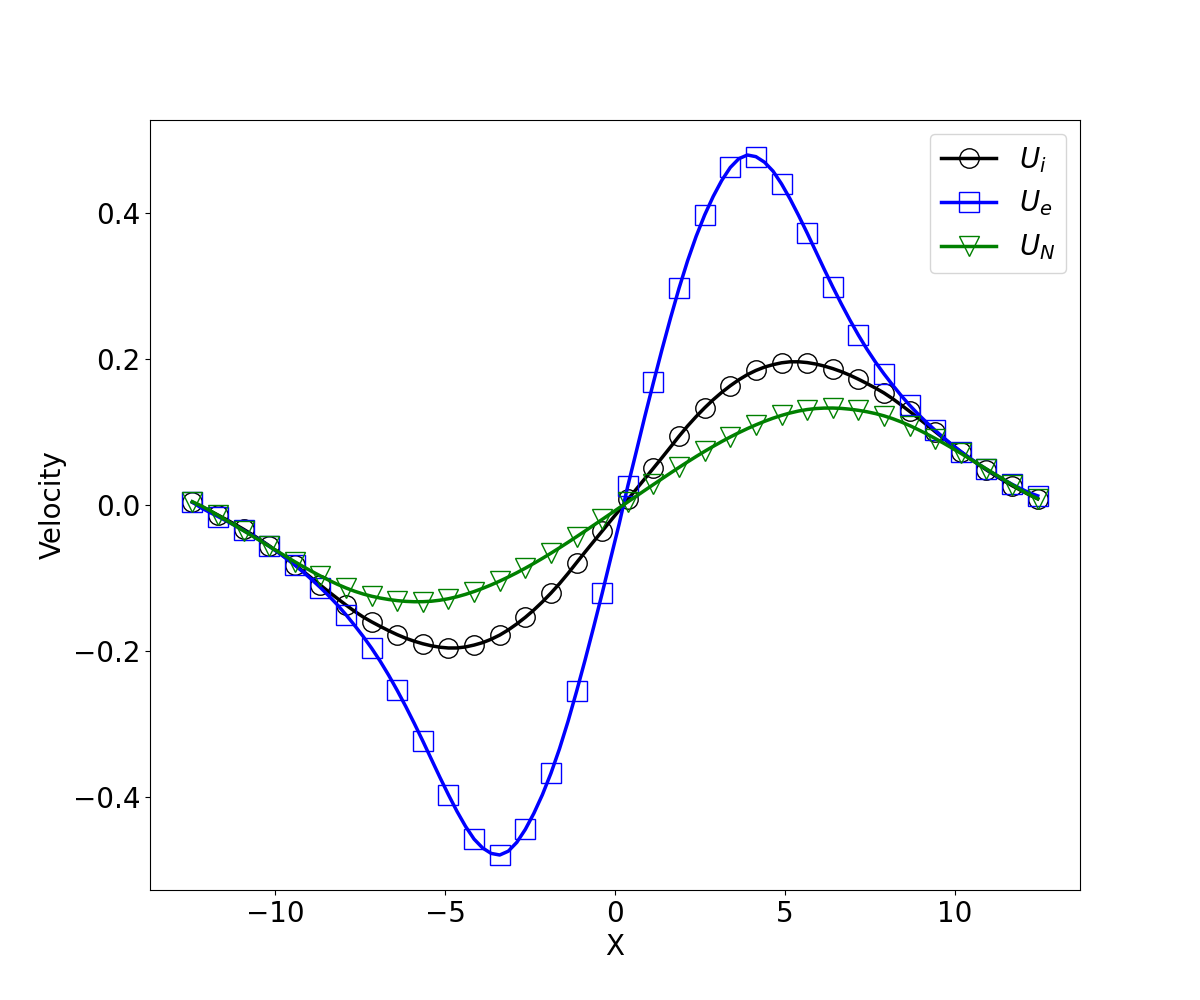}
    \caption{}
\end{subfigure}
\begin{subfigure}[b]{0.48\textwidth}
    \centering
    \includegraphics[width=0.98\textwidth]{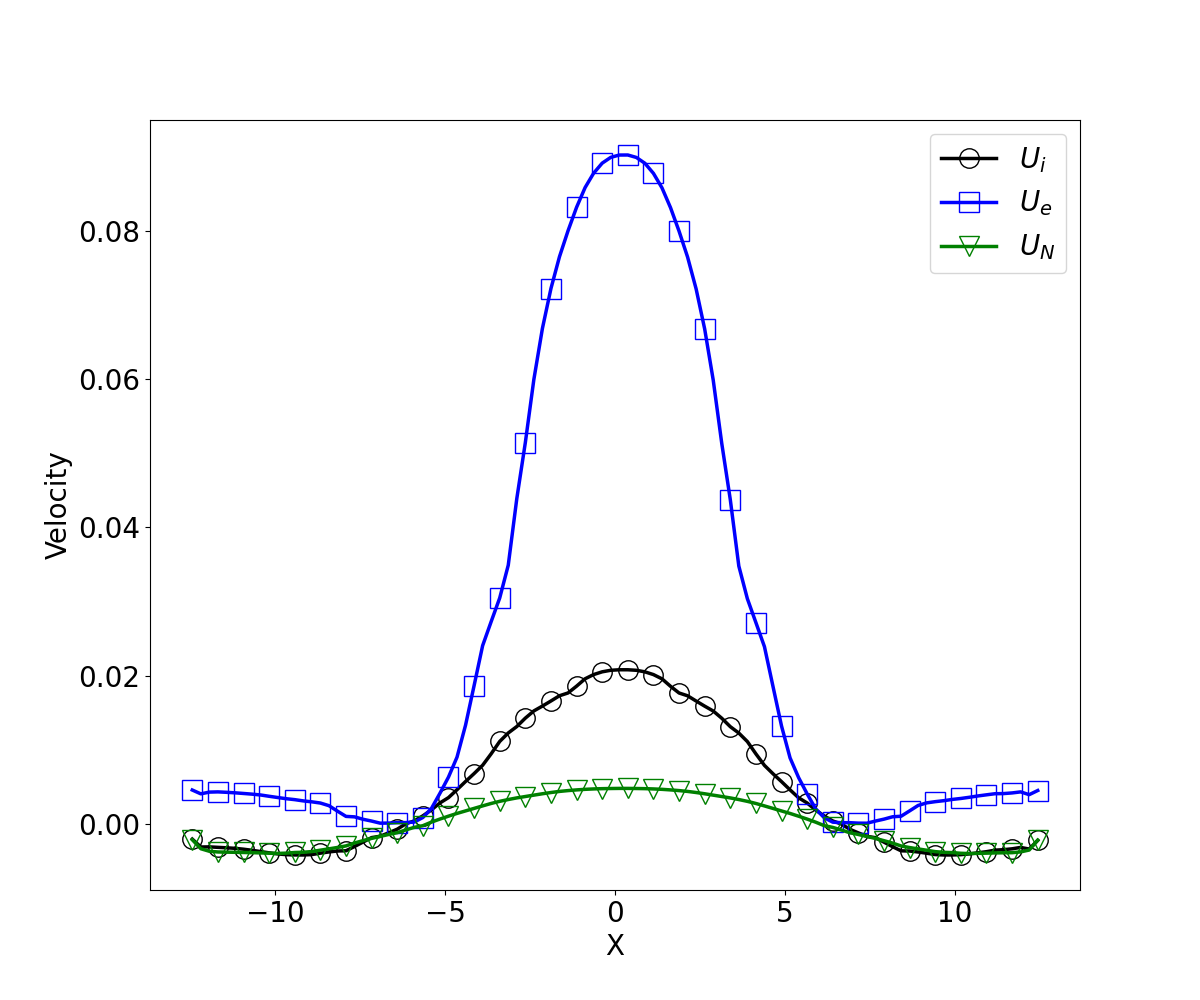}
    \caption{}
\end{subfigure}
\caption{Velocity of ions and elctrons at $Kn=0$ with uniform neutral background. (a) X-component, (b) y-component of electron and ion velocity when $\chi=1.0$. (c) X-component (d) y-component when $\chi=0.1$}
\label{fig:chicomparevelocity}
\end{figure}

Compared to Figure \ref{fig:GEM-Kn0-r1-B-streamplot}, the aspect ratio of the reconnection region is smaller in the current figure. Comparing Figure \ref{fig:chicompareflux} and Figure \ref{fig:GEM-Kn0-r1-reconnection-flux}, the reconnection rate is now much slower.
From Figure \ref{fig:chicomparevelocity}, it can be seen that in the strongly coupled case (a)(b), the velocity separation between species becomes very small, and thus the Hall effect is negligible. In this case, the reconnection process is supported more by resistivity. In the weakly coupled case (c)(d), the separation is relatively large, so both resistivity and the Hall effect contribute to the reconnection process.

\begin{figure}[H]
\centering
\begin{subfigure}[b]{0.48\textwidth}
    \centering
    \includegraphics[width=0.98\textwidth]{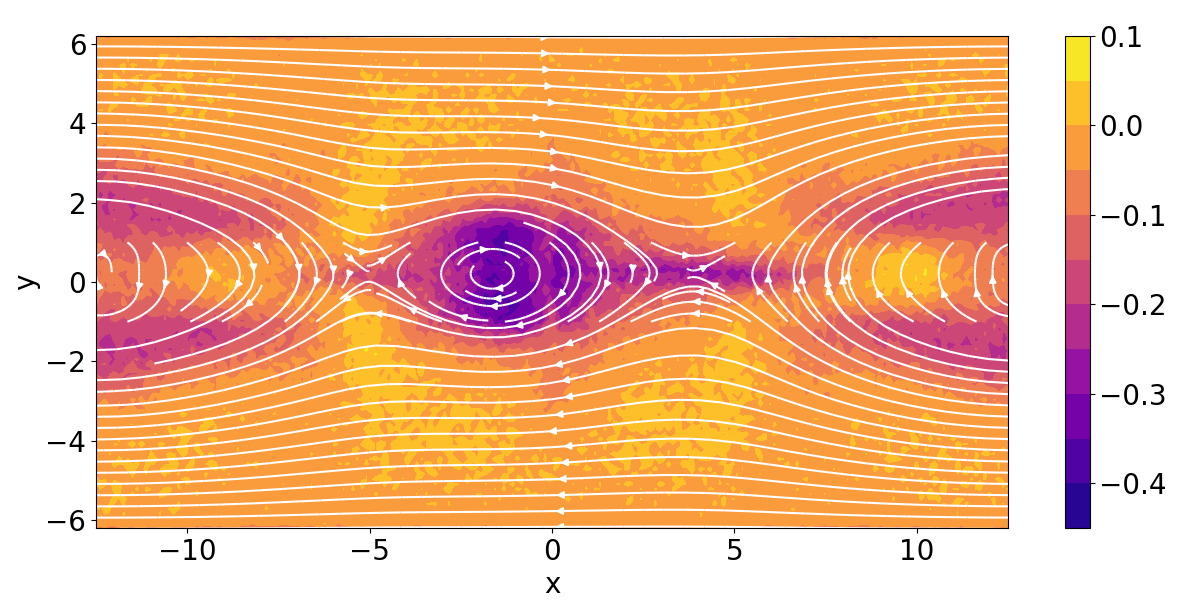}
    \caption{}
\end{subfigure}
\begin{subfigure}[b]{0.48\textwidth}
    \centering
    \includegraphics[width=0.98\textwidth]{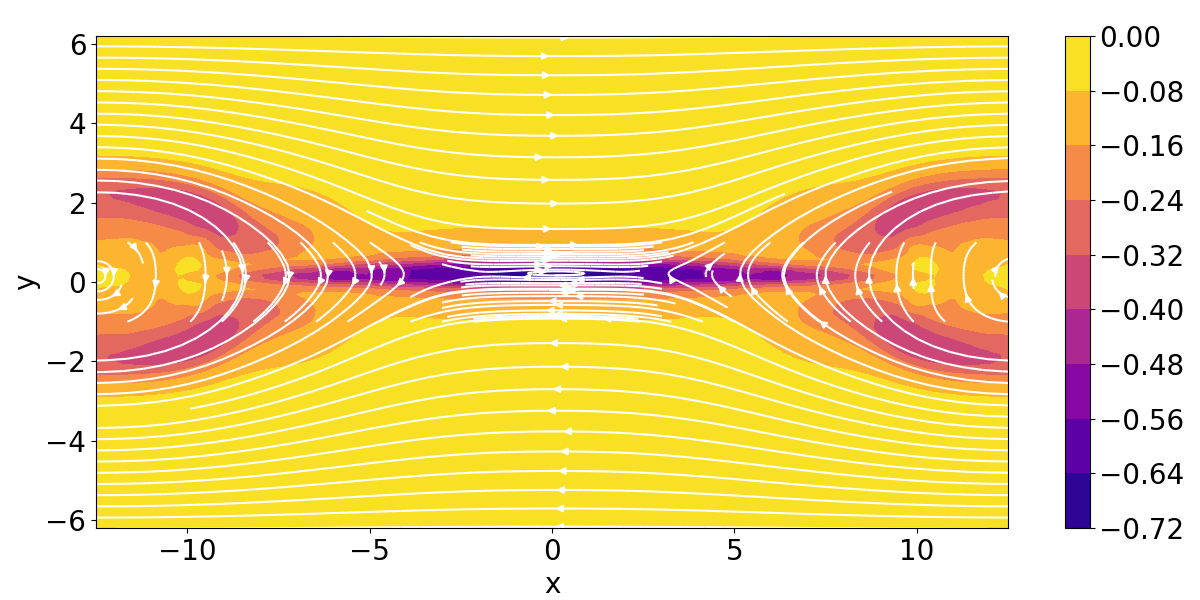}
    \caption{}
\end{subfigure}
\caption{(Magnetic lines (white solid line) and contours of current density $J_z$ of (a) $Kn=0.001$ (b) $Kn=0.0001$ at $\omega_{pi}\Delta t = 30$. A magnetic island is observed in the (a) which will accelerate the magnetic reconnection.}
\label{fig:Knreconnectioncompare}
\end{figure}

\begin{figure}
    \centering
    \includegraphics[width=0.6\textwidth]{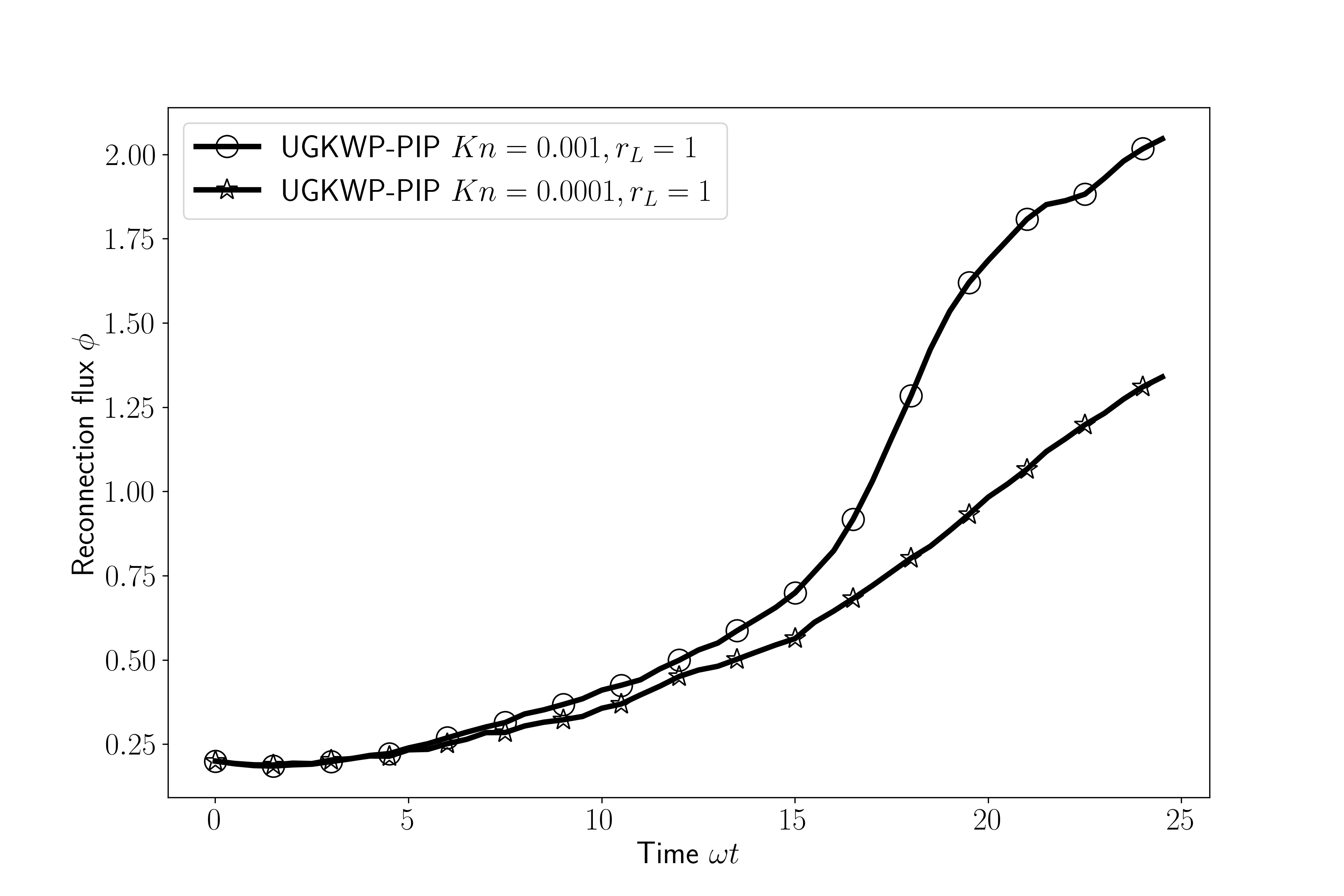}
    \caption{Magnetic reconnection flux of $Kn=0.001$ and $Kn=0.0001$. The magnetic island in $Kn=0.001$ case accelerate the reconnection rate. }
    \label{fig:KnreconnectionFlux}
\end{figure}

Figure \ref{fig:Knreconnectioncompare} shows the magnetic field lines (white solid lines) and current density $J_z$ at different Knudsen numbers. A magnetic island is observed in the case of $Kn=0.001$ compared to the $Kn=0.0001$ case. Two X-points accelerate the magnetic reconnection process compared to the single X-point in the $Kn=0.0001$ case. Figure \ref{fig:KnreconnectionFlux} clearly shows the different magnetic flux evolution between the $Kn=0.001$ and $Kn=0.0001$ cases. The higher Knudsen number of 0.001 appears to result in a faster reconnection rate compared to the lower Knudsen number of 0.0001.

\section{Conclusions}
\label{conclusions}

This work presents a kinetic model for multi-species transport and interaction among neutrals, electrons, and protons and a corresponding multiscale numerical method. The research explores various plasma physics phenomena across different flow regimes, demonstrating the method's multiscale capabilities and asymptotic-preserving properties.
First, the Landau damping problem is investigated in both collisionless and collisional regimes. The results obtained through this method agree well with those from the collisional PIC approach, validating its ability to accurately capture kinetic phenomena. Notably, in the strongly collisional regime, the method exhibits a significant advantage by overcoming the limitations imposed by collision time, resulting in substantially faster computation  compared to traditional PIC simulations.
Second, the multiscale Brio-Wu problem, a benchmark test in ideal MHD, is investigated. The method successfully recovers the wave structure characteristic of the ideal MHD system, demonstrating its capability for fluid descriptions of plasma. Moreover, the Brio-Wu shock tube is analyzed for a range of Knudsen numbers, with results aligning well with PIC simulations. The investigation is pushed further to explore cases with larger Larmor radius, revealing the algorithm's proficiency in capturing phenomena at both ion and electron inertia scales. Further, the Brio-Wu shock tube with a uniform neutral species background is explored, showing the method's ability to capture non-ideal effects in PIP such as ambipolar diffusion, where the magnetic field exhibits a diffusive structure instead of the discontinuous structure observed in the ideal MHD limit.
Third, the Orszag-Tang problem is solved to study the method's capability to capture MHD shocks. The problem is also examined for larger Knudsen numbers. Additionally, the magnetic reconnection problem is investigated, serving as a test of the method's multiscale capabilities. The results successfully reproduce phenomena across various scales, from resistive-MHD and Hall-MHD to fully kinetic effects, proving the method's ability as a powerful tool for studying complex multiscale plasma dynamics.
In conclusion, this work presents a computational method that bridges the gap between kinetic and fluid descriptions of plasma. Its ability to handle a wide range of spatial and temporal scales, from collisionless to strongly collisional regimes, and from ideal MHD to kinetic scales, makes it a potential tool for multiscale simulation of PIP in the field of plasma physics.

\section*{Acknowledgements}

This current work was supported by National Key R$\&$D Program of China (Grant Nos. 2022YFA1004500), 
National Natural Science Foundation of China (12172316, 92371107),
and Hong Kong research grant council (16208021,16301222, 16208324).

\appendix
\section{Collisional PIC method}
\label{collisional PIC}

In this section, the collisional version of the Particle-In-Cell (PIC) method is introduced. First, the general concept and main components of the classical collisionless PIC method are reviewed. Then, the particle loading method and collision procedures are described. Finally, the general procedure of the collisional PIC method is outlined.

The PIC method can be viewed as a representation of the distribution function of each species by a superposition of moving computational particles, where each element represents a cloud of physical particles. The mathematical formulation of the PIC method is obtained by assuming that the distribution function of each species is given by the superposition of elements computational particles:
$$
f_s(\boldsymbol{x}, \boldsymbol{v}, t)=\sum_P f_P(\boldsymbol{x}, \boldsymbol{v}, t) ,
$$
where each computational particles $P$ represents a large number of physical particles that are near each other in phase space.The PIC method is based upon assigning to each computational particle a specific functional form for its distribution. In the classical PIC method, the functional dependence is assumed to be a tensor product of the shape in each direction of phase space \cite{hockney2021computer}:
\begin{equation}
f_P(\boldsymbol{x}, \boldsymbol{v}, t)=N_P S_{\boldsymbol{x}}\left(\boldsymbol{x}-\boldsymbol{x}_P(t)\right) S_{\boldsymbol{v}}\left(\boldsymbol{v}-\boldsymbol{v}_P(t)\right),
\label{eqn:shape function}
\end{equation}
where $S_{\mathrm{x}}$ and $S_{\mathrm{v}}$ are the shape functions for the computational particles and $N_P$ is the number of physical particles that are present in the element of phase space represented by the computational particle. For the velocity, $S_v$, the most common choice is a Dirac's delta in each direction:
$$
S_v\left(\boldsymbol{v}-\boldsymbol{v}_P\right)=\delta\left(v_x-v_{x p}\right) \delta\left(v_y-v_{y p}\right) \delta\left(v_z-v_{z p}\right) .
$$
Based on the $b$-splines, the spatial shape function of PIC methods is chosen as
\begin{equation}
S_{\boldsymbol{x}}\left(\boldsymbol{x}-\boldsymbol{x}_P\right)=\frac{1}{\Delta x_P \Delta y_P \Delta z_P} b_{\ell}\left(\frac{x-x_P}{\Delta x_P}\right) b_{\ell}\left(\frac{y-y_P}{\Delta y_P}\right) b_{\ell}\left(\frac{z-z_P}{\Delta z_P}\right),
\label{eqn:pic_shape}
\end{equation}
where $\Delta x_P, \Delta y_P$, and $\Delta z_P$ are the lengths of the computational particles in each spatial dimension.

The equation of motion for each particle $f_P$ is the Vlasov equation
\begin{equation}
\frac{\partial f_P}{\partial t}+\boldsymbol{v} \cdot \frac{\partial f_P}{\partial \boldsymbol{x}}+\frac{q_s}{m_s}(\boldsymbol{E}+\boldsymbol{v} \times \boldsymbol{B}) \cdot \frac{\partial f_P}{\partial \boldsymbol{v}}=0.
\label{eqn:vlasov}
\end{equation}
Take the moment, the evolution equations is obtained,
$$
\begin{aligned}
& \frac{\mathrm{d} N_P}{\mathrm{~d} t}=0, \\
& \frac{\mathrm{d} \boldsymbol{x}_P}{\mathrm{~d} t}=\boldsymbol{v}_P, \\
& \frac{\mathrm{d} \boldsymbol{v}_P}{\mathrm{~d} t}=\frac{q_P}{m_P}\left(\boldsymbol{E}_P+\boldsymbol{v}_P \times \boldsymbol{B}_P\right) .
\end{aligned}
$$
A great advantage of moment equations is that the evolution equations resemble the same Newtonian equations as followed by the regular physical particles. The Leap-frog scheme is used to numerically solve the above equation:
$$
\begin{aligned}
& \boldsymbol{x}_P^{n+1}=\boldsymbol{x}_P^{n}+\Delta t \boldsymbol{v}_P^{n+1/2}, \\
& \boldsymbol{v}_P^{n+1/2} =\frac{q_P \Delta t}{m_P}\left(\boldsymbol{E}_P^k+\frac{\boldsymbol{v}_P^{n+1/2}+\boldsymbol{v}_P^{n-1/2}}{2} \times \boldsymbol{B}_P\right) .
\end{aligned}
$$
Boris method is used to deal with the effect of lorentz force. The field is computed as average over the shape function based on the definition of $\boldsymbol{E}_P$ and $\boldsymbol{B}_P$ as
\begin{equation}
\begin{aligned}
& \boldsymbol{E}_P=\int S_{\boldsymbol{x}}\left(\boldsymbol{x}-\boldsymbol{x}_P\right) \boldsymbol{E}(\boldsymbol{x}) \mathrm{d} \boldsymbol{x}, \\
& \boldsymbol{B}_P=\int S_{\boldsymbol{x}}\left(\boldsymbol{x}-\boldsymbol{x}_P\right) \boldsymbol{B}(\boldsymbol{x}) \mathrm{d} \boldsymbol{x}.
\end{aligned}
\label{eqn:maxshape}
\end{equation}
The set of equations presented above provides a self-contained description for the Vlasov equation. When combined with an algorithm to solve Maxwell's equations, the full Vlasov-Maxwell system can then be solved. A wide range of numerical methods can be employed to solve the electromagnetic field equations. The crucial step is to use shape functions to deposit the charge and current densities at the grid points as follows,
$$
\begin{aligned}
& \rho_g=\sum_P \frac{1}{V_g} \int_{V_g} S_{\boldsymbol{x}}\left(\boldsymbol{x}-\boldsymbol{x}_P\right) \mathrm{d} \boldsymbol{x}, \\
& \boldsymbol{J}_g=\sum_P \boldsymbol{v}_P \frac{1}{V_g} \int_{V_g} S_{\boldsymbol{x}}\left(\boldsymbol{x}-\boldsymbol{x}_P\right) \mathrm{d} \boldsymbol{x},
\end{aligned}
$$
which then serve as the source terms for the field equations.

The particle loading procedure is closely tied to the application of the initial conditions. Generally, a random approach is adopted to load the particles, such as using the Box-Muller method to sample numbers from a Gaussian distribution. This "noise start"(NS) technique is commonly employed in the context of PIC methods. In this approach, the particle velocities are randomly sampled around the macroscopic mean velocity, with the standard deviation controlled by the temperature. Another similar method is the acceptance-rejection method, which can be used to load particles from any desired distribution function. However, these random loading techniques will inevitably introduce a certain level of numerical noise into the initial conditions. An alternative loading method is the "quiet start" (QS) approach, where the targeted distribution function is approximated by a prescribed series of numbers. This method can generate an initial particle loading that is as smooth as possible. In this work, we utilize bit-reversed numbers \cite{birdsall2018plasma} for the QS method.

This equation indicates that a particle has a probability of $1 - \exp(-\Delta t / \tau)$ to undergo a collision with others within a given cell. The following procedure is used to handle the collision process: First, the collision probability $P_j$ is calculated in each cell based on the macroscopic quantities. Then, for each particle, a random number $R_P$ is generated. If $R_P < P_j$, the particle will experience a collision and be randomly assigned a new velocity drawn from the Maxwellian distribution. After all particles have been processed in this manner, a correction procedure is applied, as described in \cite{liu2020discrete}.

\section{Ohm's law in weakly ionized limit}
\label{ohms law in weakly ionized limit}

First introduce the velocity of a charged component $c$ relative to the neutrals \cite{nakano1986dissipation}
$$
\boldsymbol{V}_{c} = \boldsymbol{U}_{c} - \boldsymbol{U}_n,
$$
and the viscous damping time of motion relative to the neutrals
$$
\tau_{c} = \frac{\rho_c}{2\mu_{cn} n_c n_n \chi_{cn}} = \frac{m_n+m_c}{m_n}\frac{1}{2\chi_{cn}},
$$
we can also write
$$
\boldsymbol{h}_{\alpha} \equiv -\nabla\boldsymbol{P}_\alpha - \frac{d\rho_\alpha \boldsymbol{U}_\alpha}{dt}.
$$
Then for charged components and neutral component, we have
\begin{equation}
    \begin{aligned}
    &q_cn_c(\boldsymbol{E}+\boldsymbol{V}_c\times\boldsymbol{B}) - \frac{\rho_c}{\tau_c}\boldsymbol{V}_c + \boldsymbol{h}_c = 0,\\
    &\sum_c \frac{\rho_c}{\tau_c} + \boldsymbol{h}_n = 0.
    \end{aligned}
    \label{eq:h description}
\end{equation}
From the equation of motion for electrons, we have
$$
\boldsymbol{E}=-\boldsymbol{V}_e\times \boldsymbol{B} - \frac{m_e}{q_e\tau_e}\boldsymbol{V}_e + \frac{1}{q_en_e}\boldsymbol{h}_e,
$$
Substitute $\boldsymbol{E}$ into Eq.\eqref{eq:h description} for a charged species other than the electron, we have
\begin{equation}
    \omega_i\boldsymbol{V}_i\times \boldsymbol{e}_{B} - \frac{1}{\tau_i}\boldsymbol{V}_i + \frac{1}{\rho_i}\boldsymbol{h}_i - \omega_i\boldsymbol{V}_e\times\boldsymbol{e}_B - \frac{q_im_e}{q_em_i}\frac{1}{\tau_e}\boldsymbol{V}_e + \frac{q_i}{q_e}\frac{1}{n_em_i}\boldsymbol{h}_e = 0,
\end{equation}
where
$
\omega_i =\frac{q_iB}{m_i}
$
is the gyrofrequency of charged ions and $\boldsymbol{e}_B = \boldsymbol{B}/B$ is the unit vector in the direction of magnetic field. Suppose $\boldsymbol{B}=B\boldsymbol{e}_z$, then the velocity in the $(x,y)$ plane can be solved as
\begin{equation}
    \boldsymbol{V}_i = \mathbb{M}_i\left(
        \frac{1}{\rho_i}\boldsymbol{h}_i - \omega_i\boldsymbol{V}_e\times\boldsymbol{e}_B - \frac{q_im_e}{q_em_i}\frac{1}{\tau_e}\boldsymbol{V}_e + \frac{q_i}{q_e}\frac{1}{n_em_i}\boldsymbol{h}_e
    \right),
    \label{eq:vieqn}
\end{equation}
where $\mathbb{I}$ is identity matrix,
$$
\mathbb{M}_i = \frac{1}{\Omega_i^2}\left(\frac{1}{\tau_i}\mathbb{I}+\omega_i\mathbb{J}\right),\quad \Omega_i^2 = \frac{1}{\tau_i^2} +\omega_i^2,\quad \mathbb{J}=\left(\begin{array}{cc}
    0 & 1 \\
    -1 & 0
\end{array}\right).
$$
Substitute the above equation into Eq.\eqref{eq:h description}, we have
$$
\begin{aligned}
\left(A_1\right. & \left.+\frac{A_2}{\tau_{\mathrm{e}} \omega_{\mathrm{e}}}\right) \boldsymbol{V}_{\mathrm{e}}-\left(A_2-\frac{A_1}{\tau_{\mathrm{e}} \omega_{\mathrm{e}}}\right) \mathbb{J} \boldsymbol{V}_{\mathrm{e}} \\
& =\boldsymbol{J} \times \boldsymbol{B}+\sum_c\left(\frac{\omega_c}{\Omega_c}\right)^2 \boldsymbol{h}_c-\sum_c \frac{\omega_c}{\tau_c \Omega_c^2} \mathbb{J} \boldsymbol{h}_c+\frac{A_2}{\rho_{\mathrm{e}} \omega_{\mathrm{e}}} \boldsymbol{h}_{\mathrm{e}}+\frac{A_1}{\rho_{\mathrm{e}} \omega_{\mathrm{e}}} \mathbb{J} \boldsymbol{h}_{\mathrm{e}},
\end{aligned}
$$
where
$$
A_1=\sum_c \frac{\rho_c \omega_c^2}{\tau_c \Omega_c^2}, \quad A_2=\sum_c \frac{\rho_c \omega_c}{\tau_c^2 \Omega_c^2} .
$$
Ignore all the inertia and pressure term, we achieve the simple form as
\begin{equation}
\left(A_1+\frac{A_2}{\tau_{\mathrm{e}} \omega_{\mathrm{e}}}\right) \boldsymbol{V}_{\mathrm{e}}-\left(A_2-\frac{A_1}{\tau_{\mathrm{e}} \omega_{\mathrm{e}}}\right) \mathbb{J} \boldsymbol{V}_{\mathrm{e}} =\boldsymbol{J} \times \boldsymbol{B}.
\label{eq:solelecV}
\end{equation}

Now neglecting the $\boldsymbol{h}_e$ in the equation of motion for electrons,
$$
\boldsymbol{E} = -(\boldsymbol{V}_e  + \frac{1}{\tau_e\omega_e}\mathbb{J}\boldsymbol{V}_e)\times\boldsymbol{B}
$$
and substitute Eq.\eqref{eq:solelecV} into the above equation, we finally have the Ohm's law as
\begin{equation}
    \boldsymbol{E} = -\boldsymbol{U}_n\times \boldsymbol{B} + \frac{1}{\sigma_c}\boldsymbol{J} + \beta\boldsymbol{J}\times\boldsymbol{B}-\xi(\boldsymbol{J}\times\boldsymbol{B})\times\boldsymbol{B},
\end{equation}
where $\sigma_c$ is the electrical conductivity along the magnetic field
$$
\sigma_c = \sum_c\sigma_c = \sum_c \frac{q_c^2\tau_cn_c}{m_c},
$$
and
$$
\beta =\frac{B}{c^2} \frac{A_2}{A},\quad \xi  =\frac{1}{c^2} \frac{A_1}{A}-\frac{1}{B^2 \sigma_c},\quad A =A_1^2+A_2^2.
$$

Based on the solution, the induction equation can be written as \cite{o2006explicit}
\begin{equation}
    \frac{\partial \boldsymbol{B}}{\partial t} + \frac{\partial \boldsymbol{M}}{\partial t} = \frac{\partial}{\partial x}\mathbb{R}\frac{\partial \boldsymbol{B}}{\partial x},
    \label{eq:induction_weaklyionized}
\end{equation}
where
$$
\boldsymbol{M} = (u_nB_y-v_nB_x, u_nB_z-w_nB_x)^T,
$$
and the resistive matrix is
\begin{equation}
\begin{aligned}
& \mathbb{R}=
 \left(\begin{array}{ll}
\left(r_{\mathrm{O}}-r_{\mathrm{A}} \frac{B_2^2}{B^2}+r_{\mathrm{A}}\right. & \left(r_{\mathrm{A}}-r_{\mathrm{O}}\right) \frac{B_y B_z}{B^2}+r_{\mathrm{H}} \frac{B_x}{B} \\
\left(r_{\mathrm{A}}-r_{\mathrm{O}}\right) \frac{B_y B_z}{B^2}-r_{\mathrm{H}} \frac{B_x}{B} & \left(r_{\mathrm{O}}-r_{\mathrm{A}}\right) \frac{B_y^2}{B^2}+r_{\mathrm{A}}
\end{array}\right),
\end{aligned}
\label{eq:diffusivitycoef}
\end{equation}
where $r_{\mathrm{O}}, r_{\mathrm{H}}$ and $r_{\mathrm{A}}$ are the Ohmic, Hall and ambipolar resistivities respectively, and are defined by
$$
\begin{aligned}
& r_{\mathrm{O}}=\frac{1}{\sigma_{\mathrm{O}}}, \\
& r_{\mathrm{H}}=\frac{\sigma_{\mathrm{H}}}{\sigma_{\mathrm{H}}^2+\sigma_{\mathrm{A}}^2}, \\
& r_{\mathrm{A}}=\frac{\sigma_{\mathrm{A}}}{\sigma_{\mathrm{H}}^2+\sigma_{\mathrm{A}}^2},
\end{aligned}
$$
with conductivities
$$
\begin{aligned}
& \sigma_{\mathrm{O}}=\sum q_cn_c\beta_c, \\
& \sigma_{\mathrm{H}}=\frac{1}{\boldsymbol{B}} \sum_{c} \frac{q_cn_c}{1+\beta_c^2}, \\
& \sigma_{\mathrm{A}}=\frac{1}{B} \sum_{c} \frac{q_cn_c\beta_c}{1+\beta_c^2},
\end{aligned}
$$
where the Hall parameter for species $i$ is given by
$$
\beta_c=\frac{q_c \boldsymbol{B}}{k_{cn} m_c\rho_n}, \quad k_{cn} = \frac{2\chi_{cn}}{m_c+m_n}.
$$

\bibliographystyle{elsarticle-num}
\bibliography{UGKWP-PIC}
\end{document}